\documentclass[longauth]{aaEC}
\usepackage{graphicx}
\usepackage{natbib}
\usepackage{scalerel}
\usepackage{xspace}
\usepackage{ulem}

\usepackage[table]{xcolor}

\usepackage{txfonts}
\usepackage[pdfencoding=auto,psdextra]{hyperref}
\hypersetup{
    colorlinks=true,
    linkcolor=blue,
    filecolor=magenta,      
    urlcolor=blue,
    citecolor=blue
}
\makeatletter
\renewcommand*\aa@pageof{, page \thepage{} of \pageref*{LastPage}}
\makeatother

\usepackage[utf8]{inputenc}

\usepackage[switch, modulo]{lineno}

\usepackage{euclid}

\newcommand{\vx}{\ensuremath{\vec{x}}\xspace}

\newcommand{\mc}[1]{\ensuremath{\mathcal{#1}}\xspace}
\newcommand{\nh}{\ensuremath{\hat{\vec n}}\xspace}

\newcommand{\hmpc}{\ensuremath{h^{-1}\,\text{Mpc}}\xspace}
\newcommand{\mpc}{\ensuremath{\text{Mpc}}\xspace}
\newcommand{\kmpc}{\ensuremath{h\,\text{Mpc}^{-1}}\xspace}

\newcommand{\cgpc}{\ensuremath{h^{-3}\,\mathrm{Gpc}^3}\xspace}

\newcommand{\Comp}{\ensuremath{{\cal{C}}}\xspace}
\newcommand{\Compa}{\ensuremath{\hat{\cal C}}\xspace}
\newcommand{\flux}{erg s$^{-1}$ cm$^{-2}$\xspace}
\newcommand{\pypelid}{\texttt{pypelid}\xspace}
\newcommand{\pinocchio}{\texttt{pinocchio}\xspace}
\newcommand{\healpix}{\texttt{healpix}\xspace}
\newcommand{\tdeg}{30\degree\xspace}
\newcommand{\epixs}{$e$ pix$^{-1}$ s$^{-1}$\xspace}
\newcommand{\sqdeg}{deg$^2$\xspace}

\newcommand{\oiii}{[\ion{O}{III}]\xspace}
\newcommand{\oiiib}{[\ion{O}{III}]b\xspace}
\newcommand{\siii}{[\ion{S}{III}]\xspace}

\newcommand{\nii}{[\ion{N}{II}]\xspace}
\renewcommand{\micron}{$\mu$m\xspace}

\newcommand{\citepype}{Euclid Collaboration: Granett et al. (in prep.)\xspace}

\newcommand{\citepk}{Euclid Collaboration: Salvalaggio et al. (in prep.)\xspace}
\newcommand{\citeppk}{(Euclid Collaboration: Salvalaggio et al., in prep.)\xspace}
\newcommand{\citelee}{Euclid Collaboration: Lee et al. (in prep.)\xspace}

\newcommand{\citebruton}{Euclid Collaboration: Bruton et al. (in prep.)\xspace}
\newcommand{\citepbruton}{(Euclid Collaboration: Bruton et al., in prep.)\xspace}
\newcommand{\citepassa}{Euclid Collaboration: Passalacqua et al. (in prep.)\xspace}
\newcommand{\citeppassa}{(Euclid Collaboration: Passalacqua et al., in prep.)\xspace}
\newcommand{\citeapassa}{Euclid Collaboration: Passalacqua et al., in prep.\xspace}

\begin{document}
\title{\Euclid preparation}
\subtitle{XCII. Controlling angular systematics in the Euclid spectroscopic galaxy sample}

\newcommand{\orcid}[1]{} 
\author{Euclid Collaboration: P.~Monaco\orcid{0000-0003-2083-7564}\thanks{\email{pierluigi.monaco@inaf.it}}\inst{\ref{aff1},\ref{aff2},\ref{aff3},\ref{aff4},\ref{aff5}}
\and M.~Y.~Elkhashab\orcid{0000-0001-9306-2603}\inst{\ref{aff2},\ref{aff3},\ref{aff1},\ref{aff4}}
\and B.~R.~Granett\orcid{0000-0003-2694-9284}\inst{\ref{aff6}}
\and J.~Salvalaggio\orcid{0000-0002-1431-5607}\inst{\ref{aff2},\ref{aff4},\ref{aff1},\ref{aff3}}
\and E.~Sefusatti\orcid{0000-0003-0473-1567}\inst{\ref{aff2},\ref{aff4},\ref{aff3}}
\and C.~Scarlata\orcid{0000-0002-9136-8876}\inst{\ref{aff7}}
\and B.~Zabelle\orcid{0000-0002-7830-363X}\inst{\ref{aff7}}
\and M.~Bethermin\orcid{0000-0002-3915-2015}\inst{\ref{aff8}}
\and S.~Bruton\orcid{0000-0002-6503-5218}\inst{\ref{aff9}}
\and C.~Carbone\orcid{0000-0003-0125-3563}\inst{\ref{aff10}}
\and S.~de~la~Torre\inst{\ref{aff11}}
\and S.~Dusini\orcid{0000-0002-1128-0664}\inst{\ref{aff12}}
\and A.~Eggemeier\orcid{0000-0002-1841-8910}\inst{\ref{aff13}}
\and L.~Guzzo\orcid{0000-0001-8264-5192}\inst{\ref{aff14},\ref{aff6},\ref{aff15}}
\and G.~Lavaux\orcid{0000-0003-0143-8891}\inst{\ref{aff16}}
\and S.~Lee\orcid{0000-0002-8289-740X}\inst{\ref{aff17}}
\and K.~Markovic\orcid{0000-0001-6764-073X}\inst{\ref{aff17}}
\and K.~S.~McCarthy\orcid{0000-0001-6857-018X}\inst{\ref{aff17},\ref{aff18}}
\and M.~Moresco\orcid{0000-0002-7616-7136}\inst{\ref{aff19},\ref{aff20}}
\and F.~Passalacqua\orcid{0000-0002-8606-4093}\inst{\ref{aff21},\ref{aff12}}
\and W.~J.~Percival\orcid{0000-0002-0644-5727}\inst{\ref{aff22},\ref{aff23},\ref{aff24}}
\and I.~Risso\orcid{0000-0003-2525-7761}\inst{\ref{aff6},\ref{aff25}}
\and A.~G.~S\'anchez\orcid{0000-0003-1198-831X}\inst{\ref{aff26}}
\and D.~Scott\orcid{0000-0002-6878-9840}\inst{\ref{aff27}}
\and C.~Sirignano\orcid{0000-0002-0995-7146}\inst{\ref{aff21},\ref{aff12}}
\and Y.~Wang\orcid{0000-0002-4749-2984}\inst{\ref{aff28}}
\and B.~Altieri\orcid{0000-0003-3936-0284}\inst{\ref{aff29}}
\and S.~Andreon\orcid{0000-0002-2041-8784}\inst{\ref{aff6}}
\and N.~Auricchio\orcid{0000-0003-4444-8651}\inst{\ref{aff20}}
\and C.~Baccigalupi\orcid{0000-0002-8211-1630}\inst{\ref{aff4},\ref{aff2},\ref{aff3},\ref{aff30}}
\and M.~Baldi\orcid{0000-0003-4145-1943}\inst{\ref{aff31},\ref{aff20},\ref{aff32}}
\and S.~Bardelli\orcid{0000-0002-8900-0298}\inst{\ref{aff20}}
\and A.~Biviano\orcid{0000-0002-0857-0732}\inst{\ref{aff2},\ref{aff4}}
\and E.~Branchini\orcid{0000-0002-0808-6908}\inst{\ref{aff33},\ref{aff25},\ref{aff6}}
\and M.~Brescia\orcid{0000-0001-9506-5680}\inst{\ref{aff34},\ref{aff35}}
\and J.~Brinchmann\orcid{0000-0003-4359-8797}\inst{\ref{aff36},\ref{aff37},\ref{aff38}}
\and S.~Camera\orcid{0000-0003-3399-3574}\inst{\ref{aff39},\ref{aff40},\ref{aff41}}
\and G.~Ca\~nas-Herrera\orcid{0000-0003-2796-2149}\inst{\ref{aff42},\ref{aff43}}
\and V.~Capobianco\orcid{0000-0002-3309-7692}\inst{\ref{aff41}}
\and V.~F.~Cardone\inst{\ref{aff44},\ref{aff45}}
\and J.~Carretero\orcid{0000-0002-3130-0204}\inst{\ref{aff46},\ref{aff47}}
\and S.~Casas\orcid{0000-0002-4751-5138}\inst{\ref{aff48},\ref{aff49}}
\and F.~J.~Castander\orcid{0000-0001-7316-4573}\inst{\ref{aff50},\ref{aff51}}
\and M.~Castellano\orcid{0000-0001-9875-8263}\inst{\ref{aff44}}
\and G.~Castignani\orcid{0000-0001-6831-0687}\inst{\ref{aff20}}
\and S.~Cavuoti\orcid{0000-0002-3787-4196}\inst{\ref{aff35},\ref{aff52}}
\and A.~Cimatti\inst{\ref{aff53}}
\and C.~Colodro-Conde\inst{\ref{aff54}}
\and G.~Congedo\orcid{0000-0003-2508-0046}\inst{\ref{aff55}}
\and C.~J.~Conselice\orcid{0000-0003-1949-7638}\inst{\ref{aff56}}
\and L.~Conversi\orcid{0000-0002-6710-8476}\inst{\ref{aff57},\ref{aff29}}
\and Y.~Copin\orcid{0000-0002-5317-7518}\inst{\ref{aff58}}
\and F.~Courbin\orcid{0000-0003-0758-6510}\inst{\ref{aff59},\ref{aff60},\ref{aff61}}
\and H.~M.~Courtois\orcid{0000-0003-0509-1776}\inst{\ref{aff62}}
\and H.~Degaudenzi\orcid{0000-0002-5887-6799}\inst{\ref{aff63}}
\and G.~De~Lucia\orcid{0000-0002-6220-9104}\inst{\ref{aff2}}
\and H.~Dole\orcid{0000-0002-9767-3839}\inst{\ref{aff64}}
\and F.~Dubath\orcid{0000-0002-6533-2810}\inst{\ref{aff63}}
\and C.~A.~J.~Duncan\orcid{0009-0003-3573-0791}\inst{\ref{aff55}}
\and X.~Dupac\inst{\ref{aff29}}
\and S.~Escoffier\orcid{0000-0002-2847-7498}\inst{\ref{aff65}}
\and M.~Farina\orcid{0000-0002-3089-7846}\inst{\ref{aff66}}
\and R.~Farinelli\inst{\ref{aff20}}
\and S.~Ferriol\inst{\ref{aff58}}
\and N.~Fourmanoit\orcid{0009-0005-6816-6925}\inst{\ref{aff65}}
\and M.~Frailis\orcid{0000-0002-7400-2135}\inst{\ref{aff2}}
\and E.~Franceschi\orcid{0000-0002-0585-6591}\inst{\ref{aff20}}
\and M.~Fumana\orcid{0000-0001-6787-5950}\inst{\ref{aff10}}
\and S.~Galeotta\orcid{0000-0002-3748-5115}\inst{\ref{aff2}}
\and K.~George\orcid{0000-0002-1734-8455}\inst{\ref{aff67}}
\and W.~Gillard\orcid{0000-0003-4744-9748}\inst{\ref{aff65}}
\and B.~Gillis\orcid{0000-0002-4478-1270}\inst{\ref{aff55}}
\and C.~Giocoli\orcid{0000-0002-9590-7961}\inst{\ref{aff20},\ref{aff32}}
\and J.~Gracia-Carpio\inst{\ref{aff26}}
\and A.~Grazian\orcid{0000-0002-5688-0663}\inst{\ref{aff68}}
\and F.~Grupp\inst{\ref{aff26},\ref{aff69}}
\and S.~V.~H.~Haugan\orcid{0000-0001-9648-7260}\inst{\ref{aff70}}
\and W.~Holmes\inst{\ref{aff17}}
\and F.~Hormuth\inst{\ref{aff71}}
\and A.~Hornstrup\orcid{0000-0002-3363-0936}\inst{\ref{aff72},\ref{aff73}}
\and K.~Jahnke\orcid{0000-0003-3804-2137}\inst{\ref{aff74}}
\and M.~Jhabvala\inst{\ref{aff75}}
\and B.~Joachimi\orcid{0000-0001-7494-1303}\inst{\ref{aff76}}
\and E.~Keih\"anen\orcid{0000-0003-1804-7715}\inst{\ref{aff77}}
\and S.~Kermiche\orcid{0000-0002-0302-5735}\inst{\ref{aff65}}
\and A.~Kiessling\orcid{0000-0002-2590-1273}\inst{\ref{aff17}}
\and B.~Kubik\orcid{0009-0006-5823-4880}\inst{\ref{aff58}}
\and M.~K\"ummel\orcid{0000-0003-2791-2117}\inst{\ref{aff69}}
\and M.~Kunz\orcid{0000-0002-3052-7394}\inst{\ref{aff78}}
\and H.~Kurki-Suonio\orcid{0000-0002-4618-3063}\inst{\ref{aff79},\ref{aff80}}
\and A.~M.~C.~Le~Brun\orcid{0000-0002-0936-4594}\inst{\ref{aff81}}
\and S.~Ligori\orcid{0000-0003-4172-4606}\inst{\ref{aff41}}
\and P.~B.~Lilje\orcid{0000-0003-4324-7794}\inst{\ref{aff70}}
\and V.~Lindholm\orcid{0000-0003-2317-5471}\inst{\ref{aff79},\ref{aff80}}
\and I.~Lloro\orcid{0000-0001-5966-1434}\inst{\ref{aff82}}
\and G.~Mainetti\orcid{0000-0003-2384-2377}\inst{\ref{aff83}}
\and D.~Maino\inst{\ref{aff14},\ref{aff10},\ref{aff15}}
\and E.~Maiorano\orcid{0000-0003-2593-4355}\inst{\ref{aff20}}
\and O.~Mansutti\orcid{0000-0001-5758-4658}\inst{\ref{aff2}}
\and S.~Marcin\inst{\ref{aff84}}
\and O.~Marggraf\orcid{0000-0001-7242-3852}\inst{\ref{aff13}}
\and M.~Martinelli\orcid{0000-0002-6943-7732}\inst{\ref{aff44},\ref{aff45}}
\and N.~Martinet\orcid{0000-0003-2786-7790}\inst{\ref{aff11}}
\and F.~Marulli\orcid{0000-0002-8850-0303}\inst{\ref{aff19},\ref{aff20},\ref{aff32}}
\and R.~J.~Massey\orcid{0000-0002-6085-3780}\inst{\ref{aff85}}
\and E.~Medinaceli\orcid{0000-0002-4040-7783}\inst{\ref{aff20}}
\and S.~Mei\orcid{0000-0002-2849-559X}\inst{\ref{aff86},\ref{aff87}}
\and Y.~Mellier\thanks{Deceased}\inst{\ref{aff88},\ref{aff16}}
\and M.~Meneghetti\orcid{0000-0003-1225-7084}\inst{\ref{aff20},\ref{aff32}}
\and E.~Merlin\orcid{0000-0001-6870-8900}\inst{\ref{aff44}}
\and G.~Meylan\inst{\ref{aff89}}
\and A.~Mora\orcid{0000-0002-1922-8529}\inst{\ref{aff90}}
\and L.~Moscardini\orcid{0000-0002-3473-6716}\inst{\ref{aff19},\ref{aff20},\ref{aff32}}
\and C.~Neissner\orcid{0000-0001-8524-4968}\inst{\ref{aff91},\ref{aff47}}
\and S.-M.~Niemi\orcid{0009-0005-0247-0086}\inst{\ref{aff42}}
\and C.~Padilla\orcid{0000-0001-7951-0166}\inst{\ref{aff91}}
\and S.~Paltani\orcid{0000-0002-8108-9179}\inst{\ref{aff63}}
\and F.~Pasian\orcid{0000-0002-4869-3227}\inst{\ref{aff2}}
\and K.~Pedersen\inst{\ref{aff92}}
\and V.~Pettorino\orcid{0000-0002-4203-9320}\inst{\ref{aff42}}
\and S.~Pires\orcid{0000-0002-0249-2104}\inst{\ref{aff93}}
\and G.~Polenta\orcid{0000-0003-4067-9196}\inst{\ref{aff94}}
\and M.~Poncet\inst{\ref{aff95}}
\and L.~A.~Popa\inst{\ref{aff96}}
\and L.~Pozzetti\orcid{0000-0001-7085-0412}\inst{\ref{aff20}}
\and F.~Raison\orcid{0000-0002-7819-6918}\inst{\ref{aff26}}
\and A.~Renzi\orcid{0000-0001-9856-1970}\inst{\ref{aff21},\ref{aff12}}
\and J.~Rhodes\orcid{0000-0002-4485-8549}\inst{\ref{aff17}}
\and G.~Riccio\inst{\ref{aff35}}
\and E.~Romelli\orcid{0000-0003-3069-9222}\inst{\ref{aff2}}
\and M.~Roncarelli\orcid{0000-0001-9587-7822}\inst{\ref{aff20}}
\and E.~Rossetti\orcid{0000-0003-0238-4047}\inst{\ref{aff31}}
\and R.~Saglia\orcid{0000-0003-0378-7032}\inst{\ref{aff69},\ref{aff26}}
\and Z.~Sakr\orcid{0000-0002-4823-3757}\inst{\ref{aff97},\ref{aff98},\ref{aff99}}
\and D.~Sapone\orcid{0000-0001-7089-4503}\inst{\ref{aff100}}
\and B.~Sartoris\orcid{0000-0003-1337-5269}\inst{\ref{aff69},\ref{aff2}}
\and P.~Schneider\orcid{0000-0001-8561-2679}\inst{\ref{aff13}}
\and T.~Schrabback\orcid{0000-0002-6987-7834}\inst{\ref{aff101}}
\and M.~Scodeggio\inst{\ref{aff10}}
\and A.~Secroun\orcid{0000-0003-0505-3710}\inst{\ref{aff65}}
\and G.~Seidel\orcid{0000-0003-2907-353X}\inst{\ref{aff74}}
\and S.~Serrano\orcid{0000-0002-0211-2861}\inst{\ref{aff51},\ref{aff102},\ref{aff50}}
\and P.~Simon\inst{\ref{aff13}}
\and G.~Sirri\orcid{0000-0003-2626-2853}\inst{\ref{aff32}}
\and A.~Spurio~Mancini\orcid{0000-0001-5698-0990}\inst{\ref{aff103}}
\and L.~Stanco\orcid{0000-0002-9706-5104}\inst{\ref{aff12}}
\and J.~Steinwagner\orcid{0000-0001-7443-1047}\inst{\ref{aff26}}
\and C.~Surace\orcid{0000-0003-2592-0113}\inst{\ref{aff11}}
\and P.~Tallada-Cresp\'{i}\orcid{0000-0002-1336-8328}\inst{\ref{aff46},\ref{aff47}}
\and A.~N.~Taylor\inst{\ref{aff55}}
\and H.~I.~Teplitz\orcid{0000-0002-7064-5424}\inst{\ref{aff28}}
\and I.~Tereno\orcid{0000-0002-4537-6218}\inst{\ref{aff104},\ref{aff105}}
\and N.~Tessore\orcid{0000-0002-9696-7931}\inst{\ref{aff106}}
\and S.~Toft\orcid{0000-0003-3631-7176}\inst{\ref{aff107},\ref{aff108}}
\and R.~Toledo-Moreo\orcid{0000-0002-2997-4859}\inst{\ref{aff109}}
\and F.~Torradeflot\orcid{0000-0003-1160-1517}\inst{\ref{aff47},\ref{aff46}}
\and I.~Tutusaus\orcid{0000-0002-3199-0399}\inst{\ref{aff50},\ref{aff51},\ref{aff98}}
\and L.~Valenziano\orcid{0000-0002-1170-0104}\inst{\ref{aff20},\ref{aff110}}
\and J.~Valiviita\orcid{0000-0001-6225-3693}\inst{\ref{aff79},\ref{aff80}}
\and T.~Vassallo\orcid{0000-0001-6512-6358}\inst{\ref{aff2}}
\and A.~Veropalumbo\orcid{0000-0003-2387-1194}\inst{\ref{aff6},\ref{aff25},\ref{aff33}}
\and D.~Vibert\orcid{0009-0008-0607-631X}\inst{\ref{aff11}}
\and J.~Weller\orcid{0000-0002-8282-2010}\inst{\ref{aff69},\ref{aff26}}
\and A.~Zacchei\orcid{0000-0003-0396-1192}\inst{\ref{aff2},\ref{aff4}}
\and G.~Zamorani\orcid{0000-0002-2318-301X}\inst{\ref{aff20}}
\and F.~M.~Zerbi\inst{\ref{aff6}}
\and E.~Zucca\orcid{0000-0002-5845-8132}\inst{\ref{aff20}}
\and V.~Allevato\orcid{0000-0001-7232-5152}\inst{\ref{aff35}}
\and M.~Ballardini\orcid{0000-0003-4481-3559}\inst{\ref{aff111},\ref{aff112},\ref{aff20}}
\and M.~Bolzonella\orcid{0000-0003-3278-4607}\inst{\ref{aff20}}
\and A.~Boucaud\orcid{0000-0001-7387-2633}\inst{\ref{aff86}}
\and E.~Bozzo\orcid{0000-0002-8201-1525}\inst{\ref{aff63}}
\and C.~Burigana\orcid{0000-0002-3005-5796}\inst{\ref{aff113},\ref{aff110}}
\and R.~Cabanac\orcid{0000-0001-6679-2600}\inst{\ref{aff98}}
\and M.~Calabrese\orcid{0000-0002-2637-2422}\inst{\ref{aff114},\ref{aff10}}
\and A.~Cappi\inst{\ref{aff115},\ref{aff20}}
\and J.~A.~Escartin~Vigo\inst{\ref{aff26}}
\and G.~Fabbian\orcid{0000-0002-3255-4695}\inst{\ref{aff64}}
\and L.~Gabarra\orcid{0000-0002-8486-8856}\inst{\ref{aff116}}
\and W.~G.~Hartley\inst{\ref{aff63}}
\and R.~Maoli\orcid{0000-0002-6065-3025}\inst{\ref{aff117},\ref{aff44}}
\and J.~Mart\'{i}n-Fleitas\orcid{0000-0002-8594-569X}\inst{\ref{aff118}}
\and S.~Matthew\orcid{0000-0001-8448-1697}\inst{\ref{aff55}}
\and N.~Mauri\orcid{0000-0001-8196-1548}\inst{\ref{aff53},\ref{aff32}}
\and R.~B.~Metcalf\orcid{0000-0003-3167-2574}\inst{\ref{aff19},\ref{aff20}}
\and A.~Pezzotta\orcid{0000-0003-0726-2268}\inst{\ref{aff6}}
\and M.~P\"ontinen\orcid{0000-0001-5442-2530}\inst{\ref{aff79}}
\and V.~Scottez\orcid{0009-0008-3864-940X}\inst{\ref{aff88},\ref{aff119}}
\and M.~Sereno\orcid{0000-0003-0302-0325}\inst{\ref{aff20},\ref{aff32}}
\and M.~Tenti\orcid{0000-0002-4254-5901}\inst{\ref{aff32}}
\and M.~Viel\orcid{0000-0002-2642-5707}\inst{\ref{aff4},\ref{aff2},\ref{aff30},\ref{aff3},\ref{aff5}}
\and M.~Wiesmann\orcid{0009-0000-8199-5860}\inst{\ref{aff70}}
\and Y.~Akrami\orcid{0000-0002-2407-7956}\inst{\ref{aff120},\ref{aff121}}
\and I.~T.~Andika\orcid{0000-0001-6102-9526}\inst{\ref{aff122},\ref{aff123}}
\and S.~Anselmi\orcid{0000-0002-3579-9583}\inst{\ref{aff12},\ref{aff21},\ref{aff124}}
\and M.~Archidiacono\orcid{0000-0003-4952-9012}\inst{\ref{aff14},\ref{aff15}}
\and F.~Atrio-Barandela\orcid{0000-0002-2130-2513}\inst{\ref{aff125}}
\and S.~Avila\orcid{0000-0001-5043-3662}\inst{\ref{aff46}}
\and D.~Bertacca\orcid{0000-0002-2490-7139}\inst{\ref{aff21},\ref{aff68},\ref{aff12}}
\and A.~Blanchard\orcid{0000-0001-8555-9003}\inst{\ref{aff98}}
\and L.~Blot\orcid{0000-0002-9622-7167}\inst{\ref{aff126},\ref{aff81}}
\and M.~Bonici\orcid{0000-0002-8430-126X}\inst{\ref{aff22},\ref{aff10}}
\and S.~Borgani\orcid{0000-0001-6151-6439}\inst{\ref{aff1},\ref{aff4},\ref{aff2},\ref{aff3},\ref{aff5}}
\and M.~L.~Brown\orcid{0000-0002-0370-8077}\inst{\ref{aff56}}
\and A.~Calabro\orcid{0000-0003-2536-1614}\inst{\ref{aff44}}
\and B.~Camacho~Quevedo\orcid{0000-0002-8789-4232}\inst{\ref{aff4},\ref{aff30},\ref{aff2}}
\and F.~Caro\inst{\ref{aff44}}
\and C.~S.~Carvalho\inst{\ref{aff105}}
\and T.~Castro\orcid{0000-0002-6292-3228}\inst{\ref{aff2},\ref{aff3},\ref{aff4},\ref{aff5}}
\and F.~Cogato\orcid{0000-0003-4632-6113}\inst{\ref{aff19},\ref{aff20}}
\and S.~Conseil\orcid{0000-0002-3657-4191}\inst{\ref{aff58}}
\and A.~R.~Cooray\orcid{0000-0002-3892-0190}\inst{\ref{aff127}}
\and O.~Cucciati\orcid{0000-0002-9336-7551}\inst{\ref{aff20}}
\and S.~Davini\orcid{0000-0003-3269-1718}\inst{\ref{aff25}}
\and G.~Desprez\orcid{0000-0001-8325-1742}\inst{\ref{aff128}}
\and A.~D\'iaz-S\'anchez\orcid{0000-0003-0748-4768}\inst{\ref{aff129}}
\and J.~J.~Diaz\orcid{0000-0003-2101-1078}\inst{\ref{aff54}}
\and S.~Di~Domizio\orcid{0000-0003-2863-5895}\inst{\ref{aff33},\ref{aff25}}
\and J.~M.~Diego\orcid{0000-0001-9065-3926}\inst{\ref{aff130}}
\and A.~Enia\orcid{0000-0002-0200-2857}\inst{\ref{aff20}}
\and Y.~Fang\orcid{0000-0002-0334-6950}\inst{\ref{aff69}}
\and A.~G.~Ferrari\orcid{0009-0005-5266-4110}\inst{\ref{aff32}}
\and A.~Finoguenov\orcid{0000-0002-4606-5403}\inst{\ref{aff79}}
\and A.~Fontana\orcid{0000-0003-3820-2823}\inst{\ref{aff44}}
\and A.~Franco\orcid{0000-0002-4761-366X}\inst{\ref{aff131},\ref{aff132},\ref{aff133}}
\and J.~Garc\'ia-Bellido\orcid{0000-0002-9370-8360}\inst{\ref{aff120}}
\and T.~Gasparetto\orcid{0000-0002-7913-4866}\inst{\ref{aff44}}
\and V.~Gautard\inst{\ref{aff134}}
\and E.~Gaztanaga\orcid{0000-0001-9632-0815}\inst{\ref{aff50},\ref{aff51},\ref{aff135}}
\and F.~Giacomini\orcid{0000-0002-3129-2814}\inst{\ref{aff32}}
\and F.~Gianotti\orcid{0000-0003-4666-119X}\inst{\ref{aff20}}
\and G.~Gozaliasl\orcid{0000-0002-0236-919X}\inst{\ref{aff136},\ref{aff79}}
\and M.~Guidi\orcid{0000-0001-9408-1101}\inst{\ref{aff31},\ref{aff20}}
\and C.~M.~Gutierrez\orcid{0000-0001-7854-783X}\inst{\ref{aff137}}
\and A.~Hall\orcid{0000-0002-3139-8651}\inst{\ref{aff55}}
\and C.~Hern\'andez-Monteagudo\orcid{0000-0001-5471-9166}\inst{\ref{aff138},\ref{aff54}}
\and H.~Hildebrandt\orcid{0000-0002-9814-3338}\inst{\ref{aff139}}
\and J.~Hjorth\orcid{0000-0002-4571-2306}\inst{\ref{aff92}}
\and S.~Joudaki\orcid{0000-0001-8820-673X}\inst{\ref{aff46}}
\and J.~J.~E.~Kajava\orcid{0000-0002-3010-8333}\inst{\ref{aff140},\ref{aff141}}
\and Y.~Kang\orcid{0009-0000-8588-7250}\inst{\ref{aff63}}
\and V.~Kansal\orcid{0000-0002-4008-6078}\inst{\ref{aff142},\ref{aff143}}
\and D.~Karagiannis\orcid{0000-0002-4927-0816}\inst{\ref{aff111},\ref{aff144}}
\and K.~Kiiveri\inst{\ref{aff77}}
\and J.~Kim\orcid{0000-0003-2776-2761}\inst{\ref{aff116}}
\and C.~C.~Kirkpatrick\inst{\ref{aff77}}
\and S.~Kruk\orcid{0000-0001-8010-8879}\inst{\ref{aff29}}
\and M.~Lattanzi\orcid{0000-0003-1059-2532}\inst{\ref{aff112}}
\and V.~Le~Brun\orcid{0000-0002-5027-1939}\inst{\ref{aff11}}
\and L.~Legrand\orcid{0000-0003-0610-5252}\inst{\ref{aff145},\ref{aff146}}
\and M.~Lembo\orcid{0000-0002-5271-5070}\inst{\ref{aff16},\ref{aff111},\ref{aff112}}
\and F.~Lepori\orcid{0009-0000-5061-7138}\inst{\ref{aff147}}
\and G.~Leroy\orcid{0009-0004-2523-4425}\inst{\ref{aff148},\ref{aff85}}
\and G.~F.~Lesci\orcid{0000-0002-4607-2830}\inst{\ref{aff19},\ref{aff20}}
\and J.~Lesgourgues\orcid{0000-0001-7627-353X}\inst{\ref{aff48}}
\and L.~Leuzzi\orcid{0009-0006-4479-7017}\inst{\ref{aff20}}
\and T.~I.~Liaudat\orcid{0000-0002-9104-314X}\inst{\ref{aff149}}
\and S.~J.~Liu\orcid{0000-0001-7680-2139}\inst{\ref{aff66}}
\and A.~Loureiro\orcid{0000-0002-4371-0876}\inst{\ref{aff150},\ref{aff151}}
\and J.~Macias-Perez\orcid{0000-0002-5385-2763}\inst{\ref{aff152}}
\and M.~Magliocchetti\orcid{0000-0001-9158-4838}\inst{\ref{aff66}}
\and F.~Mannucci\orcid{0000-0002-4803-2381}\inst{\ref{aff153}}
\and C.~J.~A.~P.~Martins\orcid{0000-0002-4886-9261}\inst{\ref{aff154},\ref{aff36}}
\and L.~Maurin\orcid{0000-0002-8406-0857}\inst{\ref{aff64}}
\and M.~Miluzio\inst{\ref{aff29},\ref{aff155}}
\and C.~Moretti\orcid{0000-0003-3314-8936}\inst{\ref{aff2},\ref{aff4},\ref{aff3},\ref{aff30}}
\and G.~Morgante\inst{\ref{aff20}}
\and S.~Nadathur\orcid{0000-0001-9070-3102}\inst{\ref{aff135}}
\and K.~Naidoo\orcid{0000-0002-9182-1802}\inst{\ref{aff135},\ref{aff76}}
\and A.~Navarro-Alsina\orcid{0000-0002-3173-2592}\inst{\ref{aff13}}
\and S.~Nesseris\orcid{0000-0002-0567-0324}\inst{\ref{aff120}}
\and D.~Paoletti\orcid{0000-0003-4761-6147}\inst{\ref{aff20},\ref{aff110}}
\and K.~Paterson\orcid{0000-0001-8340-3486}\inst{\ref{aff74}}
\and L.~Patrizii\inst{\ref{aff32}}
\and A.~Pisani\orcid{0000-0002-6146-4437}\inst{\ref{aff65}}
\and D.~Potter\orcid{0000-0002-0757-5195}\inst{\ref{aff147}}
\and S.~Quai\orcid{0000-0002-0449-8163}\inst{\ref{aff19},\ref{aff20}}
\and M.~Radovich\orcid{0000-0002-3585-866X}\inst{\ref{aff68}}
\and G.~Rodighiero\orcid{0000-0002-9415-2296}\inst{\ref{aff21},\ref{aff68}}
\and S.~Sacquegna\orcid{0000-0002-8433-6630}\inst{\ref{aff156}}
\and M.~Sahl\'en\orcid{0000-0003-0973-4804}\inst{\ref{aff157}}
\and D.~B.~Sanders\orcid{0000-0002-1233-9998}\inst{\ref{aff158}}
\and E.~Sarpa\orcid{0000-0002-1256-655X}\inst{\ref{aff30},\ref{aff5},\ref{aff3}}
\and A.~Schneider\orcid{0000-0001-7055-8104}\inst{\ref{aff147}}
\and D.~Sciotti\orcid{0009-0008-4519-2620}\inst{\ref{aff44},\ref{aff45}}
\and E.~Sellentin\inst{\ref{aff159},\ref{aff43}}
\and L.~C.~Smith\orcid{0000-0002-3259-2771}\inst{\ref{aff160}}
\and K.~Tanidis\orcid{0000-0001-9843-5130}\inst{\ref{aff116}}
\and C.~Tao\orcid{0000-0001-7961-8177}\inst{\ref{aff65}}
\and G.~Testera\inst{\ref{aff25}}
\and R.~Teyssier\orcid{0000-0001-7689-0933}\inst{\ref{aff161}}
\and S.~Tosi\orcid{0000-0002-7275-9193}\inst{\ref{aff33},\ref{aff25},\ref{aff6}}
\and A.~Troja\orcid{0000-0003-0239-4595}\inst{\ref{aff21},\ref{aff12}}
\and M.~Tucci\inst{\ref{aff63}}
\and A.~Venhola\orcid{0000-0001-6071-4564}\inst{\ref{aff162}}
\and D.~Vergani\orcid{0000-0003-0898-2216}\inst{\ref{aff20}}
\and F.~Vernizzi\orcid{0000-0003-3426-2802}\inst{\ref{aff163}}
\and G.~Verza\orcid{0000-0002-1886-8348}\inst{\ref{aff164}}
\and P.~Vielzeuf\orcid{0000-0003-2035-9339}\inst{\ref{aff65}}
\and N.~A.~Walton\orcid{0000-0003-3983-8778}\inst{\ref{aff160}}}
										   
\institute{Dipartimento di Fisica - Sezione di Astronomia, Universit\`a di Trieste, Via Tiepolo 11, 34131 Trieste, Italy\label{aff1}
\and
INAF-Osservatorio Astronomico di Trieste, Via G. B. Tiepolo 11, 34143 Trieste, Italy\label{aff2}
\and
INFN, Sezione di Trieste, Via Valerio 2, 34127 Trieste TS, Italy\label{aff3}
\and
IFPU, Institute for Fundamental Physics of the Universe, via Beirut 2, 34151 Trieste, Italy\label{aff4}
\and
ICSC - Centro Nazionale di Ricerca in High Performance Computing, Big Data e Quantum Computing, Via Magnanelli 2, Bologna, Italy\label{aff5}
\and
INAF-Osservatorio Astronomico di Brera, Via Brera 28, 20122 Milano, Italy\label{aff6}
\and
Minnesota Institute for Astrophysics, University of Minnesota, 116 Church St SE, Minneapolis, MN 55455, USA\label{aff7}
\and
Universit\'e de Strasbourg, CNRS, Observatoire astronomique de Strasbourg, UMR 7550, 67000 Strasbourg, France\label{aff8}
\and
California Institute of Technology, 1200 E California Blvd, Pasadena, CA 91125, USA\label{aff9}
\and
INAF-IASF Milano, Via Alfonso Corti 12, 20133 Milano, Italy\label{aff10}
\and
Aix-Marseille Universit\'e, CNRS, CNES, LAM, Marseille, France\label{aff11}
\and
INFN-Padova, Via Marzolo 8, 35131 Padova, Italy\label{aff12}
\and
Universit\"at Bonn, Argelander-Institut f\"ur Astronomie, Auf dem H\"ugel 71, 53121 Bonn, Germany\label{aff13}
\and
Dipartimento di Fisica "Aldo Pontremoli", Universit\`a degli Studi di Milano, Via Celoria 16, 20133 Milano, Italy\label{aff14}
\and
INFN-Sezione di Milano, Via Celoria 16, 20133 Milano, Italy\label{aff15}
\and
Institut d'Astrophysique de Paris, UMR 7095, CNRS, and Sorbonne Universit\'e, 98 bis boulevard Arago, 75014 Paris, France\label{aff16}
\and
Jet Propulsion Laboratory, California Institute of Technology, 4800 Oak Grove Drive, Pasadena, CA, 91109, USA\label{aff17}
\and
Kavli Institute for the Physics and Mathematics of the Universe (WPI), University of Tokyo, Kashiwa, Chiba 277-8583, Japan\label{aff18}
\and
Dipartimento di Fisica e Astronomia "Augusto Righi" - Alma Mater Studiorum Universit\`a di Bologna, via Piero Gobetti 93/2, 40129 Bologna, Italy\label{aff19}
\and
INAF-Osservatorio di Astrofisica e Scienza dello Spazio di Bologna, Via Piero Gobetti 93/3, 40129 Bologna, Italy\label{aff20}
\and
Dipartimento di Fisica e Astronomia "G. Galilei", Universit\`a di Padova, Via Marzolo 8, 35131 Padova, Italy\label{aff21}
\and
Waterloo Centre for Astrophysics, University of Waterloo, Waterloo, Ontario N2L 3G1, Canada\label{aff22}
\and
Department of Physics and Astronomy, University of Waterloo, Waterloo, Ontario N2L 3G1, Canada\label{aff23}
\and
Perimeter Institute for Theoretical Physics, Waterloo, Ontario N2L 2Y5, Canada\label{aff24}
\and
INFN-Sezione di Genova, Via Dodecaneso 33, 16146, Genova, Italy\label{aff25}
\and
Max Planck Institute for Extraterrestrial Physics, Giessenbachstr. 1, 85748 Garching, Germany\label{aff26}
\and
Department of Physics and Astronomy, University of British Columbia, Vancouver, BC V6T 1Z1, Canada\label{aff27}
\and
Infrared Processing and Analysis Center, California Institute of Technology, Pasadena, CA 91125, USA\label{aff28}
\and
ESAC/ESA, Camino Bajo del Castillo, s/n., Urb. Villafranca del Castillo, 28692 Villanueva de la Ca\~nada, Madrid, Spain\label{aff29}
\and
SISSA, International School for Advanced Studies, Via Bonomea 265, 34136 Trieste TS, Italy\label{aff30}
\and
Dipartimento di Fisica e Astronomia, Universit\`a di Bologna, Via Gobetti 93/2, 40129 Bologna, Italy\label{aff31}
\and
INFN-Sezione di Bologna, Viale Berti Pichat 6/2, 40127 Bologna, Italy\label{aff32}
\and
Dipartimento di Fisica, Universit\`a di Genova, Via Dodecaneso 33, 16146, Genova, Italy\label{aff33}
\and
Department of Physics "E. Pancini", University Federico II, Via Cinthia 6, 80126, Napoli, Italy\label{aff34}
\and
INAF-Osservatorio Astronomico di Capodimonte, Via Moiariello 16, 80131 Napoli, Italy\label{aff35}
\and
Instituto de Astrof\'isica e Ci\^encias do Espa\c{c}o, Universidade do Porto, CAUP, Rua das Estrelas, PT4150-762 Porto, Portugal\label{aff36}
\and
Faculdade de Ci\^encias da Universidade do Porto, Rua do Campo de Alegre, 4150-007 Porto, Portugal\label{aff37}
\and
European Southern Observatory, Karl-Schwarzschild-Str.~2, 85748 Garching, Germany\label{aff38}
\and
Dipartimento di Fisica, Universit\`a degli Studi di Torino, Via P. Giuria 1, 10125 Torino, Italy\label{aff39}
\and
INFN-Sezione di Torino, Via P. Giuria 1, 10125 Torino, Italy\label{aff40}
\and
INAF-Osservatorio Astrofisico di Torino, Via Osservatorio 20, 10025 Pino Torinese (TO), Italy\label{aff41}
\and
European Space Agency/ESTEC, Keplerlaan 1, 2201 AZ Noordwijk, The Netherlands\label{aff42}
\and
Leiden Observatory, Leiden University, Einsteinweg 55, 2333 CC Leiden, The Netherlands\label{aff43}
\and
INAF-Osservatorio Astronomico di Roma, Via Frascati 33, 00078 Monteporzio Catone, Italy\label{aff44}
\and
INFN-Sezione di Roma, Piazzale Aldo Moro, 2 - c/o Dipartimento di Fisica, Edificio G. Marconi, 00185 Roma, Italy\label{aff45}
\and
Centro de Investigaciones Energ\'eticas, Medioambientales y Tecnol\'ogicas (CIEMAT), Avenida Complutense 40, 28040 Madrid, Spain\label{aff46}
\and
Port d'Informaci\'{o} Cient\'{i}fica, Campus UAB, C. Albareda s/n, 08193 Bellaterra (Barcelona), Spain\label{aff47}
\and
Institute for Theoretical Particle Physics and Cosmology (TTK), RWTH Aachen University, 52056 Aachen, Germany\label{aff48}
\and
Deutsches Zentrum f\"ur Luft- und Raumfahrt e. V. (DLR), Linder H\"ohe, 51147 K\"oln, Germany\label{aff49}
\and
Institute of Space Sciences (ICE, CSIC), Campus UAB, Carrer de Can Magrans, s/n, 08193 Barcelona, Spain\label{aff50}
\and
Institut d'Estudis Espacials de Catalunya (IEEC),  Edifici RDIT, Campus UPC, 08860 Castelldefels, Barcelona, Spain\label{aff51}
\and
INFN section of Naples, Via Cinthia 6, 80126, Napoli, Italy\label{aff52}
\and
Dipartimento di Fisica e Astronomia "Augusto Righi" - Alma Mater Studiorum Universit\`a di Bologna, Viale Berti Pichat 6/2, 40127 Bologna, Italy\label{aff53}
\and
Instituto de Astrof\'{\i}sica de Canarias, E-38205 La Laguna, Tenerife, Spain\label{aff54}
\and
Institute for Astronomy, University of Edinburgh, Royal Observatory, Blackford Hill, Edinburgh EH9 3HJ, UK\label{aff55}
\and
Jodrell Bank Centre for Astrophysics, Department of Physics and Astronomy, University of Manchester, Oxford Road, Manchester M13 9PL, UK\label{aff56}
\and
European Space Agency/ESRIN, Largo Galileo Galilei 1, 00044 Frascati, Roma, Italy\label{aff57}
\and
Universit\'e Claude Bernard Lyon 1, CNRS/IN2P3, IP2I Lyon, UMR 5822, Villeurbanne, F-69100, France\label{aff58}
\and
Institut de Ci\`{e}ncies del Cosmos (ICCUB), Universitat de Barcelona (IEEC-UB), Mart\'{i} i Franqu\`{e}s 1, 08028 Barcelona, Spain\label{aff59}
\and
Instituci\'o Catalana de Recerca i Estudis Avan\c{c}ats (ICREA), Passeig de Llu\'{\i}s Companys 23, 08010 Barcelona, Spain\label{aff60}
\and
Institut de Ciencies de l'Espai (IEEC-CSIC), Campus UAB, Carrer de Can Magrans, s/n Cerdanyola del Vall\'es, 08193 Barcelona, Spain\label{aff61}
\and
UCB Lyon 1, CNRS/IN2P3, IUF, IP2I Lyon, 4 rue Enrico Fermi, 69622 Villeurbanne, France\label{aff62}
\and
Department of Astronomy, University of Geneva, ch. d'Ecogia 16, 1290 Versoix, Switzerland\label{aff63}
\and
Universit\'e Paris-Saclay, CNRS, Institut d'astrophysique spatiale, 91405, Orsay, France\label{aff64}
\and
Aix-Marseille Universit\'e, CNRS/IN2P3, CPPM, Marseille, France\label{aff65}
\and
INAF-Istituto di Astrofisica e Planetologia Spaziali, via del Fosso del Cavaliere, 100, 00100 Roma, Italy\label{aff66}
\and
University Observatory, LMU Faculty of Physics, Scheinerstr.~1, 81679 Munich, Germany\label{aff67}
\and
INAF-Osservatorio Astronomico di Padova, Via dell'Osservatorio 5, 35122 Padova, Italy\label{aff68}
\and
Universit\"ats-Sternwarte M\"unchen, Fakult\"at f\"ur Physik, Ludwig-Maximilians-Universit\"at M\"unchen, Scheinerstr.~1, 81679 M\"unchen, Germany\label{aff69}
\and
Institute of Theoretical Astrophysics, University of Oslo, P.O. Box 1029 Blindern, 0315 Oslo, Norway\label{aff70}
\and
Felix Hormuth Engineering, Goethestr. 17, 69181 Leimen, Germany\label{aff71}
\and
Technical University of Denmark, Elektrovej 327, 2800 Kgs. Lyngby, Denmark\label{aff72}
\and
Cosmic Dawn Center (DAWN), Denmark\label{aff73}
\and
Max-Planck-Institut f\"ur Astronomie, K\"onigstuhl 17, 69117 Heidelberg, Germany\label{aff74}
\and
NASA Goddard Space Flight Center, Greenbelt, MD 20771, USA\label{aff75}
\and
Department of Physics and Astronomy, University College London, Gower Street, London WC1E 6BT, UK\label{aff76}
\and
Department of Physics and Helsinki Institute of Physics, Gustaf H\"allstr\"omin katu 2, University of Helsinki, 00014 Helsinki, Finland\label{aff77}
\and
Universit\'e de Gen\`eve, D\'epartement de Physique Th\'eorique and Centre for Astroparticle Physics, 24 quai Ernest-Ansermet, CH-1211 Gen\`eve 4, Switzerland\label{aff78}
\and
Department of Physics, P.O. Box 64, University of Helsinki, 00014 Helsinki, Finland\label{aff79}
\and
Helsinki Institute of Physics, Gustaf H{\"a}llstr{\"o}min katu 2, University of Helsinki, 00014 Helsinki, Finland\label{aff80}
\and
Laboratoire d'etude de l'Univers et des phenomenes eXtremes, Observatoire de Paris, Universit\'e PSL, Sorbonne Universit\'e, CNRS, 92190 Meudon, France\label{aff81}
\and
SKAO, Jodrell Bank, Lower Withington, Macclesfield SK11 9FT, UK\label{aff82}
\and
Centre de Calcul de l'IN2P3/CNRS, 21 avenue Pierre de Coubertin 69627 Villeurbanne Cedex, France\label{aff83}
\and
University of Applied Sciences and Arts of Northwestern Switzerland, School of Computer Science, 5210 Windisch, Switzerland\label{aff84}
\and
Department of Physics, Institute for Computational Cosmology, Durham University, South Road, Durham, DH1 3LE, UK\label{aff85}
\and
Universit\'e Paris Cit\'e, CNRS, Astroparticule et Cosmologie, 75013 Paris, France\label{aff86}
\and
CNRS-UCB International Research Laboratory, Centre Pierre Bin\'etruy, IRL2007, CPB-IN2P3, Berkeley, USA\label{aff87}
\and
Institut d'Astrophysique de Paris, 98bis Boulevard Arago, 75014, Paris, France\label{aff88}
\and
Institute of Physics, Laboratory of Astrophysics, Ecole Polytechnique F\'ed\'erale de Lausanne (EPFL), Observatoire de Sauverny, 1290 Versoix, Switzerland\label{aff89}
\and
Telespazio UK S.L. for European Space Agency (ESA), Camino bajo del Castillo, s/n, Urbanizacion Villafranca del Castillo, Villanueva de la Ca\~nada, 28692 Madrid, Spain\label{aff90}
\and
Institut de F\'{i}sica d'Altes Energies (IFAE), The Barcelona Institute of Science and Technology, Campus UAB, 08193 Bellaterra (Barcelona), Spain\label{aff91}
\and
DARK, Niels Bohr Institute, University of Copenhagen, Jagtvej 155, 2200 Copenhagen, Denmark\label{aff92}
\and
Universit\'e Paris-Saclay, Universit\'e Paris Cit\'e, CEA, CNRS, AIM, 91191, Gif-sur-Yvette, France\label{aff93}
\and
Space Science Data Center, Italian Space Agency, via del Politecnico snc, 00133 Roma, Italy\label{aff94}
\and
Centre National d'Etudes Spatiales -- Centre spatial de Toulouse, 18 avenue Edouard Belin, 31401 Toulouse Cedex 9, France\label{aff95}
\and
Institute of Space Science, Str. Atomistilor, nr. 409 M\u{a}gurele, Ilfov, 077125, Romania\label{aff96}
\and
Institut f\"ur Theoretische Physik, University of Heidelberg, Philosophenweg 16, 69120 Heidelberg, Germany\label{aff97}
\and
Institut de Recherche en Astrophysique et Plan\'etologie (IRAP), Universit\'e de Toulouse, CNRS, UPS, CNES, 14 Av. Edouard Belin, 31400 Toulouse, France\label{aff98}
\and
Universit\'e St Joseph; Faculty of Sciences, Beirut, Lebanon\label{aff99}
\and
Departamento de F\'isica, FCFM, Universidad de Chile, Blanco Encalada 2008, Santiago, Chile\label{aff100}
\and
Universit\"at Innsbruck, Institut f\"ur Astro- und Teilchenphysik, Technikerstr. 25/8, 6020 Innsbruck, Austria\label{aff101}
\and
Satlantis, University Science Park, Sede Bld 48940, Leioa-Bilbao, Spain\label{aff102}
\and
Department of Physics, Royal Holloway, University of London, Surrey TW20 0EX, UK\label{aff103}
\and
Departamento de F\'isica, Faculdade de Ci\^encias, Universidade de Lisboa, Edif\'icio C8, Campo Grande, PT1749-016 Lisboa, Portugal\label{aff104}
\and
Instituto de Astrof\'isica e Ci\^encias do Espa\c{c}o, Faculdade de Ci\^encias, Universidade de Lisboa, Tapada da Ajuda, 1349-018 Lisboa, Portugal\label{aff105}
\and
Mullard Space Science Laboratory, University College London, Holmbury St Mary, Dorking, Surrey RH5 6NT, UK\label{aff106}
\and
Cosmic Dawn Center (DAWN)\label{aff107}
\and
Niels Bohr Institute, University of Copenhagen, Jagtvej 128, 2200 Copenhagen, Denmark\label{aff108}
\and
Universidad Polit\'ecnica de Cartagena, Departamento de Electr\'onica y Tecnolog\'ia de Computadoras,  Plaza del Hospital 1, 30202 Cartagena, Spain\label{aff109}
\and
INFN-Bologna, Via Irnerio 46, 40126 Bologna, Italy\label{aff110}
\and
Dipartimento di Fisica e Scienze della Terra, Universit\`a degli Studi di Ferrara, Via Giuseppe Saragat 1, 44122 Ferrara, Italy\label{aff111}
\and
Istituto Nazionale di Fisica Nucleare, Sezione di Ferrara, Via Giuseppe Saragat 1, 44122 Ferrara, Italy\label{aff112}
\and
INAF, Istituto di Radioastronomia, Via Piero Gobetti 101, 40129 Bologna, Italy\label{aff113}
\and
Astronomical Observatory of the Autonomous Region of the Aosta Valley (OAVdA), Loc. Lignan 39, I-11020, Nus (Aosta Valley), Italy\label{aff114}
\and
Universit\'e C\^{o}te d'Azur, Observatoire de la C\^{o}te d'Azur, CNRS, Laboratoire Lagrange, Bd de l'Observatoire, CS 34229, 06304 Nice cedex 4, France\label{aff115}
\and
Department of Physics, Oxford University, Keble Road, Oxford OX1 3RH, UK\label{aff116}
\and
Dipartimento di Fisica, Sapienza Universit\`a di Roma, Piazzale Aldo Moro 2, 00185 Roma, Italy\label{aff117}
\and
Aurora Technology for European Space Agency (ESA), Camino bajo del Castillo, s/n, Urbanizacion Villafranca del Castillo, Villanueva de la Ca\~nada, 28692 Madrid, Spain\label{aff118}
\and
ICL, Junia, Universit\'e Catholique de Lille, LITL, 59000 Lille, France\label{aff119}
\and
Instituto de F\'isica Te\'orica UAM-CSIC, Campus de Cantoblanco, 28049 Madrid, Spain\label{aff120}
\and
CERCA/ISO, Department of Physics, Case Western Reserve University, 10900 Euclid Avenue, Cleveland, OH 44106, USA\label{aff121}
\and
Technical University of Munich, TUM School of Natural Sciences, Physics Department, James-Franck-Str.~1, 85748 Garching, Germany\label{aff122}
\and
Max-Planck-Institut f\"ur Astrophysik, Karl-Schwarzschild-Str.~1, 85748 Garching, Germany\label{aff123}
\and
Laboratoire Univers et Th\'eorie, Observatoire de Paris, Universit\'e PSL, Universit\'e Paris Cit\'e, CNRS, 92190 Meudon, France\label{aff124}
\and
Departamento de F{\'\i}sica Fundamental. Universidad de Salamanca. Plaza de la Merced s/n. 37008 Salamanca, Spain\label{aff125}
\and
Center for Data-Driven Discovery, Kavli IPMU (WPI), UTIAS, The University of Tokyo, Kashiwa, Chiba 277-8583, Japan\label{aff126}
\and
Department of Physics \& Astronomy, University of California Irvine, Irvine CA 92697, USA\label{aff127}
\and
Kapteyn Astronomical Institute, University of Groningen, PO Box 800, 9700 AV Groningen, The Netherlands\label{aff128}
\and
Departamento F\'isica Aplicada, Universidad Polit\'ecnica de Cartagena, Campus Muralla del Mar, 30202 Cartagena, Murcia, Spain\label{aff129}
\and
Instituto de F\'isica de Cantabria, Edificio Juan Jord\'a, Avenida de los Castros, 39005 Santander, Spain\label{aff130}
\and
INFN, Sezione di Lecce, Via per Arnesano, CP-193, 73100, Lecce, Italy\label{aff131}
\and
Department of Mathematics and Physics E. De Giorgi, University of Salento, Via per Arnesano, CP-I93, 73100, Lecce, Italy\label{aff132}
\and
INAF-Sezione di Lecce, c/o Dipartimento Matematica e Fisica, Via per Arnesano, 73100, Lecce, Italy\label{aff133}
\and
CEA Saclay, DFR/IRFU, Service d'Astrophysique, Bat. 709, 91191 Gif-sur-Yvette, France\label{aff134}
\and
Institute of Cosmology and Gravitation, University of Portsmouth, Portsmouth PO1 3FX, UK\label{aff135}
\and
Department of Computer Science, Aalto University, PO Box 15400, Espoo, FI-00 076, Finland\label{aff136}
\and
 Instituto de Astrof\'{\i}sica de Canarias, E-38205 La Laguna; Universidad de La Laguna, Dpto. Astrof\'\i sica, E-38206 La Laguna, Tenerife, Spain\label{aff137}
\and
Universidad de La Laguna, Dpto. Astrof\'\i sica, E-38206 La Laguna, Tenerife, Spain\label{aff138}
\and
Ruhr University Bochum, Faculty of Physics and Astronomy, Astronomical Institute (AIRUB), German Centre for Cosmological Lensing (GCCL), 44780 Bochum, Germany\label{aff139}
\and
Department of Physics and Astronomy, Vesilinnantie 5, University of Turku, 20014 Turku, Finland\label{aff140}
\and
Serco for European Space Agency (ESA), Camino bajo del Castillo, s/n, Urbanizacion Villafranca del Castillo, Villanueva de la Ca\~nada, 28692 Madrid, Spain\label{aff141}
\and
ARC Centre of Excellence for Dark Matter Particle Physics, Melbourne, Australia\label{aff142}
\and
Centre for Astrophysics \& Supercomputing, Swinburne University of Technology,  Hawthorn, Victoria 3122, Australia\label{aff143}
\and
Department of Physics and Astronomy, University of the Western Cape, Bellville, Cape Town, 7535, South Africa\label{aff144}
\and
DAMTP, Centre for Mathematical Sciences, Wilberforce Road, Cambridge CB3 0WA, UK\label{aff145}
\and
Kavli Institute for Cosmology Cambridge, Madingley Road, Cambridge, CB3 0HA, UK\label{aff146}
\and
Department of Astrophysics, University of Zurich, Winterthurerstrasse 190, 8057 Zurich, Switzerland\label{aff147}
\and
Department of Physics, Centre for Extragalactic Astronomy, Durham University, South Road, Durham, DH1 3LE, UK\label{aff148}
\and
IRFU, CEA, Universit\'e Paris-Saclay 91191 Gif-sur-Yvette Cedex, France\label{aff149}
\and
Oskar Klein Centre for Cosmoparticle Physics, Department of Physics, Stockholm University, Stockholm, SE-106 91, Sweden\label{aff150}
\and
Astrophysics Group, Blackett Laboratory, Imperial College London, London SW7 2AZ, UK\label{aff151}
\and
Univ. Grenoble Alpes, CNRS, Grenoble INP, LPSC-IN2P3, 53, Avenue des Martyrs, 38000, Grenoble, France\label{aff152}
\and
INAF-Osservatorio Astrofisico di Arcetri, Largo E. Fermi 5, 50125, Firenze, Italy\label{aff153}
\and
Centro de Astrof\'{\i}sica da Universidade do Porto, Rua das Estrelas, 4150-762 Porto, Portugal\label{aff154}
\and
HE Space for European Space Agency (ESA), Camino bajo del Castillo, s/n, Urbanizacion Villafranca del Castillo, Villanueva de la Ca\~nada, 28692 Madrid, Spain\label{aff155}
\and
INAF - Osservatorio Astronomico d'Abruzzo, Via Maggini, 64100, Teramo, Italy\label{aff156}
\and
Theoretical astrophysics, Department of Physics and Astronomy, Uppsala University, Box 516, 751 37 Uppsala, Sweden\label{aff157}
\and
Institute for Astronomy, University of Hawaii, 2680 Woodlawn Drive, Honolulu, HI 96822, USA\label{aff158}
\and
Mathematical Institute, University of Leiden, Einsteinweg 55, 2333 CA Leiden, The Netherlands\label{aff159}
\and
Institute of Astronomy, University of Cambridge, Madingley Road, Cambridge CB3 0HA, UK\label{aff160}
\and
Department of Astrophysical Sciences, Peyton Hall, Princeton University, Princeton, NJ 08544, USA\label{aff161}
\and
Space physics and astronomy research unit, University of Oulu, Pentti Kaiteran katu 1, FI-90014 Oulu, Finland\label{aff162}
\and
Institut de Physique Th\'eorique, CEA, CNRS, Universit\'e Paris-Saclay 91191 Gif-sur-Yvette Cedex, France\label{aff163}
\and
Center for Computational Astrophysics, Flatiron Institute, 162 5th Avenue, 10010, New York, NY, USA\label{aff164}}

\abstract{We present the strategy used to identify and mitigate potential sources of angular
  systematics in the \textit{Euclid} spectroscopic galaxy survey, and we quantify their
  impact on galaxy clustering measurements and cosmological parameter estimation. We first
  surveyed the \textit{Euclid} processing pipeline to identify all evident, potential
  sources of systematics, and classified them into two broad classes: angular systematics,
  which modulate the galaxy number density across the sky, and catastrophic redshift
  errors, which lead to interlopers in the galaxy sample. We then used simulated
  spectroscopic surveys to test our ability to mitigate angular systematics by
  constructing a random catalogue that represents the `visibility mask' of the survey; this
  is a dense set of intrinsically unclustered objects, subject to the same selection
  effects as the data catalogue. The construction of this random catalogue relies on a
  detection model, which gives the probability of reliably measuring the galaxy redshift
  as a function of the signal-to-noise ratio (S/N) of its emission lines. We demonstrate
  that, in the ideal case of a perfect knowledge of the visibility mask, the galaxy power
  spectrum in the presence of systematics is recovered, to within sub-per cent accuracy, by
  convolving a theory power spectrum with a window function obtained from the random
  catalogue itself. In the case of only approximate knowledge of the visibility mask, we
  test the stability of power spectrum measurements and cosmological parameter posteriors
  by using perturbed versions of the random catalogue. We find that significant effects
  are limited to very large scales, and parameter estimation remains robust; the most
  impacting effects are connected to the calibration of the detection model.}


\titlerunning{Controlling angular systematics}
\authorrunning{Euclid Collaboration: P.~Monaco et al.}
\nolinenumbers

\maketitle

\section{Introduction}
\label{sec:Intro}

\Euclid \citep{EuclidSkyOverview} is a space telescope of the European Space Agency
designed to survey the Universe out to redshift $z\sim$\,2, with the main aim of measuring
weak gravitational lensing and galaxy clustering. The goal is to use these measurements to constrain the
late-time evolution of the Universe, when dark energy becomes dominant and potential
signatures of modified gravity become detectable in the growth of structure. With its
visible imager \citep[VIS;][]{EuclidSkyVIS} and near-infrared photometer and spectrometer
\citep[NISP;][]{EuclidSkyNISP}, it will provide images for billions of galaxies and
slitless spectra for tens of millions of emission-line galaxies (ELGs).

The ongoing Euclid Wide Survey \citep[EWS;][]{Scaramella-EP1} will eventually cover a sky
area of 14\,000 \sqdeg, split into four connected regions separated by the Milky Way (MW)
zone of avoidance and by the ecliptic, where zodiacal light hampers deep observations. The
EWS is currently tiling the sky in a shift-and-stare mode, where each tile is characterised
by a reference observing sequence (ROS) consisting of four dithers, each shifted following
an S-shaped pattern to minimise the impact of detector gaps \citep{Markovic2017}. Each
dither includes the simultaneous parallel observation of VIS photometric (\IE band) and
NISP spectroscopic observations, followed by three shorter near-infrared photometric
exposures in the \JE, \HE, and \YE filters. Spectroscopy is performed in slitless mode
with red grisms, which are sensitive in the wavelength range $\lambda\in[1.2,1.9]$
\micron. To minimise the effect of spectral confusion, a different orientation for the
dispersion of light is used in each dither of the ROS, corresponding to 0\degree,
$-4$\degree, 184\degree, and 180{\degree} rotations with respect to the focal plane
reference orientation. This scanning strategy delivers images with an AB magnitude depth
of $24.5$ (10\,$\sigma$) and $24$ (5\,$\sigma$), respectively, in the \IE and near-infrared
bands, and spectra deep enough to measure emission lines with a line flux of $f_{\rm
  line}\ga2\times10^{-16}$ \flux. Because {\ha} is the most prominent emission line in the
spectrum of an ELG in the optical/near-infrared wavelength range, the main sample will
consist of {\ha} emitters in the redshift range from $0.85$ to $1.88$. This redshift range
is rounded to $[0.9,1.8]$ in the rest of the paper.

A smaller area of 53 {\sqdeg}, split into three separate regions of the sky, is also being
observed, accumulating what eventually will be $15$ independent wide-like passes. To these
$25$ further passes will be added using a blue grism that covers the wavelength range
$\lambda\in[0.9,1.3]$ {\micron}. Thanks to its larger depth and increased spectral
wavelength coverage, the EDS will allow us to extract a 99\% pure and complete sample of
the `target' ELGs visible in the EWS. Moreover, the 15 repeated passes assess the
probability that a galaxy is detected with a correct or catastrophically incorrect redshift,
thus quantifying the purity and completeness of the sample (defined in
Sect.~\ref{sec:P&C}).

With the huge volume surveyed by the EWS, the error budget, on both the measured summary
statistics and the inferred cosmological parameters, will be dominated by
systematics.\footnote{
Following the conventional jargon, we use here the word systematics as a shorthand for
the more correct but longer systematic effects.
} For measured summary statistics, corresponding to a data vector $\vec{D}$, fitted with a
model that depends on a set of cosmological and nuisance parameters $\vec{\theta}$, the
posterior distribution on the parameters is evaluated in a Bayesian framework as

\be P({\vec{\theta}}|{\vec{D}}) = \frac{ {\cal L}({\vec{D}} | {\vec{\theta}} )\,
  P({\vec{\theta}})} {P({\vec{D}})}\, ,
\label{eq:likelihood}\ee

\noindent
where ${\cal L}({\vec{D}} | {\vec{\theta}})$ is the likelihood of having the data given a
set of parameter values, $P({\vec{\theta}})$ are the priors on parameters, and the
evidence $P({\vec{D}})$ enters as a normalisation. The crucial ingredient of this equation
is the likelihood.

We can classify systematics as (1) data systematics, namely all those effects that
perturb the data vector ${\vec{D}}$; (2) theory systematics, namely the inaccuracy in
the model to be compared with the data; and (3) likelihood systematics, namely all the
issues that affect the likelihood $\cal L$ and the process of parameter inference.
We concentrated on the data systematics of spectroscopic galaxy clustering, and for the data
vector $\vec{D}$ we used the first three even multipoles of the galaxy power spectrum. In
Sect.~\ref{sec:formalism} we present a formalism that describes the main observable, the
galaxy density, and show that the data systematics in a spectroscopic sample can be broadly
divided into angular systematics, which originate from fluctuations of the survey depth on
the sky, and those arising from errors in redshift measurements. The latter effect is
addressed in \cite{EP-Risso} and \citelee, respectively in configuration and Fourier
spaces. In Sect.~\ref{sec:classification} we identify all evident potential sources of
systematics along the \Euclid data reduction pipeline (details given in
Appendix~\ref{app:entrypoints}) and group them into a few classes, for which we discuss
the adopted mitigation strategy, assesses the impact on the survey (how strongly the
systematics affect the clustering measurement), and estimate its risk factor (how
difficult it is to mitigate it). After describing in Sect.~\ref{sec:tools} the tools used
to model the effect of angular systematics on the measured power spectrum of simulated
(mock) galaxy catalogues, we quantify in Sect.~\ref{sec:results} their effect and
propagate the uncertainty of its mitigation to cosmological parameters. Finally,
Sect.~\ref{sec:conclusions} gives the conclusions of the paper. For convenience, we
report in Table~\ref{table:definitions} some definitions that are used throughout the
text.

\section{Formalism}
\label{sec:formalism}

\subsection{The observed galaxy density}
\label{sec:no}

The key observable for extracting cosmological information from a spectroscopic galaxy
redshift survey is the galaxy density field $n_{\rm g}(\vx)$, where $\vx$ is the comoving
position in redshift space (assuming here to be an exact redshift measurement). Here we do not specify how the galaxy sample is selected. Once the galaxy density has
been defined, its density contrast reads

\be
\delta_{\rm g}(\vx) := \frac{n_{\rm g}(\vx) - \overline{n}_{\rm g}(z)}{\overline{n}_{\rm g}(z)}\, ,
\label{eq:density_contrast}\ee

\noindent
where averages are performed over the survey volume and the average density
$\overline{n}_{\rm g}(z)$ is, in principle, redshift dependent, due to the observational
selection function. The density contrast $\delta_{\rm g}(\vx)$ can be used to obtain
summary statistic such as the galaxy two-point correlation function in configuration space
or the galaxy power spectrum in Fourier space. We formalise the effect of systematics by
focusing on the galaxy number density, distinguishing between the target galaxy density
$n_{\rm t}$ of an ideal reference sample from the observed one, $n_{\rm o}$, which
includes systematics. As a consequence, our mitigation strategy will hold for any summary
statistics based on $\delta_{\rm g}$. The formalism presented here follows that of
\cite{Colavincenzo2017} and \cite{Monaco2019}.

To keep the formalism simple, we call $f$ the flux of the {\ha} line. The spectroscopic
catalogue of the EWS will roughly be limited in $f$, and while the spectra of bright
galaxies will be measured with high reliability, most galaxies will be detected in
spectroscopy only through a single emission line with relatively low S/N. The
spectroscopic sample will faithfully trace the underlying galaxy density field as long as
(i) the flux limit does not vary across the sky; (ii) the catalogue is free of interlopers
with a catastrophically wrong redshift. Neither assumption is expected to be correct, so
these two effects must be carefully characterised and mitigated.

We consider a volume around a comoving position $\vx$ along our past light cone, large enough
to contain many galaxies but small enough to neglect evolution effects inside it. Observed
galaxy positions consist of a redshift $z$ and a line-of-sight direction $\nh$,
represented by two angles, such as the Galactic coordinates. Assuming a fiducial
cosmology, they can be translated into three (redshift-space) comoving coordinates,
measured in {\hmpc}; we  use the two coordinate systems (redshift and $\nh$, or
comoving position $\vx$) interchangeably. We call $\phi_{\rm local}(f|\vx)$ the local
{\ha} luminosity function\footnote{
Here we loosely call luminosity function a quantity that is defined in terms of the
observed flux; at fixed $z$, flux and luminosity are proportional so $\phi_{\rm local}$ and
$\Phi$ can be trivially reformulated in terms of proper luminosity functions.}
(LF) of that volume, expressed in terms of $f$ and having units of $h^3$ Mpc$^{-3}$
$($\flux$)^{-1}$. We define a fiducial limiting flux $f_0:=2\times10^{-16}$ {\flux} as the
nominal flux limit of the galaxy sample we aim to measure. The `target' galaxy density
is then defined as

\be
n_{\rm t}(\vx) := \int_{f_0}^\infty\, \phi_{\rm local}(f|\vx) \, {\rm d}f\, .
\label{eq:target_ng}\ee

We can define a global {\ha} LF $\Phi(f|z)$ as the volume average of $\phi_{\rm local}$ in
a small redshift bin at a median redshift $z$,

\be
\Phi(f|z) := \langle \phi_{\rm local}(f|\vx) \rangle_z\, .
\label{eq:global_Phi}\ee

\noindent
This global LF is related to the redshift-dependent average galaxy density of the survey
as

\be
\overline{n}_{\rm t}(z) =  \int_{f_0}^\infty\, \Phi(f|z) \, {\rm d}f\, .
\label{eq:average_nt}\ee

\noindent
Here both bars and $\langle\rangle_z$ brackets denote volume averages. To first
approximation, the local LF is equal to the global one $\Phi$ rescaled by the galaxy
density contrast $\delta_{\rm g}$,

\be
\phi_{\rm local}(f|\vx) \simeq [1+\delta_{\rm g}(\vx)]\, \Phi(f|z)\, ,
\label{eq:universal_LF}\ee

\noindent
where $\delta_{\rm g}$ is independent of $f_0$. This relation is exact as long as the
shape of the local LF is universal, independent of environment (meaning here the value of
the local galaxy density on some scale larger than the size of a typical halo). We know
that this is not true because the halo bias depends on the halo mass and the galaxy
luminosity correlates with the latter, so that brighter galaxies are more clustered than
fainter ones. For simplicity, we lay down our formalism by assuming that
Eq.~(\ref{eq:universal_LF}) is true, so that we can define a generic galaxy density
contrast that is independent of the galaxy sample definition. We show in
Sect.~\ref{sec:mocks} how to force this behaviour in the mock galaxy catalogue, and use it
for testing our mitigation strategies in Sect.~\ref{sec:shuffled}, while
Appendix~\ref{app:flux_modulations} briefly discusses how the formalism laid down here
changes when this simplification is removed.

Defining an observed sample that is exactly flux-limited would require a reliable
measurement of line flux, which is possible only at high S/N, but this may not lead to an
optimal sample for cosmology, where number density is crucial. We  then assume that
the observed sample contains all galaxies that have a reliable redshift measurement. To
formalise the `observed' galaxy density, we consider the probability $P_{\rm det}$ that
an ELG enters the observed sample with a correct redshift, meaning that its {\ha} line is
detected and correctly interpreted 
(see also Sect.~\ref{sec:redshift_errors}).
This probability will depend on several parameters; on
the one hand, it will depend on the pixel-level properties of the image in the stripe
where the galaxy spectrum is dispersed, and more generally on several details of the
observation and data reduction process. These details can be represented by a set of
nuisance maps $\{N_i(\nh,z)\}$, which mostly depend on the sky position $\nh$ and may have
a secondary dependence on redshift.

Detection will also depend on the intrinsic properties of the galaxy beyond its {\ha} flux
$f$ and redshift $z$. At the NISP spectral resolution, {\ha} is blended with the
{\nii} doublet, so the signal will be given by the whole {\ha}+{\nii} complex; then the
list of galaxy properties that determine its detection will include the {\nii} line
luminosity, which is related to galaxy metallicity. Other factors that modulate detection
probability are the line fluxes of all other relevant emission lines, the galaxy
magnitudes {\IE}, {\YE}, {\JE}, and {\HE}, the galaxy size, morphology, ellipticity, and
inclination (see the discussion in Appendix~\ref{app:galaxyproperties}). We define the
completeness function $\Comp$ as the galaxy detection probability marginalised over all
galaxy intrinsic properties; calling for simplicity $\vec{p}$ the $m$-dimensional vector
of these properties, we have that

\be
\Comp(f,z,\{N_i\}) := \int {\rm d}^m p\, P_{\rm det}(f,z,\{N_i\}\, |\, \vec{p})\; .
\label{eq:detection_probability}
\ee

\noindent
This completeness function $\Comp(f,z,\{N_i\})$ gives the fraction of galaxies with flux
$f$ at redshift $z$ that enter the catalogue with a correct measurement of redshift. As
such, it will automatically account for Malmquist bias induced by line flux or by any
other observable property. Galaxies included in $\Comp$ that have a correct redshift
measurement are called `correct' galaxies, to emphasise their difference from the
target galaxies. Within the \Euclid data reduction pipeline, the detection probability
$P_{\rm det}$ is identified as the visibility mask of the survey. We consistently
identify $\Comp$ as its marginalised visibility mask (MVM).

The effectiveness of disentangling overlapping spectra is expected to be a
  function of source number density. In principle, this adds a dependence of the MVM on
  the density field itself, that would create a coupling between systematics and
  cosmological signal. In {\citepassa} we show that this coupling is expected to be weak,
  because contamination is due to the spectral continuum and galaxies at $z>0.9$ are
  typically too faint to have their continuum detected in the EWS. This is discussed in
  more detail in Appendix~\ref{app:confusion}.

The MVM will drop to zero at $f\ll f_0$, but this cut will not be infinitely sharp, so the
catalogue will always contain galaxies that are fainter than the nominal flux limit $f_0$.
The observed number density of correct galaxies, $n_{\rm oc}$, will be

\begin{align}
n_{\rm oc}(\vx) &= \int_0^\infty  {\Comp}(f, z, \{N_i\})\, \phi_{\rm local}(f|\vx)\, {\rm d}f \nonumber\\
&\simeq [1+\delta_{\rm g}(\vx)]\int_0^\infty  {\Comp}(f, z, \{N_i\})\, \Phi(f|z)\, {\rm d}f\, ,
\label{eq:obsdens_correct}\end{align}

\noindent
where the nuisance maps $\{N_i\}$ are evaluated at the sky position {\nh} and the second
equation is exact if Eq.~(\ref{eq:universal_LF}) holds. The average of this density can be
computed by assuming that $\delta_{\rm g}$ is statistically independent of $\Comp$,
meaning that there are no systematics that depend on the cosmological signal. One can
further assume that $\langle\delta_{\rm g}\rangle_z=0$, to obtain

\be
\overline{n}_{\rm oc} = \int_0^\infty \overline{\Comp}(f, z)\, \Phi(f|z)\, {\rm d}f\, ,
\label{eq:mean_obsdens_correct}\ee

\noindent
where $\overline{\Comp}(f, z) := \langle \Comp(f, z, \{N_i\}) \rangle_z$.\footnote{
Forcing the density of a sample to be equal to the average density in
each redshift bin amounts to forcing a strong radial integral constraint, removing radial
modes. We can therefore assume that Eq.~(\ref{eq:mean_obsdens_correct}) holds when
averaging over many realisations of a survey.}

\subsection{Redshift errors}
\label{sec:redshift_errors}

The density in Eq.~(\ref{eq:obsdens_correct}) is a function of the true galaxy
(redshift-space) comoving coordinate $\vx$. While we can assume that the uncertainty on
the galaxy sky position $\nh$ is negligible, the galaxy observed redshift $z_{\rm o}$ will
differ from the true $z$. For a correct galaxy, where the {\ha} line is recognised and correctly interpreted, the
difference $\Delta z= z_{\rm o}-z$ will be equal to the error in the redshift measurement,
which is expected to be distributed with a standard deviation $\sigma_z \simeq 0.001
(1+z)$ \citep[see][]{Q1-TP007}.\footnote{
  This is the original requirement on the redshift error, but the error is in fact
  independent of redshift; in \cite{EP-Risso} we use a $\sigma_z=0.001$ at all redshifts.}
Calling the observed position of the galaxy $\vx_{\rm o}$, the observed
density of correct galaxies is obtained by convolving $n_{\rm oc}(\vx)$ along the line of
sight with the redshift probability distribution function (PDF) $P_z(z_{\rm o}|z)$ that
represents the uncertainty, or the random error, of redshift measurements,

\be
n_{\rm oc}(\vx_{\rm o}) = n_{\rm oc}(\vx) \ast P_z(z_{\rm o}|z)\, .
\label{eq:obsdens_correct_obsz}\ee

The process of redshift measurement can give rise to catastrophic errors, where in
principle $\Delta z\gg \sigma_z$. The galaxies that enter the spectroscopic catalogue with
a catastrophically wrong redshift are called `interlopers'. We divide them into two
classes. `Line interlopers' are emission-line galaxies for which we detect a line,
different from {\ha}, that is, however, misinterpreted as {\ha}. `Noise interlopers' are
sources that do not have emission lines but acquire a false detection due to a noise
fluctuation or an instrumental feature that is not properly masked and is mistaken for
\ha. Noise interlopers may even be stars or spurious objects left over by the source
extraction algorithm.

\begin{figure}
  \centering
  \includegraphics[width=0.45\textwidth]{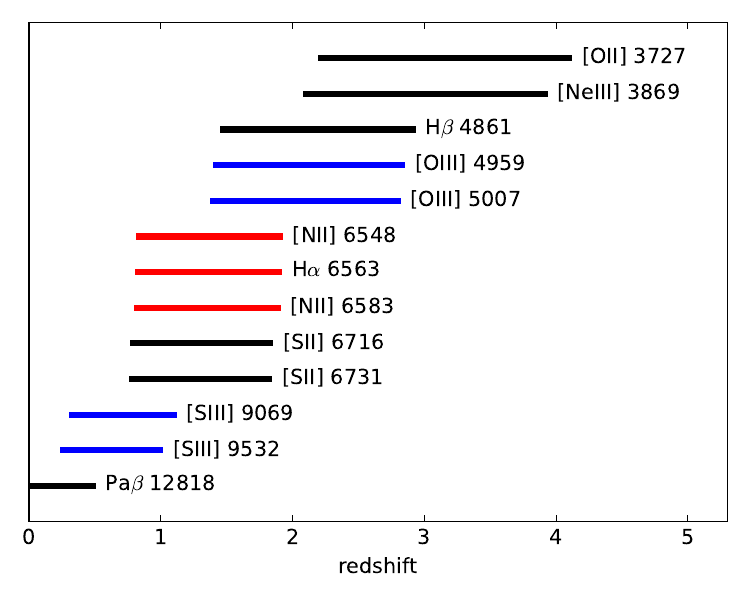}
  \caption{Main emission lines visible in the galaxy spectra, together with their
    visibility range for a detector that is sensitive in the wavelength range from 1.2 to
    1.9 $\mu$m. The red lines correspond to the {\ha} complex and the blue lines to the
    main line interlopers we expect in the EWS. In \Euclid spectra the {\ha}+{\nii}
    complex is not resolved, so the measured flux is integrated over the
    three lines.}
  \label{fig:emissionlines}
\end{figure}

Potential line interlopers are all those galaxies that have an emission line. In
Fig.~\ref{fig:emissionlines} we report the main emission lines that might be visible in
the NISP spectra, where the {\ha} complex is denoted in red; each segment denotes the
redshift range in which the line is visible. While the existence of an emission line opens
the possibility of having a line interloper, the likelihood of this to happen depends on
many crucial details, and a one-to-one mapping of lines to line interlopers yields only an
upper limit to their number.

Assuming an unspecified number of line interlopers, we call $n_{{\rm o}i}(\vx_{\rm o})$,
with $i$ an integer number, the number density of the $i$-th line interloper, and $n_{\rm
  on}$ the number density of noise interlopers. The observed galaxy number density can
thus be written as

\be
n_{\rm o}(\vx_{\rm o}) = n_{\rm oc}(\vx) \ast P_z(z_{\rm o} | z) + \Sigma_i n_{{\rm o}i}(\vx_{\rm o}) + n_{\rm on}(\vx_{\rm o})\, .
\label{eq:obsdens_all}\ee

\noindent
The role of interlopers is analysed in depth in \cite{EP-Risso} and \citelee; here the
number densities of interlopers are directly reported in the observed coordinates
$\vx_{\rm o}$, thus leaving the convolution with the redshift PDF implicit.

\subsection{Purity and completeness of a galaxy sample}
\label{sec:P&C}

The need for denser spectroscopic samples pushes us to include galaxies that have redshifts based on 
spectra with low S/N. This means that  a sample that is optimised for cosmology may still include a significant
fraction of interlopers. The quality of the sample can be quantified through the concepts
of purity and completeness, following the definitions of \citepype.

The spectroscopic sample will consist of all sources for which: (i) a reliable redshift is
provided by the data reduction pipeline; and (ii) it lies in the interval $[0.9,1.8]$. We
call $\overline{n}_{\rm o}(z)$ the average density of such sample at redshift $z$. We
define `sample purity' as the fraction of galaxies in the spectroscopic sample that: (i)
are correct galaxies (their {\ha} line is detected and their redshift is correctly
measured); and (ii) are target galaxies (their {\ha} line has a true flux $f\ge f_0$). The
numerator of this quantity, which we call $\overline{n}_{\rm ot}$, can be formalised as

\be 
\overline{n}_{\rm ot}(z) = \int_{f_0}^\infty  \overline{\Comp}(f, z)\, \Phi(f|z)\, {\rm d}f\, .
\label{eq:mean_obsdens_target}\ee

\noindent
The difference with $\overline{n}_{\rm oc}(z)$ of Eq.~(\ref{eq:mean_obsdens_correct}) lies
in the lower limit of the integral: correct galaxies with $f<f_0$ are not target galaxies.
Sample purity is then

\be
{\rm sample\ purity} := \frac{\overline{n}_{\rm ot}(z)}{\overline{n}_{\rm o}(z)}\, .
\label{eq:spurity}\ee

\noindent
This quantity is a function of redshift, but it is easy to define a sample purity for the
whole sample by computing the numerator and the denominator over the whole redshift range
of the survey.

The MVM $\Comp(f,z)$ is not expected to be a Heaviside step function in true flux $f$,
since any measure of redshift reliability would correlate with flux with some unavoidable
scatter. Due to this fact and to the steepness of the LF $\Phi(f|z)$ at $f_0$, a fair
fraction of correct galaxies will not be target galaxies and will not contribute to sample
purity. Nonetheless, these galaxies trace the same matter density field and are thus
useful for cosmology. We thus define `redshift purity' as the fraction of correct galaxies
in the spectroscopic sample,

\be
{\rm redshift\ purity} := \frac{\overline{n}_{\rm oc}(z)}{\overline{n}_{\rm o}(z)}\, .
\label{eq:zpurity}\ee

\noindent
While sample purity is of interest for assessing the performance of the instrument,
redshift purity is the quantity that is mostly relevant for cosmology.

We define `sample completeness' as the fraction of target galaxies that are present in the
sample as correct galaxies,

\be
{\rm sample\ completeness} := \frac{\overline{n}_{\rm ot}(z)}{\overline{n}_{\rm t}(z)}\, .
\label{eq:completeness}\ee

\noindent
This definition ignores galaxies below the line flux threshold $f_0$, but has the
merit of quantifying the ability of the telescope to detect target galaxies. From
Eqs.~(\ref{eq:average_nt}) and (\ref{eq:mean_obsdens_target}), it is clear that sample
completeness is the average of the completeness function $\Comp$, weighted by the LF
$\Phi$.

Finally, it is convenient to define the `redshift efficacy' of the spectroscopic sample as
its ability to include galaxies with correct redshifts compared to the target sample,

\be
{\rm redshift\ efficacy} := \frac{\overline{n}_{\rm oc}(z)}{\overline{n}_{\rm t}(z)}\, .
\label{eq:efficacy}\ee

\noindent
Contrary to what happens to the other quantities, redshift efficacy is not constrained to
be $\le1$.

If the sample is selected at lower S/N of galaxy spectra, sample completeness and redshift
efficacy increase while redshift purity decreases, and the optimal point in a redshift
purity versus sample completeness plane should be decided by maximising the figure of
merit of the final results. In \citepype it is shown that, based on simplified end-to-end
simulations of the measurement process, an optimal combination is achieved at a redshift
purity of 80\% and sample completeness of 50\%. This implies that 20\% of the
spectroscopic sample will be interlopers. These simulations also show that the main line
interlopers are expected to be detected through their {\oiii} and {\siii} lines (denoted
in blue in Fig.~\ref{fig:emissionlines}).

\subsection{The observed density contrast}
\label{sec:contrast}

In the analyses presented in this paper we do not consider redshift errors, meaning that
our simulations  have no interlopers and  consistently ignore the small difference
between the true position of the galaxy $\vx$ and the observed one $\vx_{\rm o}$. Then for
the rest of the paper

\be n_{\rm o}(\vx_{\rm o}) = n_{\rm o}(\vx) = n_{\rm oc}(\vx)\, . \label{eq:thispaper}\ee

The estimation of the galaxy average density will rely on the generation of a random
catalogue, which will represent the distribution of a dense set of unclustered galaxies
subject to known observational systematics, and will be the concrete representation of the
visibility mask (see Sect.~\ref{sec:random}). However, the construction of the random
catalogue has its uncertainty and the true visibility will be unknown; the random
catalogue will be based on an approximated visibility mask, whose marginalisation will
give the approximate MVM $\Compa(f,z,\{N_i\})$, which represents our best knowledge of the
true $\Comp(f,z,\{N_i\})$.

The number density of random galaxies will be

\be
n_{\rm r}(\vx) = N_{\rm r} \, \int_0^\infty \Compa(f,z,\{N_i\})\, \Phi(f|z)\, {\rm d}f\, ,
\label{eq:random_dens}\ee

\noindent
where $N_{\rm r}$ is the replication factor, which is planned to be $N_{\rm r}=50$ (see
\citealt{EP-DeLaTorre}; Euclid Collaboration: Salvalaggio et al., in prep.). This density
acquires a dependence on position due to the angular variation of the nuisance maps. We
call $\alpha := 1/N_{\rm r}$ the inverse of the replication factor. The random density
$n_{\rm r}$, having both an angular and a redshift dependence, is in fact a function of
the comoving position $\vx$. It is important to note that the random catalogue will be
constructed such that $\alpha\,\overline{n}_{\rm r}(z)$, averaged over angular
coordinates, reproduces $\overline{n}_{\rm o}(z)$, not $\overline{n}_{\rm t}(z)$; the two
average densities of observed and target galaxies will in general be different. The
observed galaxy density contrast will then be defined as

\be
\delta_{\rm o}(\vx) := \frac{ n_{\rm o}(\vx) - \alpha\, n_{\rm r}(\vx)}{\alpha\, n_{\rm r}(\vx)}\, .
\label{eq:delta_with_random}\ee

In Eq.~(\ref{eq:random_dens}) we have assumed an approximate knowledge of the true MVM
$\Comp$, but the integral also contains the true galaxy luminosity function $\Phi(f|z)$.
This quantity is obtained from the EDS, under the assumption that the latter provides
a highly pure and complete representation of the galaxies visible in the EWS. We 
discuss below how inaccurate knowledge of $\Phi(f|z)$, due to sample variance of the EDS,
can affect the construction of the random catalogue; here for simplicity we absorb this
uncertainty into that of $\Comp$, and assume that the equality of $n_{\rm
  r}$ and $\overline{n}_{\rm o}$ can be forced by suitably calibrating the detection
probability.

When interlopers are present in the catalogue, the random catalogue will also represent
interlopers as separate classes of unclustered galaxies subject to the relevant
systematics, which are not the same as those of correct galaxies. We return to this
point in Sect.~\ref{sec:random}, where the construction of the random catalogue is
presented in detail.

Under the approximation that Eq.~(\ref{eq:universal_LF}) is valid, the observed galaxy
density contrast (Eq.~\ref{eq:delta_with_random}) is equal to the galaxy density contrast
only if we have perfect knowledge of the MVM, $\Compa=\Comp$. Following
\cite{Colavincenzo2017} we define the angular mask $A(\nh, z)$ as

\be
A(\nh,z) := \frac{\int_0^\infty \Comp(f,z,\{N_i\})\,\Phi(f|z)\, {\rm d}f}
{\int_0^\infty \Compa(f,z,\{N_i\})\,\Phi(f|z)\, {\rm d}f} -1\, .
\label{eq:A_def} \ee

\noindent
Here the notation emphasises that $A$ will mostly depend on angular coordinates, with a
possible additional redshift dependence. This angular mask is a quantification of the
relative error on the MVM, and it vanishes if this is perfectly known. Being a
perturbation of the galaxy density, it propagates to the galaxy density contrast as

\be
1+\delta_{\rm o}(\vx) \simeq [ 1+\delta_{\rm g}(\vx) ]\, [ 1 + A(\nh, z) ]\, ,
\label{eq:deltao_vs_deltag}\ee

\noindent
and this shows that modulations of the flux limit imprint a systematic effect to
$\delta_{\rm g}$ which is neither additive nor multiplicative. The angular dependence of
$A$ will be determined by how the MVM, and our lack of knowledge of it, is sensitive to
the set of nuisance fields $\{N_i\}$. It is important to remark here that this apparently
simple relation between $\delta_{\rm o}$ and $\delta_{\rm g}$ is obtained by assuming that
galaxy bias is independent of luminosity.

In Appendix~\ref{app:flux_modulations} we discuss, both under the assumption of
luminosity-independent galaxy bias (Eq.~\ref{eq:universal_LF}) and more generally, how to
compute the response of $n_{\rm o}$ to nuisance maps. This is related in a non-trivial way
to the shape of the galaxy luminosity function, strengthening the argument of the
convenience of using mock catalogues to characterise the visibility mask.

\begin{table}
\caption{\label{table:definitions}Frequently used definitions.}
\begin{center}
\begin{small}
\begin{tabular}{ll}
\hline\hline
{\bf Target galaxy sample}    & all galaxies with $f>f_0$ \\
{\bf Observed galaxy}         & all galaxies with reliably measured \\
{\bf \ \ \ \ sample}          & redshift $z\in[0.9,1.8]$\\
{\bf Correct galaxy sample}   & the subset of observed galaxies that\\
                              & have a correct redshift measurement\\ 
{\bf (True) MVM}              & marginalised visibility mask,\\
                              & $\Comp(f,z,\{N_i\})$\\
{\bf Approximate MVM}         & our best knowledge of the MVM, \\
                              & $\Compa(f,z,\{N_i\})$\\
{\bf Tabulated detection}     & numerical representation of the MVM\\
{\bf \ \ \ \ probability}     & \\
{\bf Master data}             & 50 EuclidLargeMocks representing \\
{\bf \ \ \ \ catalogue}       & galaxies with $f>10^{-16}$\,\flux\\
{\bf Master random}           & a set of unclustered galaxies with \\
{\bf \ \ \ \ catalogue}       & number density $N_r=50$ times that of \\
                              & 1000 EuclidLargeMocks\\
{\bf Ad hoc random}           & a uniform random catalogue with no  \\
{\bf \ \ \ \ catalogue}       & angular systematics\\
{\bf Ideal mitigation}        & obtained using a random catalogue  \\
                              & selected with the true MVM\\
{\bf Realistic mitigation}    & obtained using a random catalogue   \\
                              & selected with the perturbed MVM\\
{\bf No mitigation}           & obtained with ad hoc random catalogue\\
{\bf \textit{Baseline} systematics}    & a combination of image noise, MW \\
                              & extinction and exposure time map\\
\end{tabular}
\end{small}
\end{center}
\end{table}

\section{A classification of systematics}
\label{sec:classification}

\subsection{Expected number of sources in a ROS}
\label{sec:numberofsources}

As a first step, it is useful to quantify the number of sources that are expected to be
found in a ROS; these numbers give a clear idea of the complexity of the process
  of redshift measurement and help understanding why we expect a high fraction of interlopers.
The results of this computation are reported in
Table~\ref{table:number_of_sources}. Our main interest here is to understand what happens
at the level of NISP spectroscopic detection, so we  quantify the sources that fall on
the NISP focal plane. Since the dithering pattern covers the detector gaps, we  refer
here to the whole focal plane including gaps, amounting to $0.55$ {\sqdeg}; each pixel of
18 $\mu$m corresponds to an angle of \ang{;;0.3}, so the field area translates to
$8\times10^7$ pixels. The number of sources with $\HE<24$ is obtained from the Flagship
galaxy mock catalogue, which is calibrated on a set of observations
\citep[see][]{EuclidSkyFlagship}. An assessment of the number of ELGs with line flux $f\ge
f_0$ is provided by the WISP survey \citep{Bagley2020}, which however has a smaller
wavelength coverage of $1.25\, \mu {\rm m} <\lambda<1.7\, \mu{\rm m}$. To extend WISP
number densities to \Euclid's redshift range we use the Flagship catalogue, specifically
computing the ratio between galaxy counts in the WISP and \Euclid wavelength ranges and
using this ratio to correct WISP counts.

\begin{table}
\caption{\label{table:number_of_sources} Estimates of the number of expected sources in a typical ROS.}
\begin{center}
\begin{tabular}{lr}
\hline\hline
sources in NISP & Number \\
\hline
$\HE<24$ galaxies                & 100\,000 \\
$\HE<20$ galaxies with continuum & 1900 \\
{\ha}+{\nii} & 2100 \\
{\oiii}  &  200 \\
{\siii}  &   80 \\
tot. ELGs  & 2380 \\ 
stars                   & 5000\,--\,15\,000\\ \hline
\end{tabular}
\end{center}
\tablefoot{All numbers are valid at the order-of-magnitude level. The number of ELGs
  refers to the main line visible in the grism, so that {\oiii} and {\siii} counts refer
  to galaxies where the {\ha} complex is not observed in the spectrum, while {\ha}+{\nii}
  galaxies may have other lines visible. All emission lines are counted when $f_{\rm
    line}\ge f_0$. }
\end{table}

We approximate the photometric catalogue to be limited to the magnitude $\HE<24$, where we
expect to find $1.8\times 10^5$ galaxies deg$^{-1}$; this implies $10^5$ galaxies per NISP
field. In the dispersed images, each galaxy projects a $530\times5$ pixel stripe on the
focal plane, so galaxies will nominally cover $2.7\times10^8$ pixels, more than three
times the number of actual pixels in the focal plane; this means that the focal plane is
in principle completely covered by potential source spectra. However, in a typical image
the continuum falls below the background for $\HE>20$ \citeppassa, giving $\sim$\,$1900$
galaxies per field, whose spectral tracks cover 6\% of the pixels. Taking the nominal flux
limit $f_0$ as a threshold for line detection, we expect to have 2100 {\ha}+{\nii}
complexes in the field, plus 280 other lines, mainly {\oiii} and {\siii}. The bulk of
these ELGs will have effective radii of about \ang{;;0.25}, with a very small fraction
going beyond \ang{;;1} \citeppassa. A pixel in the detector plane subtends an angle of
\ang{;;0.3}, while light is dispersed on the detector at 13.54 {\AA} pixel$^{-1}$, which
translates to a velocity from 210 to 340 km s$^{-1}$ pixel$^{-1}$. In these conditions the
flux of the {\ha} complex of a typical ELG is dispersed on an area of $5 \times 5$ pixels
\citep{Q1-TP006}. The lines will then cover a negligible fraction, less than 0.1\%, of the
NISP pixels. To these numbers we should add stars, which are quantified in
\cite{Scaramella-EP1} to range from 10\,000 to 30\,000 deg$^{-2}$, which translates to
5000 to 15\,000 per ROS. Stars will add to the contaminating sources, with tracks covering
from 16\% to 50\% of the pixels, although, as for the galaxies, their continuum will fall
below the noise level when their magnitude is $\HE>20$.

\begin{figure*}[htbp!]
\centering
\includegraphics[width=0.9\hsize]{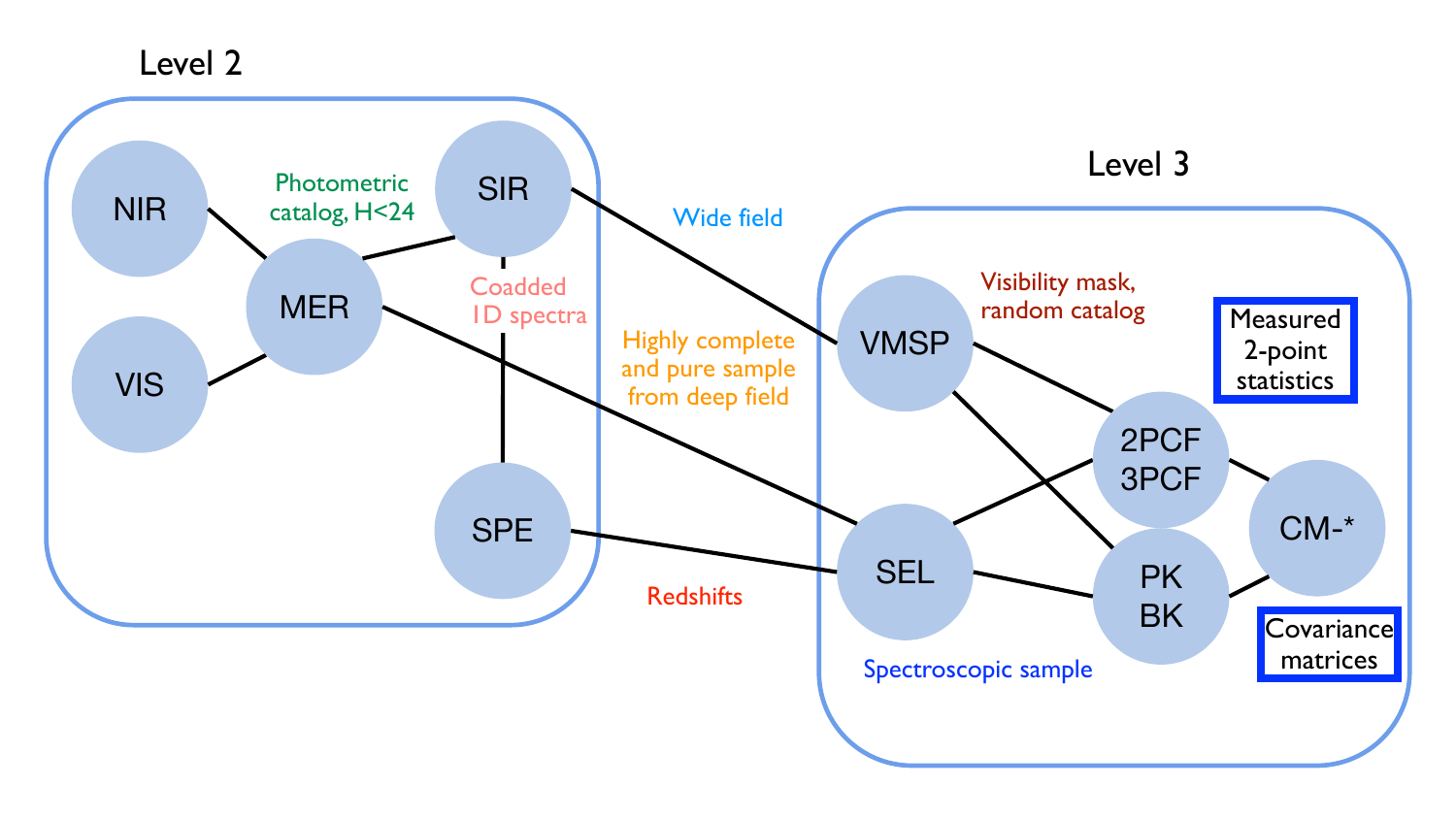}
\caption{Schematic view of the SGS pipeline for galaxy clustering. The two boxes mark
  level 2 and level 3 processing; the circles correspond to OUs in the left box
  and to OU-LE3 PFs in the right box. The black segments outline the dependences of the
  various blocks, some of which are labelled by coloured text. The text within the
  blue boxes gives the final products of the galaxy clustering pipeline.}
\label{fig:OUpipeline}
\end{figure*}

\subsection{The SGS pipeline}
\label{sec:pipeline}

The data reduction and analysis are performed using a pipeline developed by \Euclid's
Science Ground segment (SGS). This structure is organised in operational units (OUs),
which develop the data processing codes where each step is performed by a processing
function (PF).

The analysis and reconstruction of dispersed images is based on detailed input source
catalogues provided by the VIS \citep{Q1-TP002}, NIR \citep{Q1-TP003}, and MER PFs
\citep{Q1-TP004}. VIS and NIR perform the extraction of photometric sources on single and
stacked images, while the merging of their products is computed by the MER PF, which
detects objects in all the \Euclid imaging data, measures their properties, and prepares a
single multiwavelength catalogue that includes sources that are identified in either
visible or NIR images. One crucial point in MER processing is image deblending, which is
accomplished by two density-based clustering algorithms tuned on pixel-level simulations.
The MER catalogue of photometric sources is expected to have a depth in the {\HE}-band of
approximately 24.

The SIR PF \citep{Q1-TP006} processes the NISP dispersed images and produces fully
calibrated one-dimensional spectra for (potentially all) sources in the MER catalogue. SIR
performs calibration in wavelength and flux, corrects instrumental effects, subtracts
cross-contaminations of different spectra, and finally produces one-dimensional spectra
for scientific use. A crucial step of SIR is the decontamination of overlapping spectra.
As shown above (Sect.~\ref{sec:numberofsources}), galaxies with $\HE<24$ completely fill
the detector plane in the dispersed images; on average, one can expect that the first-order
spectral trace of a source of interest overlaps with 10 to 30 other spectral traces.
Contamination can also be induced by the zeroth-order image of relatively bright sources
and, to a lower extent, by other orders. Decontamination is performed starting from NIR
photometry, which gives a robust estimate of the flux to be subtracted out. Modelling of
the dispersed spectrum makes use of other photometric information from the MER catalogue,
such as source brightness profile, ellipticity, and inclination. NISP detectors are affected
by what is called the persistence of the signal (see Appendix~\ref{app:persistence}). This
can be masked out or subtracted, making use of the information from previous images in the
observing sequence, building a persistence model that estimates the level of contamination
of each pixel.

Starting from one-dimensional spectra produced by SIR, the SPE PF \citep{Q1-TP007}
performs redshift measurement, reliability estimation, and spectroscopic classification of
all the extracted sources. SPE applies a modified version of the algorithm for massive
automated $z$ evaluation and determination (\texttt{AMAZED}), which classifies the spectra
into three categories: galaxies, stars, and quasars, with an associated probability of
belonging to a specific class. Then, for each spectral class, the five best redshift
solutions are computed, each complemented with an estimate of the measurement reliability.
To maximise sample completeness, SPE applies a prior that favours solutions containing the
{\ha} line. The final SPE output is a catalogue that, for each source, contains a list of
redshift solutions, and for each redshift solution it provides line flux, S/N, central
wavelengths, and full width at half maximum for each line.

The resulting catalogue is analysed by OU-LE3, responsible for the galaxy clustering
measurements. The first step of this processing is the selection of a sample and the
construction of the random catalogue, performed respectively by SEL and VMSP. In
particular, VMSP creates a master random catalogue, where each galaxy is assigned full
galaxy properties (from the EDS) and a detection probability as described below in
Sect.~\ref{sec:random}. The selection of the master data and random catalogues is
performed by the SEL PF, following criteria that will in general be aimed at maximising
the figure of merit of cosmological parameters. In addition to selecting the data and random
catalogues, SEL estimates the (sample) purity and completeness of the data catalogue in
the areas covered by the EDS.

The selected data and random catalogues are then fed to the other LE3 PFs, named 2PCF
\citep{EP-DeLaTorre} and 3PCF (Euclid Collaboration: Veropalumbo et al., in prep.) for the
two-point and three-point correlation functions in configuration space, PK \citeppk and BK
(Euclid Collaboration: Rizzo et al., in prep.) for the corresponding power spectrum and
bispectrum in Fourier space. The PK code is described below in Sect.~\ref{sec:inference}.
At the end of LE3 pipeline, the covariance matrix PFs, CM-2PCF and CM-PK, take care of
computing the covariance of two-point estimators in configuration and Fourier space (see
Sect.~\ref{sec:covariances}).

A detailed discussion of all the potential entry points of systematics along the
pipeline, from photometry to redshift measurement, is given in
Appendix~\ref{app:entrypoints}. We focus in the main text on the last part of the
pipeline, including a discussion on the random catalogue and on the computation of
covariance matrices.

\subsection{The random catalogue as a tool for mitigating angular systematics}
\label{sec:random}

A large part of the observational systematics will be mitigated by suitably constructing a
random catalogue that represents the visibility mask $P_{\rm det}$ (introduced in
Sect.~\ref{sec:contrast}). This will be done by the VMSP PF, relying on a bona fide sample
of galaxies drawn from the EDS that represents the galaxies for which a reliable redshift
measurement is possible. It is required that the sub-sample of target galaxies
(Eq.~\ref{eq:target_ng}) is recovered with a very high purity, greater than $99$\%.

VMSP picks up galaxies from this sample and places them in random positions on the sky,
then evaluates the probability $P_{\rm det}$ that the sources are detected at the EWS
depth. In principle this step could be done by injecting the source spectra in the four
NISP dispersed dithers, then processing the ROS with SIR and SPE; this complete process
will be performed for limited areas, but it would be too costly to populate the whole EWS
this way. Using VMSP, we resort to a bypass simulation algorithm, called \pypelid and
described in \citepype. For each random galaxy, VMSP reads the noise at the location of
the main emission lines in the NISP dithers. It then simulates the emission lines to
obtain their S/N. The mapping from S/N (called $S$ in the equations) to the detection
probability, $P_{\rm det}(S)$, is then given by a `detection model', that is assumed to be
a sigmoid function,

\be P_{\rm det}(S) = \frac{c} {1 + \left(\frac{S}{S_0}\right)^{-\beta} } \, , \label{eq:sigmoid}\ee

\noindent
where $c$, $S_0$, and $\beta$ are free parameters to be calibrated; in particular, the
function crosses $c/2$ at $S_0$ with a slope determined by $\beta$, and then levels off to
$c$.

\begin{figure*}
  \centering{\includegraphics[width=0.45\textwidth]{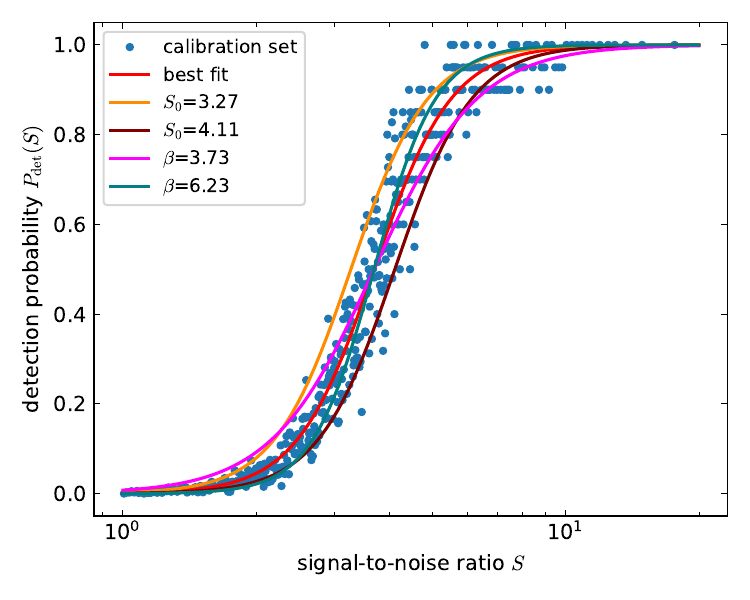}
    \includegraphics[width=0.45\textwidth]{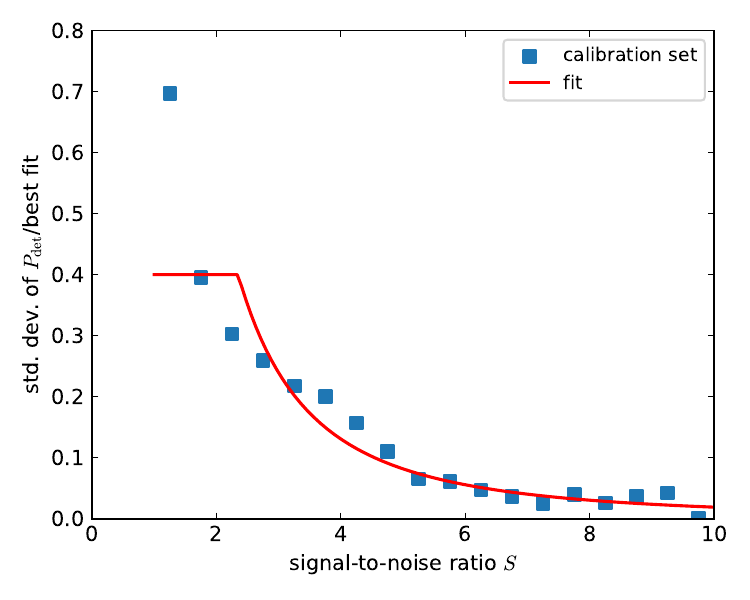}}
  \caption{\textit{Left:} Detection probability $P_{\rm det}(S)$, i.e. the probability
    of detecting a galaxy in the spectroscopic sample as a function of the S/N of its
    {\ha} line. The points give the measurements from the calibration set used to fix the
    parameters of the detection model. The five curves correspond to the best-fit
    detection model, together with those obtained by varying the two fitting parameters by
    1\,$\sigma$ (see the legend for the assumed parameter values). \textit{Right:} Standard deviation of the ratio of the measurements from the calibration set and the
    best-fit model.}
  \label{fig:detmodel}
\end{figure*}

Using the 15 wide-like passes of the EDS, it will be possible to estimate for these
galaxies both the S/N of their emission lines and their detection probability, thus
creating a calibration set for the detection model $P_{\rm det}(S)$. In \citepype we use a
calibration set obtained from spectroscopic simulations processed by SPE; the detection
probability thus relies on a choice of the threshold for spectrum reliability used in
sample selection. The $P_{\rm det}$ of this set is shown in the left panel of
Fig.~\ref{fig:detmodel}, and is fit by a sigmoid (Eq.~\ref{eq:sigmoid}) with parameter
values $S_0=3.69$, $\beta=4.98$ (the red line), while $c$ is found consistent with unity,
thus reducing the parameters to two. For later use, we adopt a simplified approach to
assign a 1\,$\sigma$ error to the parameters: we count the calibration points above and
below the curve, then vary one parameter at a time and mark when the smallest of these
numbers gets as low as 16\% of the whole set.\footnote
{To find the error in $\beta$, we count what is above the model for $S<S_0$ plus what is
  below it at higher $S$.}
This gives $S_0=3.69\pm0.42$ and $\beta=4.98\pm1.25$. Given that the calibration set is
small and the parameters may be correlated, we use this as a pessimistic estimation of the
error on the calibration of the detection model. Figure~\ref{fig:detmodel} shows four
sigmoid curves obtained by varying one of the parameters by 1\,$\sigma$. We also compute
the scatter around this relation; the right panel of Fig.~\ref{fig:detmodel} reports the
standard deviation of $P_{\rm det}$ from the calibration set divided by its best-fit
value; this scatter is reproduced by a function $\sigma_P=2.4\,S^{-2.1}$. This is capped
at the value $0.4$ to avoid divergence at very low $S$, which have little impact since
detection probabilities are very low in that range.

It is clear that such a detection model will be valid as long as the relation between $S$
and $P_{\rm det}$ is found to be tight in the calibration set, and the residuals are not
correlated to any other galaxy property; in case any secondary dependence of detection
probability is found, the algorithm will be revised to take them into account. It is also
worth noticing that the algorithm deals with the uncertainties in the dispersed images in
a realistic way, but assumes a perfect photometric measurement.

Adding line interlopers will be performed using a very similar approach: starting from all
EDS ELGs that have at least one emission line, the analysis of the 15 wide-like passes
will determine the probability that these galaxies are misclassified as {\ha} ELGs, then
{\pypelid} will compute the S/N of these spectra and a suitable (mis-)detection model will
be calibrated on these results and used to create a set of random galaxies that represent
line interlopers. For noise interlopers a different strategy will be adopted, with two
possible options: one is to add a completely unclustered set of points to the random
catalogue, another is to add a fraction of the photometric sample, equal to $N_{\rm r}$
times the small probability that a photometric source is mistaken as a noise interloper.
As a result, random galaxies will be labelled according to the kind of galaxies they are
supposed to represent, be it target, $i$-th line, or noise interloper.

Systematics coming from the EDS are addressed in \citebruton; we discuss here some
relevant points. For a $\sim$\,50 {\sqdeg} collective area of the EDS and a replication
factor $N_{\rm r}=50$, it is easy to see that each EDS galaxy will be replicated about
14\,000 times in the random catalogue. In particular, the {\ha} LF $\Phi(f|z)$ sampled by
EDS galaxies will be subject to significant fluctuations in redshift, due to sample
variance in the limited area surveyed. If propagated to the number density of the random
catalogue for the much larger EWS, these fluctuations would create spurious radial
modulations of the density contrast that, due to their anisotropy, would bias the higher
multipoles of two-point correlations. To mitigate this effect, we  fit the number
density of EDS galaxies with a smooth function, thus reducing the impact of sample
variance on $\bar{n}_{\rm r}(z)$ to the 1--2\% level \citepbruton.

The addition of interlopers  implies a further uncertainty, driven again by sample
variance. Because the interlopers come from a different region of the comoving space, the
sample variances of target galaxies and the various classes of line interlopers do not
cancel out in the interloper fractions.

If photometry is used to select the spectroscopic sample (for instance to limit
interlopers), the uncertainty in the photometry must be propagated to the random
catalogue, which in its standard version assumes exact magnitudes. While some effects of
over- or under-deblending, if significant, may require specific strategies, propagating
the error in photometry may simply require us to perturb the magnitudes of deep field
galaxies when injecting them into the master random catalogue; this alone would reproduce
any Malmquist-like bias due to selection based on photometry.

As a matter of fact, no information from the EWS is used to construct the random
catalogue, so there is no guarantee that $\overline{n}_{\rm r}(z)=N_{\rm r}\,
\overline{n}_{\rm o}(z)$ at all redshifts. A failure to achieve this equality would be a
clear sign of some problems in the pipeline and would prompt a reassessment of the
systematics budget and of the construction of the random catalogue (see
Sect.~\ref{sec:realistic}). However, this equality can be reached only up to the
uncertainty due to EDS sample variance; once this level of agreement is reached, it is
legitimate to rescale the number of random galaxies so that the random number density
$n_{\rm r}(z)$ is exactly equal to $N_{\rm r}\,\overline{n}_{\rm o}(z)$. Any effect that
perturbs the average number density by less than the sample variance of the EDS cannot be
recognised with this test.
Finally, sample variance of the EWS results in an uncertainty in the true number
  counts $\Phi(f)$, but these enter the observed density (Eq.~\ref{eq:obsdens_correct}) as
  a product with $\Comp$, so their uncertainty propagates to the uncertainty in
  the calibration of the detection model.

\subsection{Covariances}
\label{sec:covariances}

The calculation of covariance matrices is based on a brute-force computation of 2PCF and
PK of a set of thousands of mock catalogues \citep{EP-Monaco1}. The CM-2PCF and CM-PK PFs
then compute the sample average and covariance of the corresponding measurements; we 
call this the numerical covariance of the corresponding estimator. While trivial from
the technical point of view, these PFs will perform a crucial step in the analysis that
involves codes that are outside the \Euclid SGS pipeline.

Given a large set of mock galaxy catalogues, their numerical covariance allows us to take
into account sample variance due to the finite volume of the survey and to account for
mode coupling induced by non-linearities and by the window function. At the same time, the
random catalogue will mitigate the systematics induced by fluctuations of survey depth,
and in part by redshift errors. Due to the complicated (both additive and multiplicative)
relation that connects the galaxy density contrast with the observed one
(Eq.~\ref{eq:deltao_vs_deltag}, which is a simplification anyway), the contributions of
sample variance (coming through $\delta_{\rm g}$) and uncertainty on the visibility mask
(coming through the mask $A$) are intertwined in a complicated way, as shown for instance
by~\cite{Colavincenzo2017}. This makes a numerical approach to the problem unavoidable.

The uncertainty in this process may be obtained by perturbing each step in the
construction of the random catalogue and producing a set of thousands of random
catalogues, one for each mock, which represent the expected uncertainty in the mitigation
process. However, this is an inefficient way to tackle the problem; a much more convenient
option is to perturb the visibility mask applied to the galaxy mocks, while using a single
random catalogue for all clustering measurements, which represents the average visibility
mask (in principle the same random catalogue used for the observed data catalogue). It is
important to point out that imposing angular systematics to a galaxy mock catalogue is a
very similar operation to constructing the random catalogue, the main difference being
that galaxy positions are not randomly chosen. But each galaxy in the data or random
catalogue is described by several properties beyond sky coordinates and redshift,
indicated by the $\vec{p}$ vector of Eq.~(\ref{eq:detection_probability}). We 
describe in Sect.~\ref{sec:lookup_table} how it is possible to exploit the VMSP/{\pypelid}
algorithm to build an MVM, which depends only on {\ha} flux and redshift, thus enabling us
to impose angular systematics to thousands of catalogues that contain minimal information
for each galaxy.

\begin{table*}
\caption{\label{table:systematics}Table of expected systematics along the pipeline.}
\begin{small}
\begin{center}
\begin{tabular}{lllllll}
\hline\hline
{\bf Systematic effect}    & {\bf Entry} & {\bf Class} & {\bf Consequences}     & {\bf Mitigation} & {\bf Impact} & {\bf Risk} \\
                           & {\bf point} &             &                        & {\bf strategy}   &              &            \\
\hline
Fluctuation of photometric depth & MER   & angular    & loss of sources for SIR    & random        & low     & low     \\ 
+ photometric selection          & +SEL  &            & modulation of catalogue    & +photo errors &         & mild    \\ \hline
Under-deblending                 & MER   & angular    & loss of sources for SIR    & none          & v. low  & v. low  \\
                                 &       &            & mis-centred sources        &               &         &         \\ \hline
Over-deblending                  & MER   & noise int. & fake sources               & random        & low     & low     \\ \hline
Persistence: spectro to photo    & MER   & noise int. & fake sources               & random        & mild    & low     \\ \hline \hline

Zodiacal light                   & astro & angular    & increased background       & random        & v. high & v. low  \\ \hline
Nearby galaxies                  & astro & angular    & increased background       & random        & low     & low     \\ 
                                 &       &            & and confusion              &               &         &         \\ \hline
MW extinction                    & astro & angular    & decreased incoming flux    & random        & high    & v. high \\
                                 &       &            &                            & (incomplete)  &         &         \\ \hline
MW emission                      & astro & angular    & contribution to background & random        & v. low  & v. low  \\ \hline\hline

Flagged pixels                   & SIR   & angular    & degraded image quality     & random        & v. high & low     \\ 
(bad pixels, saturated stars,    &       &            &                            &               &         &         \\
cosmic rays, persistence)        &       &            &                            &               &         &         \\ \hline
Stray light                      & SIR   & angular    & increased noise            & random        & high    & low     \\ \hline
Persistence: photo to spectro    & SIR   & noise int. & fake emission lines        & random        & mild    & high    \\ \hline
Persistence: spectro to photo,   & SIR   & angular +  & biased decontamination     & none          & mild    & high    \\
photo to photo                   &       & noise int. &                            &               &         &         \\ 
spectro to spectro               &       &            &                            &               &         &         \\ \hline
Calibration of SIR noise         & SIR   & angular    & modulation of random       & none          & low     & mild    \\ \hline
Wavelength calibration           & SIR   & AP-like    & redshift bias              & none          & v. low  & v. low  \\ \hline
Confusion from foregrounds       & SIR   & angular +  & increased noise in spectra & random        & high    & low     \\
                                 &       & noise int. & potential interlopers      &               &         &         \\ \hline
Confusion from $z>0.9$ sources   & SIR   & angular    & correlated to signal       & no            & low     & v. high \\ \hline
Coaddition of dithers            & SIR   & angular    & error propagation          & random        & mild    & low     \\
                                 &       & noise int. &                            &               &         &         \\ \hline \hline

Templates for spectral fit       & SPE   & AP-like    & bias in redshift measure   & no            & low     & low     \\ \hline
Star-galaxy separation           & SPE   & noise int. & spurious sources           & random        & mild    & low     \\ \hline
Redshift uncertainty             & SPE   & theory     & anisotropic smoothing      & model         & mild    & low     \\ \hline
Line misidentification           & SPE   & line int.  & wrong redshift             & model         & high    & high    \\ \hline
Spectral reliability selection   & SPE   & noise int. & sub-optimal cleaning of    & random        & mild    & low     \\
                                 & +SEL  & + line int.& interlopers                &               &         &         \\ \hline
Photometric selection            & SPE   & noise int. & introducing photometric    & random        & high    & high    \\
                                 & +SEL  & + line int.& systematics                & +photo errors &         &         \\ \hline \hline

Galaxy properties                & astro & ---        & marginalised over by model & no            & none    & none    \\ \hline
Intrinsic alignments             & astro & theory     & detection correlated with  & no            & v. low  & mild    \\
                                 &       &            & tidal field                &               &         &         \\ \hline \hline

Sample variance in EDS           & VMSP  & radial     & fluctuations in $\overline{n}(z)$& random  & mild    & low     \\
                                 &       &            &                            & +mocks        &         &         \\ \hline
Sample variance in galaxy        & VMSP  & angular    & uncertainty in detection   & random        & low     & low     \\
properties from EDS              &       &            & probability                & +mocks        &         &         \\ \hline
Mapping of S/N to detection      & VMSP  & angular    & uncertainty in bypass      & random        & low     & low     \\
                                 &       &            &                            & +mocks        &         &         \\ \hline

\end{tabular}
\end{center}
\tablefoot{The columns report a very brief description of the systematic, its entry point,
  its classification (see the text), its consequences, the mitigation strategy, its impact
  on the overall survey, and the risk connected to its mitigation.}
\end{small}
\end{table*}

\subsection{Classification and ranking}
\label{sec:ranking}

The results of this discussion are reported in Table~\ref{table:systematics}, where each
entry corresponds to each of the systematics mentioned in this section and in Appendix
~\ref{app:entrypoints}. Each row of this table gives a very brief description of the
effect and its entry point, followed by its classification, a description of the
consequences of the effect, the implemented mitigation strategy, the impact of this effect
on the survey (i.e. how much it affects its quality), and the risk associated to its
mitigation (i.e. how difficult it to achieve an accurate mitigation).

This long list of systematics results in a classification, given in the third column of
the table: (i) angular, an effect that modulates the effective flux limit of the
spectroscopic catalogue on the sky; (ii) noise interlopers, an effect that adds spurious
objects to the catalogue that have a measured redshift unrelated to true redshift; (iii)
line interlopers, where a line is detected but misidentified; (iv) AP-like, leading to
a distortion (similar to the Alcock--Paczynski effect) in the mapping between measured
redshift and line-of-sight distance; (v) radial, leading to a contamination of the
galaxy density limited to the radial coordinate; and (vi) theory systematics, which
should be addressed by adding sophistication to the theory model.

The list of mitigation strategies is even more limited. Mitigation will mostly rely on the
random catalogue or on the theory model, with the support of a set of mock catalogues that
are accurately calibrated on the data.

\subsection{Comparison with other surveys}
\label{sec:review}

The treatment of systematics in ground-based spectroscopic galaxy surveys such as the
Baryon Oscillation Spectroscopic Survey \citep[BOSS;][]{Dawson2013}, the extended Baryon
Oscillation Spectroscopic Survey \citep[eBOSS;][]{Dawson2016}, and the Dark Energy
Spectroscopic Instrument \citep[DESI;][]{DESI2016} differs from the approach
outlined here for \Euclid. These surveys define their samples by applying selection
criteria to a pre-existing photometric parent catalogue, from which targets are selected
for spectroscopic follow up. This allows the assessment of sample completeness relative to
a well-defined parent catalogue. Moreover, the quality of such ground-based spectra makes
catastrophic redshift errors much less likely to happen. Conversely, potential angular
systematics affecting clustering statistics can be introduced in the target catalogue
(e.g. due to seeing variations in the photometric data or in the calibration of different
photometric systems) or during the subsequent spectroscopic observations (e.g. the
inability to place spectroscopic fibres on nearby targets, commonly referred to as fibre
collisions).

BOSS identified foreground stellar density as the primary source of systematics, with
additional contributions from variations in observing conditions such as seeing, sky
background, airmass, and MW extinction. These effects were mitigated using
regression-based corrections in the form of a systematics weighting scheme
\citep{Ross2012} applied to the galaxy sample. Additionally, specific weights were applied
to account for fibre collisions and redshift failures, ensuring a uniform sampling of the
underlying large-scale structure \citep{Reid2016}. eBOSS extended these methodologies
while introducing refinements to handle new galaxy tracers, including luminous red
galaxies (LRGs), ELGs, and quasars \citep{Ross2020}. Additionally, improvements were made
in fibre assignment corrections using the nearest-neighbour approach to mitigate fibre
collision effects \citep{Bautista2021, deMattia2021}.

The approach followed by DESI builds upon the methodologies developed for BOSS and eBOSS,
while incorporating more advanced techniques for systematics mitigation
\citep{DESI2024,Krolewski2025,Rosado2024,Ross2025}. One of the key innovations in DESI is the use
of machine-learning techniques to model and correct for imaging systematics
\citep{Rezaie2021}. Additionally, DESI employs improved fibre assignment algorithms to
minimise selection biases and maximise the completeness of spectroscopic observations
\citep{Bianchi2024}. In all cases, mock catalogues have played a crucial role in assessing
and validating these corrections \citep[e.g.][]{Kitaura2016}

The different observational approaches lead to a difference in the mitigation strategies,
with ground-based surveys mostly relying on a posteriori regression of the density maps
with known nuisance maps, as opposed to the forward modelling of the visibility mask,
performed on the dispersed images, that we plan to use. More specifically, here we decide
to weigh only the random galaxies, which is another way of expressing the construction of
the visibility mask, rather than weighting the data catalogue or the galaxy pairs. 
Nonetheless, the methodologies developed in
ground-based surveys provide valuable insights for {\Euclid} systematic mitigation
strategies, particularly in the treatment of angular systematics affecting the observed
galaxy number densities.

\section{Modelling angular systematics in power spectrum measurements}
\label{sec:tools}

In this section we present the methods that we use to model the impact of angular
systematics on the measured power spectrum of mock galaxy catalogues, and on the inferred
cosmological parameters. We introduce the mock galaxy catalogues, the nuisance maps, the
tabulated detection probability, and the codes to measure the power spectrum and perform
parameter inference from it.

\subsection{Mock galaxy catalogues}
\label{sec:mocks}

Several sets of thousands of mock galaxy catalogues were produced to support the study of
systematics in \Euclid's spectroscopic galaxy clustering; these are described in detail in
\cite{EP-Monaco1}. We  limit our description here to the EuclidLargeMocks set, which
is presented in that paper. This has been produced using the {\pinocchio} fast code
\citep{Monaco2002,Monaco2013,Munari2017}; this is based on excursion set theory,
ellipsoidal collapse, and Lagrangian perturbation theory (LPT), and can be seen as a halo
finder in Lagrangian space, where haloes are moved to their final position using a single
3LPT displacement. The configuration is that of a cubic box of 3380 {\hmpc} sampled by
$6144^3$ particles, for a total volume of $38.61\ h^{-3}\ {\rm Gpc}^3$ per simulation.
This allows us to reach a particle mass of $1.48\times10^{10}\ M_\odot$; being
{\pinocchio} a semi-analytic code, we can push the mass threshold down to 10 particles,
thus resolving haloes with $M_{\rm h}\ge1.48\times10^{11}\ M_\odot$. The total volume
sampled by these 1000 realisation is thus $38614\ h^{-3}\ {\rm Gpc}^3$, 11 times the
volume of the visible Universe.

Haloes were populated with galaxies using an halo occupation distribution (HOD) model
extracted from the Flagship galaxy mock catalogue \citep{EuclidSkyFlagship}, that is
calibrated on the {\ha} luminosity function given by model 3 of \cite{Pozzetti2016}. As a
result of this process, the average number density and average galaxy power spectrum of
the 1000 mocks were found to be consistent with the number density and galaxy power
spectrum of the Flagship galaxies, and this is true for different redshift bins and flux
limits.

The mocks were created using an idealised footprint of a circle of radius \tdeg, covering
2763 {\sqdeg} on the sky; this area is just larger than the 2500 {\sqdeg} originally
planned for \Euclid Data Release 1 (DR1), and can (almost) contain its north and south
islands, as described in \cite{Scaramella-EP1}. The volume subtended by a {\tdeg} circle
and lying at redshift from 0.9 to 1.8 can be fully immersed in the parent simulation box
of 3380 \hmpc, so our analysis is free from problems connected with repetitions of the
simulated box (see details in \citealt{EP-Monaco1}). This footprint was used as a
reference in the paper describing and testing the PK code \citeppk.

It is useful, for specific tests, to work with catalogues where the effect of
luminosity-dependent bias is removed. To achieve this, we binned galaxies in redshift,
with a small bin width of $\Delta z=0.01$, and for each bin we shuffled all the line
fluxes $f$ among galaxies belonging to that bin. As a consequence, a selection in flux
leads to a sparse sampling of galaxies, preserving their clustering amplitude, while the
redshift-dependent luminosity function is preserved (within the relatively fine redshift
binning). Each mock catalogues contain both standard and shuffled line fluxes.

The numerical covariance matrices used in this paper are obtained by measuring the power
spectra of all the 1000 galaxy catalogues, in a few relevant configurations. However, in
this paper we are more concerned with the average value of the power spectrum, so we 
routinely analyse a smaller number of realisations. To set the number, we define the
effective volume of a survey \citep{Tegmark1997} that covers a sky area with a solid angle
$\Omega$ from $z_1$ to $z_2$ as

\be V_{\rm eff} = \frac{\Omega }{3}\, \left[d^3_{\rm c}(z_2) - d^3_{\rm c}(z_1)\right]\,
\left[\frac{\overline{n} \,P(k)}{\overline{n}\, P(k)+1}\right]^2\, ,
\label{eq:effvol}\ee

\noindent
where $\overline{n}$ is the galaxy density in that redshift bin and the power spectrum
value is computed 
at $k=0.14$ {\kmpc} \citep[e.g. as in][]{FontRibera2014,DESI2016}.
We divide the survey into four redshift bins, bounded by the values 0.9,
1.1, 1.3, 1.5, and 1.8. For a survey area of 2763 {\sqdeg}, the effective volume of the
four redshift bins is, respectively, 0.60, 0.82, 0.87 and 1.32 \cgpc. 
We compute the
effective volume of the survey as the sum of these four values, amounting to 
3.60.
The total effective volume of the EWS is obtained by rescaling these figures to 14\,000
{\sqdeg}, and amounts to 18.7 \cgpc. 
We  mostly work with the second redshift bin,
$z\in [1.1,1.3]$, and we decided to use 50 realisations, amounting to $\sim$\,2.2 times the
EWS effective volume. Cosmological inference will be obtained by combining the four
redshift bins. In this case, five mocks cover $2763 \times 5 = 13815$ \sqdeg, just 1.3\%
lower than the 14\,000 \sqdeg expected final area of EWS, so the 50 mocks will be split
into ten groups of five mocks each, thus covering the EWS survey volume ten times.

\subsection{Nuisance maps: Exposure time, image noise, MW extinction}
\label{sec:maps}

\begin{figure}
  \includegraphics[width=0.45\textwidth]{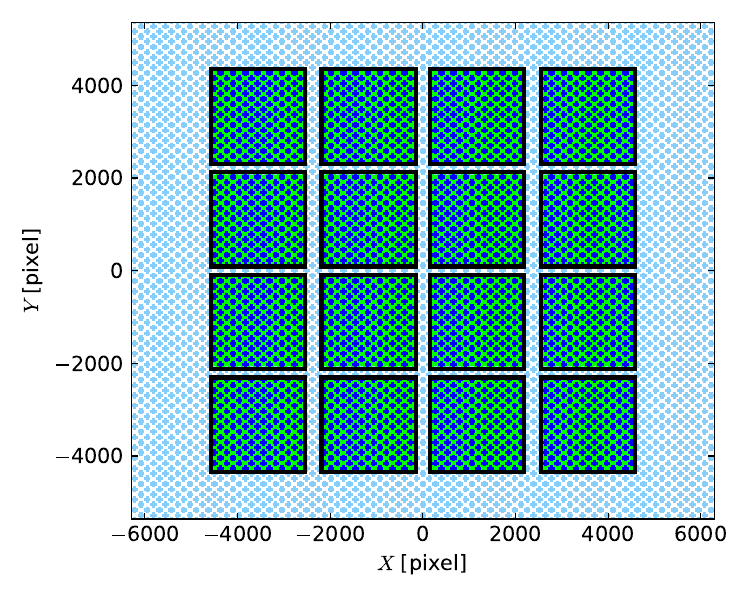}
  \caption{Adopted model for NISP detector geometry. The smaller squares denote the 16
    detectors; the crosses are the centres of \healpix sky pixels (with $N_{\rm side}=4096$)
    that are inside (blue) or outside (cyan) the detectors.}
  \label{fig:NISP}
\end{figure}

In our path towards complexity, we decided to immerse the {\tdeg} circle footprint in the
largest island of the planned final survey, thus creating a footprint with realistic holes
and dithering pattern but a regular overall shape similar to the one used in the PK paper.

To produce a realistic survey footprint, we created a model for the NISP detector
(Fig.~\ref{fig:NISP}) by tiling 16 squares of $2048^2$ pixels, exactly aligned on a grid,
where each pixel corresponds to \ang{;;0.299} \citep[see][for a very similar
  model]{Scaramella-EP1}. Gaps between detectors are identical in the $y$ direction
(amounting to 168.63 pixels), while the central gap in the $x$ direction is larger than
the other two (346.47 pixels versus 288.45 pixels). Moreover, the first eight external
pixels of the detectors were assumed to be not usable. This model was placed on the sky at
$(0,0)$ galactic angular coordinates, using a flat-sky approximation, and sampled with a
\healpix grid\footnote{
Hierarchical, Equal Area, and iso-Latitude Pixelation (\healpix) \citep{healpix} is a
code, widely used in cosmology, that performs a tessellation of the sphere into $N_{\rm
  pix} = 12\, N_{\rm side}^2$ sky pixels. Here we reserve the word `pixel' for the
detector pixels, and call `sky pixels' those related to a \healpix tessellation.}
of size $N_{\rm side}$ ranging from 512 to 8192. We stored the list of sky pixels whose
centre falls inside the detectors; Fig.~\ref{fig:NISP} reports the \healpix grid for
$N_{\rm side}=4096$. We checked that with $N_{\rm side}<2048$ the detector gaps are not
properly represented.

\begin{figure*}
  \centering
  \includegraphics[trim=25 5 25 5, clip, width=0.33\textwidth]{{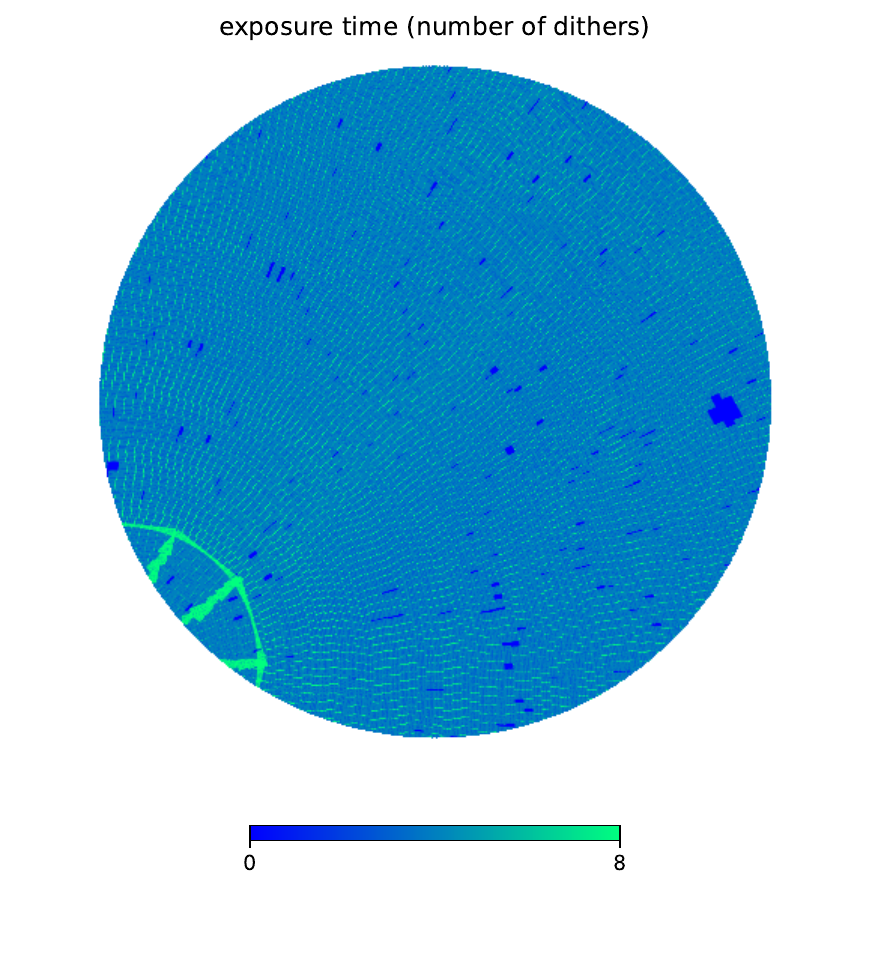}}
  \includegraphics[trim=25 5 25 5, clip, width=0.33\textwidth]{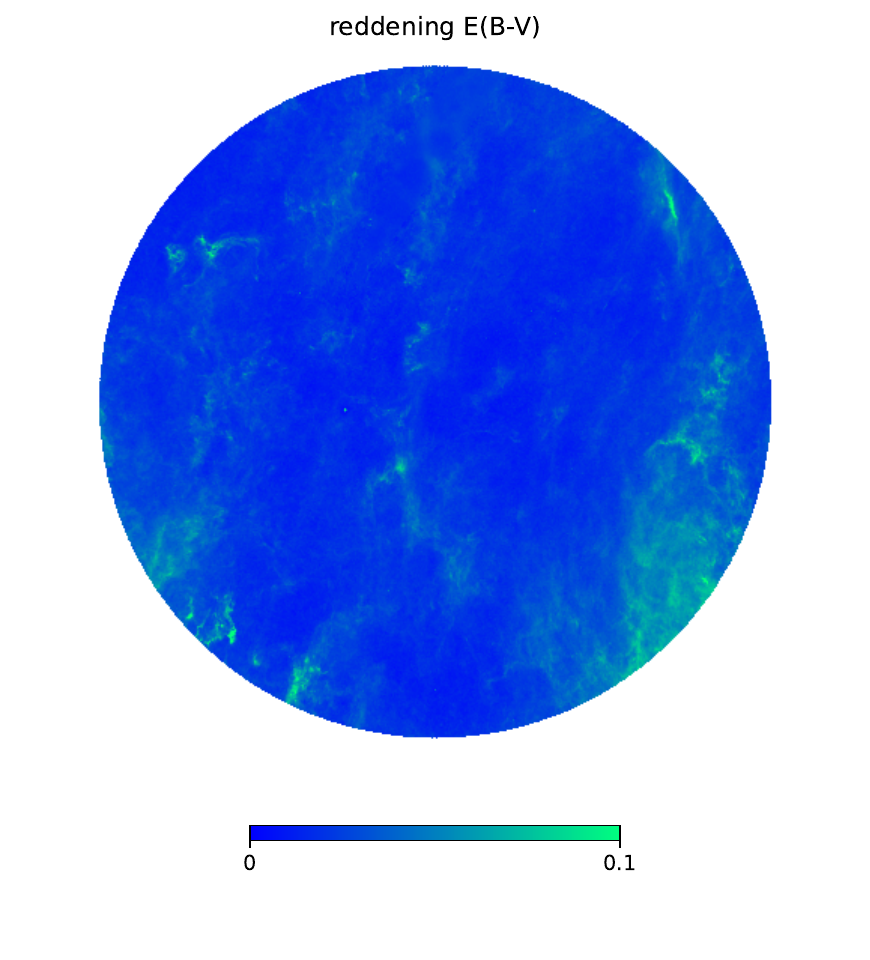}
  \includegraphics[trim=25 5 25 5, clip, width=0.33\textwidth]{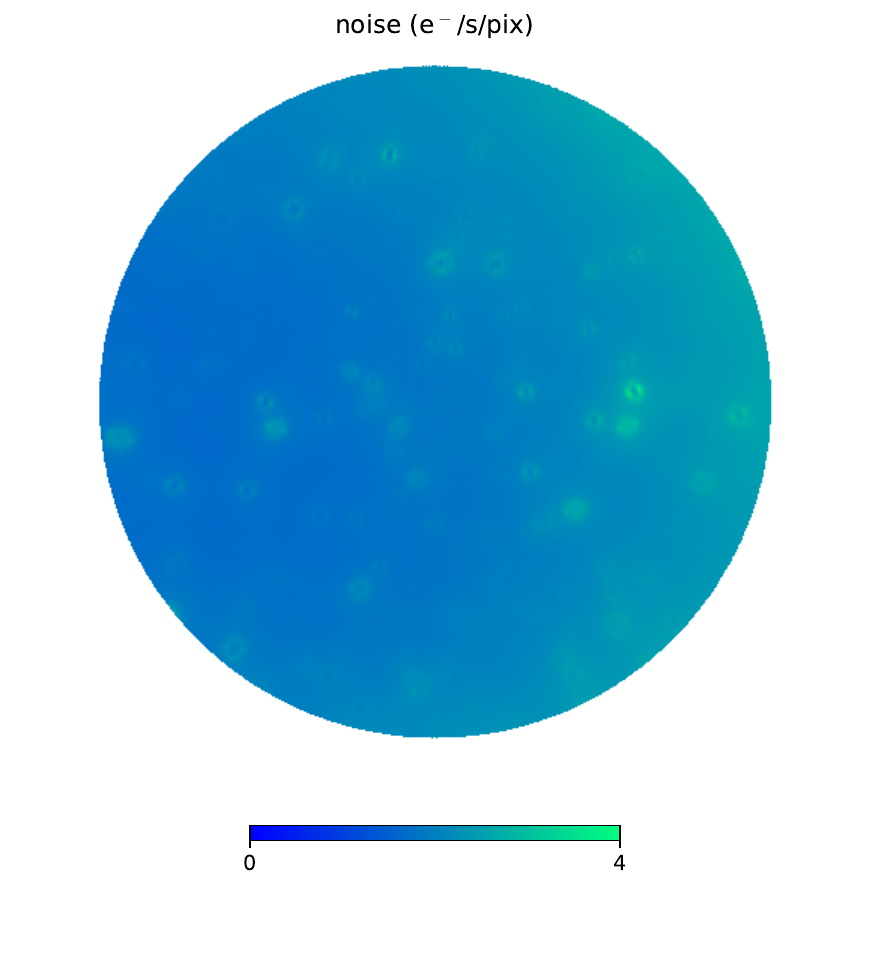}
  \caption{Gnomonic projection of maps used for the average MVM, the marginalised
    visibility mask. Left panel: Exposure time map at the photometric position of the
    galaxy, in units of the number of times (dithers) the sky pixel is visited; a full
    dithering sequence can provide up to four visits; higher values are found at pointing
    overlaps. Middle panel: $E(B-V)$ reddening map. Right panel: Noise map in units
    of \epixs.}
  \label{fig:maps}
\end{figure*}

The mapping from sky coordinates to position on the detector is well defined for a
photometric pointing, but the grism dispersion distorts the photon paths, spreading a
point source into a stripe of $\sim$\,530 pixels. If a galaxy is detected through its
{\ha} line, the position of this line is shifted along the dispersion direction by a
number of pixels equal to $(\lambda-\lambda_0)/\lambda_{\rm pix}$, where $\lambda_0$ is
the wavelength whose path is not bent by the grism; we assume $\lambda_0=1.52$~$\mu$m (so
that for a galaxy at $z=1.31$ the {\ha} line falls at the location of the photometric
source). Moreover, $\lambda_{\rm pix}=13.54$ {\AA} is the wavelength dispersion on a pixel
(as in Sect.~\ref{sec:numberofsources}). We create the exposure time maps by computing how
may times the {\ha} line of a source is observed, so these maps are redshift-dependent.

The exposure time map for the EWS was created by going through the survey timeline, as
planned before launch and presented in \cite{Scaramella-EP1}, which lists for each planned
pointing the position of its centre and its orientation on the sky. For each pointing, we
rotated the sky pixels that sample the NISP detector to the pointing position, shifted
them to compute the position of the {\ha} line, and sampled them with another \healpix
tessellation at half $N_{\rm side}$, flagging each sky pixel that contains at least one
NISP pixel; this oversampling was done to avoid that the rotation creates spurious holes
in the footprint. The survey footprint was then created as a boolean \healpix map that is
\texttt{true} in any sky pixel that is visited at least once, \texttt{false} otherwise.
The exposure time map was created by counting in how many pointings a sky pixel is
observed; this produces an integer map, but rotation or resampling can make it a floating
point map, giving the effective number of exposures in a sky pixel. The left panel of
Fig.~\ref{fig:maps} gives the resulting exposure time map in the {\tdeg} circle, at a map
resolution of $N_{\rm side}=2048$ (obtained using a NISP resolution of $N_{\rm
  side}=4096$). For this map, we simply used the photometric position of the galaxy.
Although immersed in a simple overall geometry, this map shows a complicated pattern with
a wheel-like geometry around the north ecliptic pole, where overlaps among pointings are
stronger. Moreover, several holes are present to avoid bright objects.

The middle panel of Fig.~\ref{fig:maps} gives MW reddening $E(B-V)$ in the same region of
the sky. Reddening is discussed in depth in Appendix~\ref{app:MW}. Here we use the
reddening map produced by the Planck Collaboration in their 2013 results
\citep[][hereafter P13]{PlanckDust2014}. They provide the map at $N_{\rm side}=2048$,
slightly oversampling the observational beam.

The right panel of Fig.~\ref{fig:maps} gives a noise map, again at $N_{\rm side}=2048$.
The noise map was created using a model for the zodiacal light and a model for the stray
light, produced within the Euclid Collaboration. These models were interpolated on
\healpix tessellations of varying $N_{\rm side}$, so as to have a map at the same
resolution as the exposure time map. While the exposure time and reddening maps are
realistic, the noise map is clearly optimistic and simplified, since it lacks small-scale
power coming from all the effects discussed in Sect.~\ref{sec:classification} and
Appendix~\ref{app:entrypoints}. Since these features are very hard to predict before real
images are taken, we decided to represent them by adding a detector noise term that has a
constant value of 1 {\epixs}.\footnote{
  This is a very prudent value, higher by one or two orders of magnitude than the noise
  found in detector tests (Euclid Collaboration: Cogato et al., in prep.). It is supposed
  to represent the overall effect of pixel-scale noise terms such as bad pixels, cosmic rays
  and so on.}
As a consequence, in these tests small-scale power in the random catalogue is mostly given
by exposure time variations.

We note that this modelling of survey systematics is simplistic in
  several aspects, but a modelling with precise detector layout, optical model, masked
  pixels, and fully realistic background can only be performed using the real data, which
   is beyond the scope of this paper.

\subsection{The tabulated detection probability}
\label{sec:lookup_table}

\begin{figure*}
  \centering{
  \includegraphics[width=0.45\textwidth]{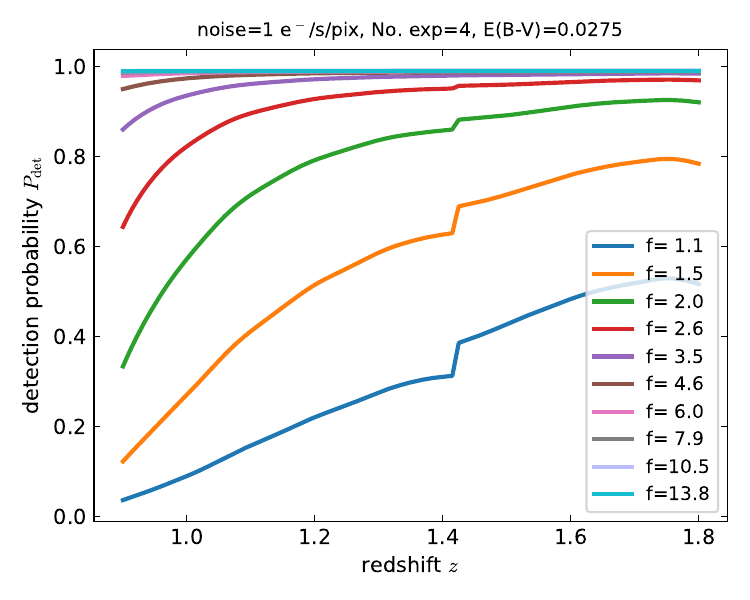}
  \includegraphics[width=0.45\textwidth]{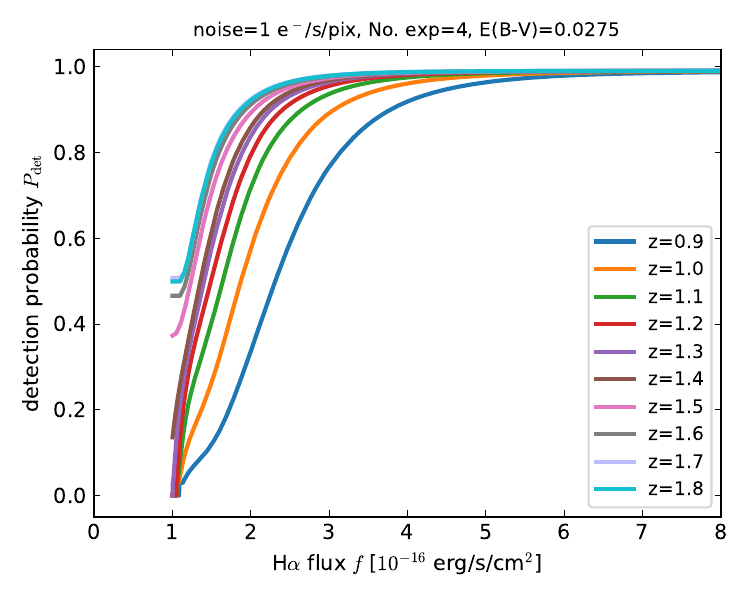}}
  \centering{
  \includegraphics[width=0.45\textwidth]{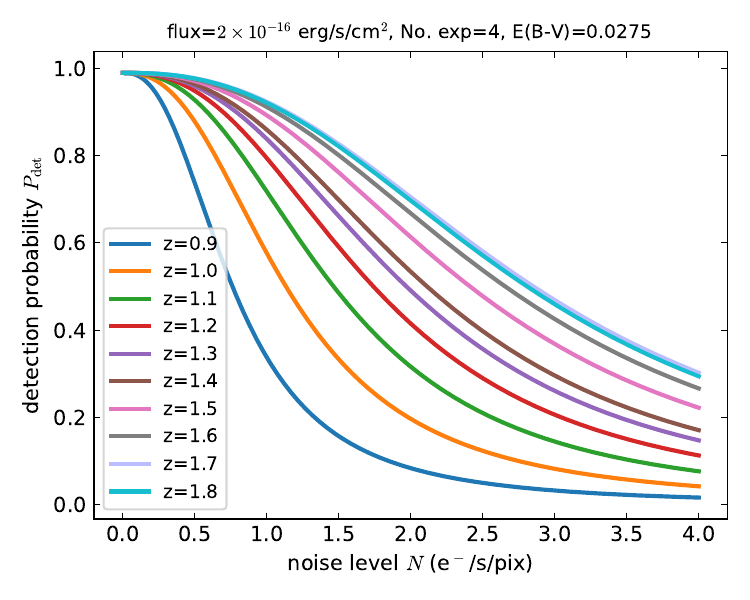}
  \includegraphics[width=0.45\textwidth]{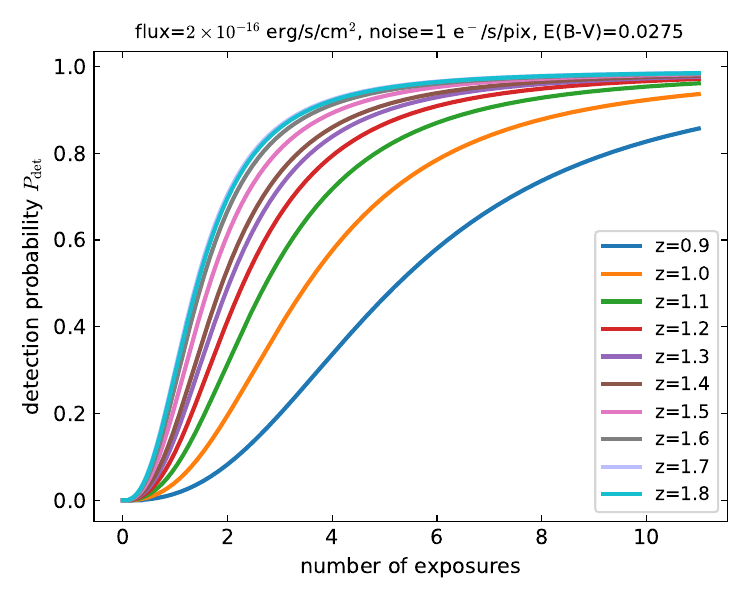}}
  \caption{Tabulated detection probability, $P_{\rm det}$, marginalised over galaxy
    properties. \textit{Upper left:} As a function of redshift $z$ for several flux limits
    (in units of $10^{-16}$ \flux). \textit{Upper right:} As a function of {\ha} flux $f$
    for several redshifts. \textit{Lower left:} As a function of noise level $N$ for
    several redshift. \textit{Lower right:} As a function of number of exposures for
    several redshifts. The values of the parameters that are kept fixed are shown above
    each panel. In all panels $E(B-V)$ is set to the value $0.0275$; the dependence on
    reddening is very mild and not very informative, so we do not show it here.}
  \label{fig:lut}
\end{figure*}

The mock catalogues described in Sect.~\ref{sec:mocks} provide a minimal set of properties
for each galaxy, namely flux $f$, redshift $z$, sky coordinates, and few other
non-observable properties (for example host halo mass, redshift without peculiar velocity
contribution, central or satellite galaxy). To decide if a galaxy is part of the observed
catalogue (to compute $P_{\rm det}$) we could use the {\pypelid} bypass of VMSP, but this
requires knowledge of many other galaxy properties, called $\vec{p}$ in
Eq.~(\ref{eq:detection_probability}), so we should find a way to assign these properties
to our galaxies. Since the mocks were constructed to reproduce the Flagship mock
catalogue, one could assign $\vec{p}$ by randomly extracting that of a Flagship galaxy
with very similar flux and redshift. An equivalent and more convenient procedure is to
directly use the MVM $\Comp(f,z,\{N_i\})$, obtained by marginalising $P_{\rm
  det}(f,z,\{N_i\}\, |\, \vec{p})$ over galaxy properties
(Eq.~\ref{eq:detection_probability}). This can be computed by applying \pypelid to a large
set of galaxies, namely those of the Flagship mock; marginalisation will simply consist of
computing an overall detection probability for galaxies in bins of $z$, $f$, and nuisance
map values (exposure time, reddening, and noise).

We thus run {\pypelid} on the full Flagship catalogue, placing the octant on the sky so as
to minimise the area heavily obscured by the MW. We assumed constant noise and exposure
time, then computed the average galaxy detection probability in 10 bins in $f$, 90 bins in
$z$, and 3 bins in $E(B-V)$ MW reddening, and then constructed a table that gives the detection
probability already marginalised over galaxy properties, using the detection model of
Eq.~(\ref{eq:sigmoid}) shown in Fig.~\ref{fig:detmodel}. The tabulated detection
probability is thus a numerical representation of the MVM. Here it is convenient to use
for $f$ the flux of the {\ha} line after correction for extinction; in this way the
effect of reddening is only to effectively shift the multivariate distribution of galaxy
properties we are marginalising over. Since the dependence of S/N on exposure time $t_{\rm
  exp}$ and noise level $N$ is predictable, the detection probability can be rescaled to
any value of these variables.

The fine binning in redshift was applied to be able to follow sharp features connected to
the entry of a line in the red grism; the only significant feature we find is a jump in
detection probability at $z=1.42$, when the {\oiiib} line enters the red grism
(this causes the jump in number density of correct galaxies seen later in Sect.~\ref{sec:shuffled}).
To avoid
sample variance being propagated to the MVM, for each combination of values of $f$,
$t_{\rm exp}$, $N$, and $E(B-V)$, the detection probability was smoothed in redshift by
applying a Gaussian smoothing over five $\Delta z=0.01$ bins,\footnote{
To avoid border effects, the data vector to be smoothed was linearly extrapolated beyond
the nominal redshift range by 15 bins before being smoothed.
} and this was done separately below and above $z=1.42$, so as not to smooth the jump.
Figure~\ref{fig:lut} shows the tabulated detection probability as a function of $z$, $f$,
$t_{\rm exp}$, and $N$, for relevant combinations of the other parameters; in this figure
the exposure time is given as a multiple of the exposure time of a single dispersed
dither, 547 s.

The procedure described here relies on a mock galaxy catalogue, for example the Flagship mock,
which provides knowledge of the `true universe'. When dealing with real data, this
role will be played by the EDS. Constructing the tabulated detection probability will not,
in principle, require more work than constructing the random catalogue: VMSP will use
{\pypelid} for assigning a detection probability to galaxies, so it will be possible to
collect the results of this task into a table, while constructing the relevant noise map.
Moreover, the very large number of galaxies in the master random catalogue will ensure
that the table is properly populated.

\subsection{Measurements and cosmological inference}
\label{sec:inference}

The power spectra of galaxy catalogues were computed using \Euclid's PK PF \citeppk. The
code computes the galaxy density from a catalogue using a piecewise cubic spline
interpolation on a grid, then computes its power spectrum using fast Fourier transforms
(FFTs) and the technique of interlacing to suppress aliasing near the Nyquist frequency
\citep{Sefusatti2016}. It implements a Yamamoto--Bianchi estimator
\citep{Yamamoto2006,Bianchi2015,Scoccimarro2015} to measure the first five multipoles of
the power spectrum.

The Fourier-space estimator by construction provides a measurement that represents the
convolution of the true power spectrum with a window function, defined as

\be
W(\vx) = W_{\rm fp}(\nh)\, w_{\rm fkp}(z)\, \alpha\, n_{\rm r}(\vx)\, .
\label{eq:PKwindow}\ee

\noindent
Here $W_{\rm fp}(\nh)$ gives the angular footprint of the survey and is $1$ where there
are observations and $0$ otherwise, $w_{\rm fkp}$ is a weight defined in \cite{FKP}, and
$n_{\rm r}(\vx)$ is the density of the random catalogue. A theoretical prediction of the
measured power spectrum can then be obtained by convolving the theory power spectrum,
valid for an infinite volume, with the Fourier transform of the window. This convolution
was performed using a mixing matrix computed as follows. We used the same PK code to
compute the power spectrum of the window function, $P^{\rm W}_{\ell} (k)$. This was then
Hankel-transformed to obtain the two-point correlation function of the window and from it
the mixing matrix $\mathcal{W}_{\ell \ell'}$, such that the  expectation value of the
power spectrum measurement can be related to the theoretical power spectrum $P^{\rm
  Th}_\ell$ as

\be
    \label{e:convo_master}
    \langle \hat{P}_{\ell}(k)\rangle = \sum_{\ell' =0}^{4}\,\, \sum_{k' = k_1}^{k_2}  
    [\mathbb{W}(k,k')]_{\ell,\ell'}\,\,P^{\rm Th}_{\ell'}(k')\, ,
\ee
where 
\be [\mathbb{W}(k,k^\prime)]_{\ell, \ell'} = \Delta \logten k^\prime \, \,(k^\prime)^{3}
\,\left[ \mathcal{W}_{\ell L}(k,k^\prime)- \mathcal{W}_{\ell L}(0,k^\prime)\right]\,.\ee
Here $\Delta \logten k^\prime$ (meaning $k^\prime$ over \kmpc) denotes the logarithmic
$k^\prime$ bin size and the limits of the summation operator, $(k_1, k_2)$, are determined
by the sampling of the power spectrum window function $P^{\rm W}_\ell (k)$ over the
Fourier-space grid.

For all the measurements shown in this work, we embedded the catalogues in a box with side
length $L_{\rm fft} = 7000$ \hmpc, with $N^3=1200^3$ cells. This allowed us to measure the
power spectrum multipoles from the fundamental frequency $k_{\rm f} = 2\pi/(7000\ \hmpc) =
8.98\times 10^{-4}$ \kmpc to the Nyquist frequency $k_{\rm ny} = Nk_{\rm f} = 1.08$ \kmpc.
To limit the data vector size with minimal information loss, measurements were binned in
bins of $6k_{\rm f}$. For the window function power spectrum measurements, we computed ten
spectra for each mixing matrix, varying only the measurement box size and, consequently,
the fundamental frequency of each measurement. Using this multiscale approach, we could
sample the window function power spectrum over the range
$7.8\times10^{-6}\,\kmpc<k<0.91\,\kmpc$. The resulting spectra were then used to compute
the mixing matrix.

For the theoretical model of the power spectrum, we considered two perturbative models
that differ in their treatment of redshift-space distortions; the performance of these two
models is analysed in the context of the \Euclid survey in Euclid Collaboration:
Camacho-Quevedo et al. (in prep.). The first model implements the effective field theory
(EFT) of large-scale structure to describe galaxy clustering observables through
perturbative expansions (at one loop) of the non-linear density and velocity fields
\citep[for a recent review, see][]{Ivanov+2022}. The second model, called
velocity-difference generator (VDG), also relies on a one-loop perturbative expansion of
the non-linear density field, but replaces the perturbative treatment of
real-to-redshift-space mapping with a non-perturbative approach that effectively captures
the impact of small-scale velocity divergences on large-scale structures
\citep{Scoccimarro2004,Eggemeier2025}. Both models decompose the theoretical power
spectrum into three main contributions. The first originates from the dynamical, bias, and
redshift-space distortion terms at leading and next-to-leading order within standard
perturbation theory (SPT). The second contribution arises from the stochastic components
of the density and velocity fields. Finally, the third component accounts for small-scale
corrections missing in the SPT terms, modelled via a set of parameters known as
counter-terms.

\begin{table}
\caption{\label{table:Priors_fit} Prior distributions used for parameter inference in
  the EFT and VDG models.}
\def\arraystretch{1.12}
\begin{center}
\begin{tabular}{lcr}
\hline\hline
Parameter & EFT  & VDG   \\
\hline
$h$              &$\mc{U} (0.5,1.0)$& $\star$ \\
$\omega_{\rm c}$  &$\mc{U} (0.085, 0.155)$& $\star$ \\
$A_{\rm s}$       &$\mc{U} (1.4, 2.6)$& $\star$ \\
$n_{\rm s}$       &$\mc{N} (0.96, 0.0041)$& $\star$ \\
$\omega_{\rm b}$  &$\mc{N} (0.02218, 0.00055)$& $\star$\\
\hline
$b_1$            & $\mc{U}(0.25,4.0)$& $\star$ \\
$b_2$            & $\mc{U}(-5,5)$& $\star$ \\
$\gamma_2$       & $\mc{U}(-5,5)$& $\star$ \\
$\gamma_{21}$    & Coevolution & $\star$ \\
$c_0\,[\mpc^2]$              &$\mc{N}(0,200)$& $\mc{U}(-400,400)$\\
$c_2\,[\mpc^2]$              &$\mc{N}(0,200)$& $\mc{U}(-400,400)$\\
$c_4\,[\mpc^2]$              &$\mc{N}(0,200)$& $\mc{U}(-400,400)$\\
$c_{\rm nlo}\,[\mpc^4]$      &$\mc{N}(0,200)$& --             \\
$a_{\rm vir}\,[\mpc]$      & --            & $\mc{U}(0,20)$ \\
$N^P_0$              & $\mc{N}(1,3)$& $\mc{U}(0,3)$ \\
$N^P_{20}\,[\mpc^2]$              & $\mc{N}(0,3)$& $\mc{U}(-3,3)$ \\
$N^P_{22}\,[\mpc^2]$              & --& --\\
\hline
\end{tabular}
\end{center}
\tablefoot{Cosmological parameters are separated by a horizontal line from the bias and
  nuisance parameters. In fits performed on multiple redshift bins, the nuisance
  parameters are repeated for each redshift bin. The symbols $\mc{N}$ and $\mc{U}$
  represent Gaussian and uniform distributions, respectively. Finally, a star indicates a
  shared prior between the two models, while a dash denotes an unused parameter. }
\end{table}

\begin{figure*}
  \includegraphics[width=0.9\textwidth]{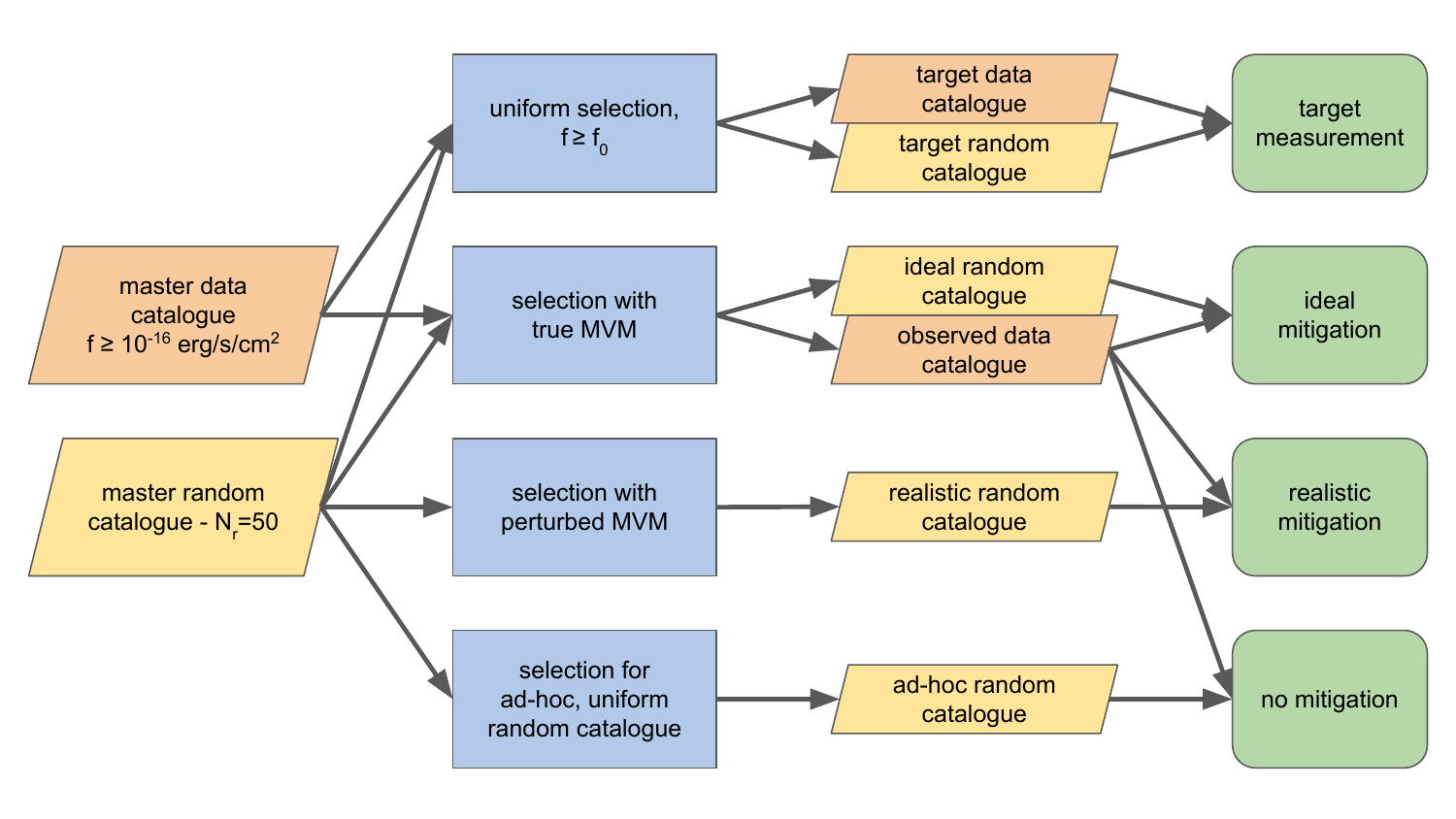}
  \caption{Scheme of the strategy adopted for the processing of the mock catalogues (see
    Sect.~\ref{sec:strategy} for details).}
  \label{fig:strategy}
\end{figure*}

We can categorise the parameters of both models into four distinct subsets (see the cited
papers for detail). The first subset comprises the cosmological parameters, $\Theta_{\rm
  cos}=\{h,\omega_{\rm c},\,\omega_{\rm b},\, A_{\rm s},\,n_{\rm s}\}$. The second subset
includes the bias parameters, which relate the galaxy overdensity field to the matter
density field: $\Theta_{\rm bias}=\{b_1,b_2,\gamma_2,\gamma_{21}\}$. The third subset
consists of the stochastic shot-noise parameters: $\Theta_{\rm
  sn}=\{N^p_0,N^p_{20},N^p_{22}\}$. Finally, the fourth subset contains the counter-term
parameters: $\Theta_{\rm ctr}=\{c_0,c_2,c_4\}$. Each model introduces an additional
model-specific degree of freedom. In the EFT model, this is the parameter $c_{\rm nlo}$,
an extra counter-term that captures next-to-leading-order corrections to the perturbative
expansion sourced by the velocity-difference generating function. In the VDG model, this
term is unnecessary as the velocity-difference distribution is modelled analytically with
an additional parameter, $a_{\rm vir}$, for its kurtosis.

For both models, we use the Gaussian-process emulator \texttt{comet} \citep{Eggemeier2023}
to compute model predictions. Following that paper, all parameters listed above are varied
in the analysis, except for $\gamma_{21}$ and $N^p_{22}$. The former is determined using
the coevolution relation,

\begin{equation} \gamma_{21} = \frac{2}{21}(b_1 -1 ) + \frac{6}{7}\gamma_2\,, \label{eq:g21Coevol} \end{equation}

\noindent
and the latter, that is highly degenerate with $c_{\rm nlo}$, is set to zero. Finally, for
parameter inference, we employ the importance nested-sampling algorithm for Bayesian
posterior reconstruction, as implemented in the \texttt{Nautilus} package
\citep{Lange2023} and use the python package \texttt{getdist} \citep{getdist} for the
plotting and analysis of the parameter chains.

We want to stress here that the analyses described in this paper are in continuation with
those we presented in \citepk, where the consistency of power spectrum measurements and
window function modelling was tested in three steps. Firstly, a `faux-cone' wedge-like
survey was extracted from a set of 10\,000 periodic boxes at $z=1$, and its mixing matrix
measured from its random catalogue; the convolution of the theory model, consisting in the
average power spectrum measured on the periodic boxes, and the window function was
consistent with the average measurement of the faux-cone power spectrum to better than
0.1\%. Secondly, we measured the power spectra of the EuclidLargeMocks, without
systematics, and fitted them with a model, checking that the original cosmological
parameters used to run the simulation were recovered. Thirdly, we imposed simple angular
systematics, caused by the varying exposure time due to the tiling of pointings, and
checked that the fitting model, convolved with the new window function (which has
small-scale power), fits the measurements at the same level as the previous case without
angular systematics. In this paper we use the same approach, with the same mock catalogues
and theory modelling, but with angular systematics represented by more complicated MVMs.
Results presented in Sect.~\ref{sec:results} are, as a matter of fact, an extended testing
of the power spectrum estimator in the case of a realistic random catalogue.

\subsection{Strategy for the tests}
\label{sec:strategy}

We start from a set of 50 master data catalogues, taken from the EuclidLargeMocks, which
are limited at $f\ge10^{-16}$ \flux. From these we construct a master random catalogue by
picking up a number of galaxies equal to $N_{\rm r}=50$ times the galaxy density $n_{\rm
  g}(z)$ averaged over the full set of 1000 mocks, which we regard as our theoretical
number density. Of these galaxies we keep the observed redshifts and {\ha} fluxes, so as
to preserve the redshift-dependent galaxy LF, but we create random angular coordinates
that sample the {\tdeg} circle footprint. This single master random catalogue is used for
all the mock catalogues.

Then, as detailed in Fig.~\ref{fig:strategy}, both master data and random catalogues are
subject to a selection, which may be as simple as selecting (target) galaxies with $f\ge
f_0$ (uniform selection), or may be based on an MVM. The data and random catalogues may
be selected using the same MVM, or the random catalogue may be selected with a perturbed
MVM that accounts for the uncertainty in its construction. The ad hoc selection of the
random catalogue is based on sparse-sampling it to reproduce ($N_{\rm r}$ times) the
number density of the data catalogue subject to the MVM (averaged over 50 realisations).
The resulting random catalogue will have a uniform density on the sky, so we use this
selection to show what happens if a given angular systematics is not mitigated.

\begin{table*}
\caption{\label{table:cases}List of cases of combinations of systematics used in the analysis.}
\begin{small}
\begin{center}
\begin{tabular}{l|ccccccc}
\hline\hline
{\bf Case reference}& {\bf Zodiacal} & {\bf Stray light}& {\bf Detector} & {\bf MW}        & {\bf Exposure} & {\bf Detection} & {\bf Calibration} \\
{\bf name}          & {\bf light}    &                  & {\bf noise}    & {\bf extinction}& {\bf time}     & {\bf model}     & {\bf error} \\
\hline
{\it zodiacal}      &  {\bf yes}     &  no              & no             & no              &   4            & standard        & no \\
{\it stray\_light}  &  no            &  {\bf yes}       & no             & no              &   4            & standard        & no \\
{\it img\_noise}    &  {\bf yes}     &  {\bf yes}       & {\bf yes}      & no              &   4            & standard        & no \\
{\it MW\_P13}       &  no            &  no              & no             & {\bf P13}       &   4            & standard        & no \\
{\it tiling}        &  no            &  no              & no             & no              &   {\bf map}    & standard        & no \\
{\it baseline}      &  {\bf yes}     &  {\bf yes}       & {\bf yes}      & {\bf P13}       &   {\bf map}    & standard        & no \\
\hline
{\it calib\_2\%}    &  {\bf yes}     &  {\bf yes}       & {\bf yes}      & {\bf P13}       &   {\bf map}    & standard        & {\bf 2\%} \\
{\it calib\_10\%}   &  {\bf yes}     &  {\bf yes}       & {\bf yes}      & {\bf P13}       &   {\bf map}    & standard        & {\bf 10\%} \\
{\it MW\_P15}       &  no            &  no              & no             & {\bf P15}       &   4            & standard        & no \\
{\it MW\_SFD}       &  no            &  no              & no             & {\bf SFD}       &   4            & standard        & no \\
{\it detmodel1}     &  {\bf yes}     &  {\bf yes}       & {\bf yes}      & {\bf P13}       &   {\bf map}    & $\mathbf{S_0=3.27}$     & no \\
{\it detmodel2}     &  {\bf yes}     &  {\bf yes}       & {\bf yes}      & {\bf P13}       &   {\bf map}    & $\mathbf{S_0=4.11}$     & no \\
{\it detmodel3}     &  {\bf yes}     &  {\bf yes}       & {\bf yes}      & {\bf P13}       &   {\bf map}    & $\mathbf{\boldsymbol\beta=3.73}$   & no \\
{\it detmodel4}     &  {\bf yes}     &  {\bf yes}       & {\bf yes}      & {\bf P13}       &   {\bf map}    & $\mathbf{\boldsymbol\beta=6.23}$   & no \\
\hline
\end{tabular}
\end{center}
\end{small}
\tablefoot{Each case is defined by a set of choices, declared in each column. The first
  six cases are discussed in Sect.~\ref{sec:ideal}, the remaining eight in
  Sect.~\ref{sec:realistic}. The contributions to image noise (zodiacal light, stray
  light, detector noise) can be present (yes) or absent (no); MW extinction can be applied
  using the P13 (see Sect.~\ref{sec:maps}), P15, or SFD reddening maps (see
  Sect.~\ref{sec:realistic}); the exposure time can be set to four dithers, equal for all sky
  pixels, or to the (redshift-dependent) exposure time map discussed in
  Sect.~\ref{sec:maps} assuming a redshift of $z=1.2$. A realistic mitigation is obtained
  by perturbing the detection model or by adding a calibration error; the relative columns
  report the value of the perturbed parameter of the detection model or the value of the
  largest calibration error. In all columns, values in bold face highlight the non-trivial
  choices.}
\end{table*}

The study of the effect of systematics will make use of these combinations. (i) Target
measurement: this is obtained by selecting the master data and random catalogues with a
uniform selection $f\ge f_0$ (specific of target galaxies), with no angular dependence.
(ii) Ideal mitigation: the master data and random catalogues are selected by applying the
same MVM, so that the random catalogue corrects the angular systematics in the best
possible way. (iii) Realistic mitigation: while the master data catalogues are selected
with an MVM, the master random catalogue is selected with an approximated one. (iv) No
mitigation: the master data catalogues are selected with an MVM, but the master random
catalogue is subject to the ad hoc selection.

\section{Results}
\label{sec:results}

We test here how systematics impact the measurement of the galaxy power spectrum and the
inferred cosmological parameters. The analysis is carried out in two steps. In the first
step we use shuffled fluxes to remove luminosity-dependent bias; in this case the galaxy
power spectrum is not expected to change (apart from the different window and level of
shot noise) when a different selection is applied. This allows us on the one hand to
quantify the effect of specific systematics, and on the other hand to stress-test our
ability to correct the effects of an MVM with a random catalogue. Limiting ourselves to
the analysis of the redshift bin $[1.1,1.3]$, we first fit the measurement of the power
spectrum of target galaxies (that have $f\ge f_0$) with a model, using the mixing matrix
computed for the \tdeg cone without angular systematics. We then consider a generic MVM,
apply it to the master catalogues and to the random catalogue, and measure the power
spectra for this ideal mitigation. With this random catalogue we compute the window
function and its mixing matrix, then use it to obtain a convolved model, where the theory
model is the one we obtained with the fit of the target measurement. We then compare the
measurements that have ideal mitigation with the convolved model, and test that their
agreement is as good as that obtained when fitting the target measurements.

The second step addresses the stability of inferred cosmological parameters in the
presence of systematics, both with ideal and with realistic (or even ad hoc) mitigation.
In this case we use the standard fluxes in place of the shuffled ones, and process the
mock catalogues dividing them into four redshift bins, bounded by 0.9, 1.1, 1.3, 1.5, and
1.8 (see Sect.~\ref{sec:mocks}), and fit their measured power spectra at the same time,
using a numerical covariance computed on the 1000 mocks with baseline systematics defined
below. We present results for cosmological parameter posteriors and check their stability
with respect to using an approximate MVM.

\subsection{Testing measurements with shuffled fluxes}
\label{sec:shuffled}

\begin{figure}
  \centering{\includegraphics[width=0.45\textwidth]{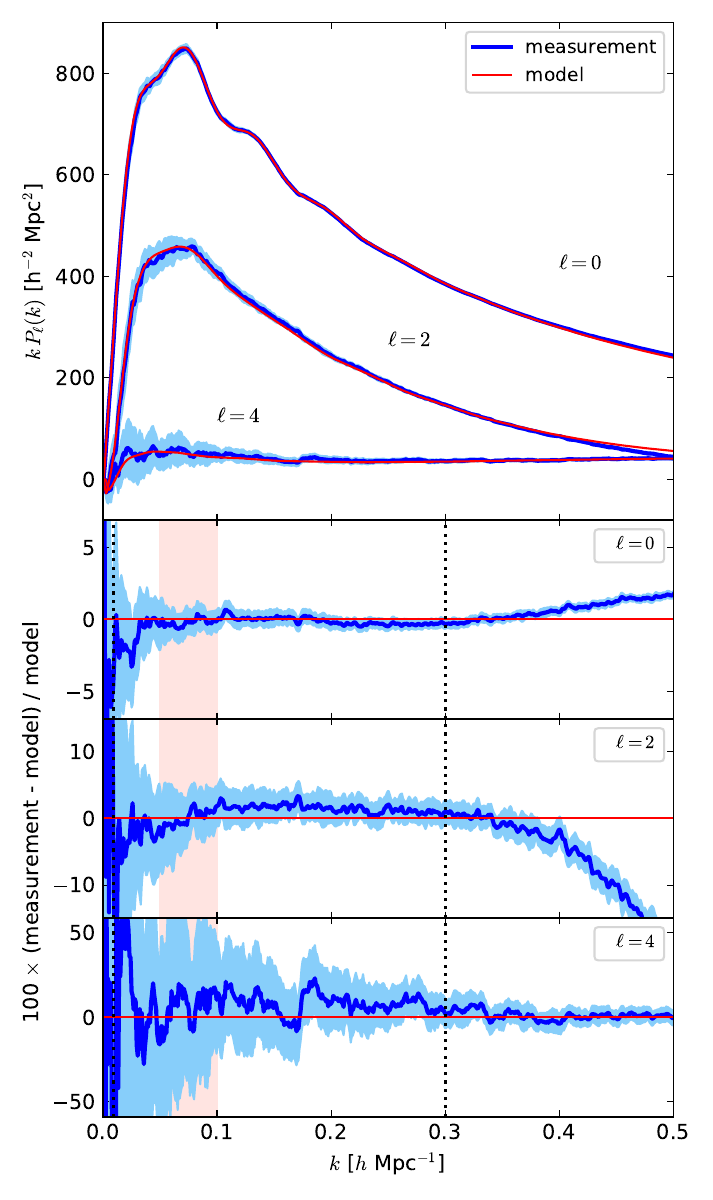}}
  \caption{ \label{fig:pkmodel} \textit{Upper panel}: Power spectra even multipoles ($\ell=0$, 2,
    4) of target galaxies in the $z\in[1.1,1.3]$ redshift bin, averaged over 50
    realisations, with shuffled fluxes to remove luminosity-dependent bias. The blue lines
    give the target measurement, while the lighter blue shaded areas give the error on an
    average of five measurements, covering an area similar to the final EWS. The red line
    gives the best-fit model, convolved with the 30$^\circ$ cone window function. \textit{Lower
    panels}: For the three multipoles, the relative difference with respect to the
    model fit to the master mocks, convolved with the relative window. The pink shaded
    area highlights the region of the first BAO, while the dotted vertical lines mark the
    range of scales used for the fit.}
\end{figure}

\begin{figure}
  \centering{
  \includegraphics[width=0.45\textwidth]{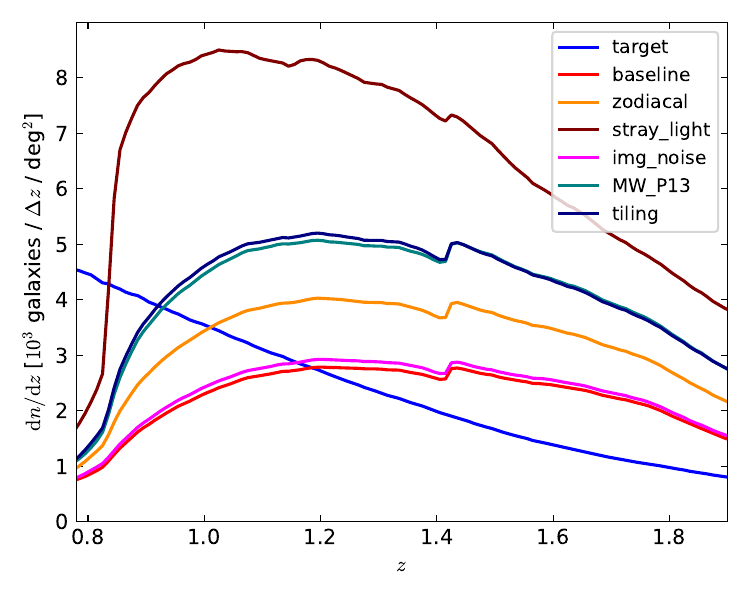}
  \includegraphics[width=0.45\textwidth]{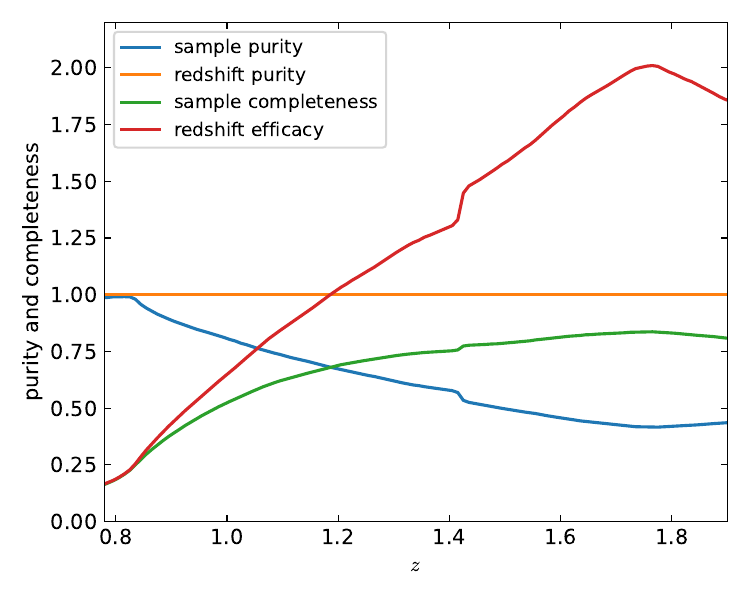}}
  \caption{\textit{Upper panel}: Number density ${\rm d}n/{\rm d}z$, in units of galaxies per
    redshift interval per \sqdeg, of the target galaxies and of the galaxies subject to
    the six cases of systematics listed in Table~\ref{table:cases}. \textit{Lower panel}: Metrics of sample purity and completeness defined in Sect.~\ref{sec:P&C}
    for the {\it baseline} case.}
  \label{fig:nz}
\end{figure}

We fit the master catalogues without angular systematics using the VDG model (see
Sect.~\ref{sec:inference}), which has the advantage, with respect to the EFT model, of
being more stable beyond the fitting range. For the numerical covariance, we compute it
from 1000 measurements of all the EuclidLargeMocks, already presented in
\cite{EP-Monaco1}. We fit the monopole and quadrupole power spectra in the range
$z\in[1.1,1.3]$ up to $k_{\rm max} = 0.3$ \kmpc, in bins of size $\Delta k = 6 k_{\rm f}$,
using the parameter priors shown in Table~\ref{table:Priors_fit}. With the exception of
the $\omega_{\rm b}$ and $n_{\rm s}$ parameters, for which we adopt informative Gaussian
priors, all the varied parameters have wide uniform priors.

We group the 50 mocks in ten groups of five mocks each (Sect.~\ref{sec:mocks}), so that
the error on the fit is relative to the area of the final survey. Figure~\ref{fig:pkmodel}
shows the results of this fit; the upper panel shows the first three even multipoles of
the power spectrum (here and in all similar plots shot noise has been subtracted), with
the blue line giving the measurement and the red line the (convolved) fitting model. The
lighter blue shaded area gives the variance of the measurements, computed for the 50 mocks
and rescaled to represent the uncertainty on five measurements. The lower panels give, for
the three multipoles, the residuals of the measurements with respect to the model, in
per cent, the vertical pink area highlights the first BAO region, and the vertical dotted
lines show the range of scales used for the fit, from $k_{\rm min}=0.009$ {\kmpc} to $0.3$ {\kmpc}.
The model is in excellent agreement with the measurements in the fitting range, then it
slowly departs from the data at higher wavenumbers, showing a 2\% difference at
$k$\,$\sim$\,0.5 {\kmpc} for the monopole, while the agreement with the quadrupole worsens
in a more marked way.

\subsubsection{Ideal mitigation}
\label{sec:ideal}

We now analyse six cases of systematics, obtained by gradually switching on the various
contributions to the noise map, the MW extinction, and the exposure time map; this last
map is either set to a constant exposure value of 4 dithers, or to the exposure time map
computed, as explained in Sect.~\ref{sec:maps}, at the median redshift $z=1.2$. These six
cases are reported in the upper part of Table~\ref{table:cases}, where the value of each
nuisance term is reported. For these six cases, the detection model is perfectly known,
and no calibration error is present. The cases tagged as {\it zodiacal}, \textit{stray\_light}, {\it MW\_P13} and {\it tiling} have just one nuisance active, {\it
  img\_noise} is a combination of the three contributions to image noise (zodiacal light,
stray light and detector noise), while for {\it baseline} all the nuisances are present.

The upper panel of Fig.~\ref{fig:nz} shows the galaxy number density for the sample of
target galaxies and for the six cases of systematics. This is quantified in observational
units of galaxies per redshift interval per {\sqdeg}. The number density of target
galaxies decreases in redshift, following the evolution of the {\ha} luminosity function
of \cite{Pozzetti2016}, while the other number densities have a peak at a redshift around
1.2. This different behaviour is the result of the increase in the detection probability
with redshift shown in Fig.~\ref{fig:lut}. Also, number densities with systematics can be
higher than those of target galaxies, since at low noise levels the detection of galaxies
with $f<f_0$ is more probable. From this figure we see that, when taken alone, stray light
and zodiacal light have respectively the weakest and the highest impact on number density.
The {\it baseline} case has the lowest number density, and this is expected since it is
the combination of all other cases. The lower panel of Fig.~\ref{fig:nz} shows, for the
{\it baseline} case, the purity and completeness metrics defined in Sect.~\ref{sec:P&C};
here redshift purity is unity because there are no interlopers, while redshift efficacy
gets values larger than unity thanks to the high probability of detecting a galaxy at high
redshift.

\begin{figure*}
  \centering{
  \includegraphics[trim=25 5 25 5, clip, width=0.33\textwidth]{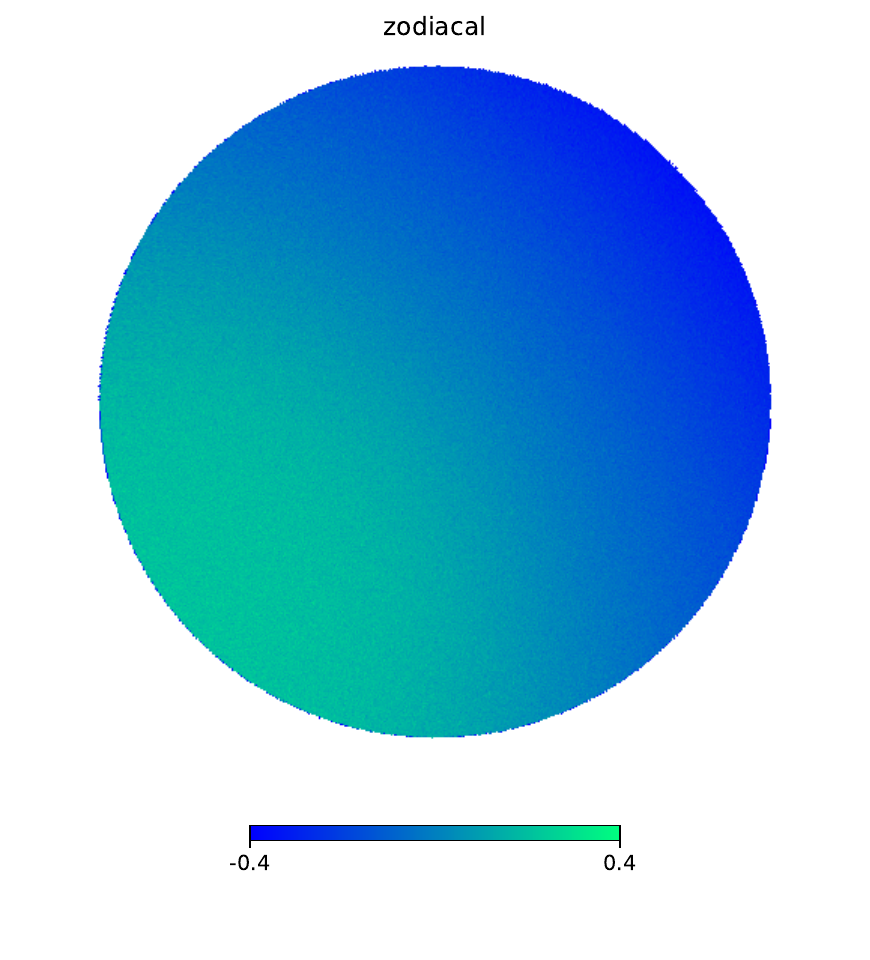}
  \includegraphics[trim=25 5 25 5, clip, width=0.33\textwidth]{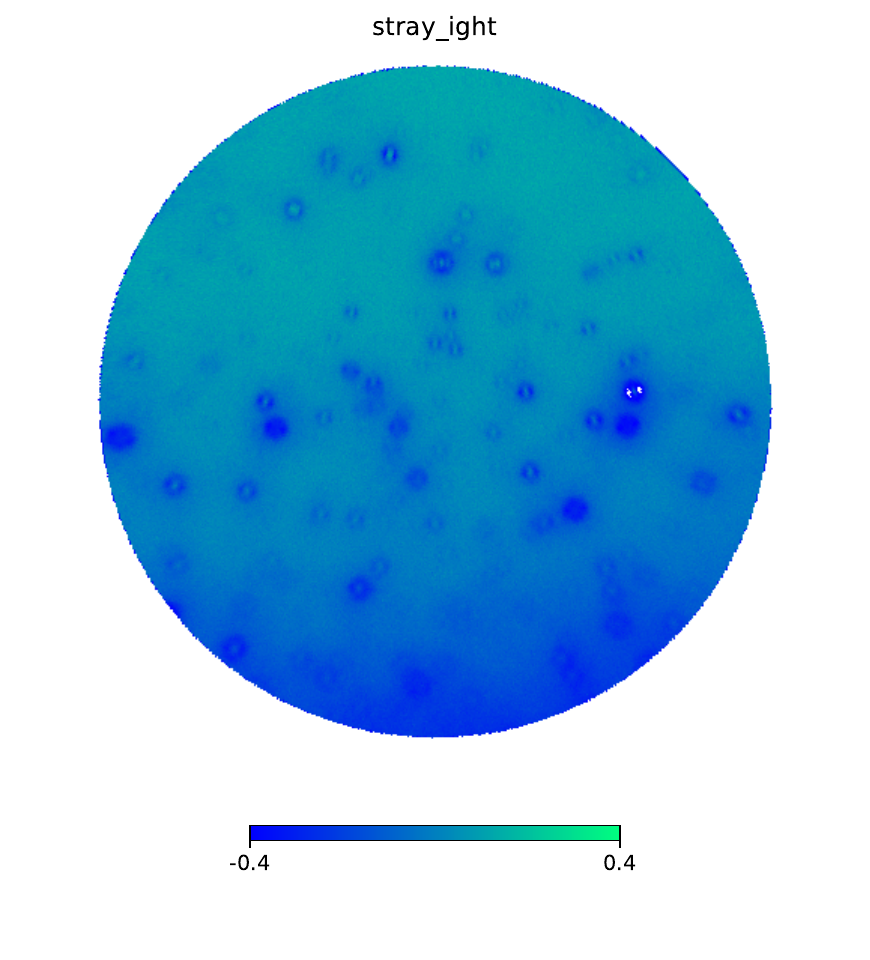}
  \includegraphics[trim=25 5 25 5, clip, width=0.33\textwidth]{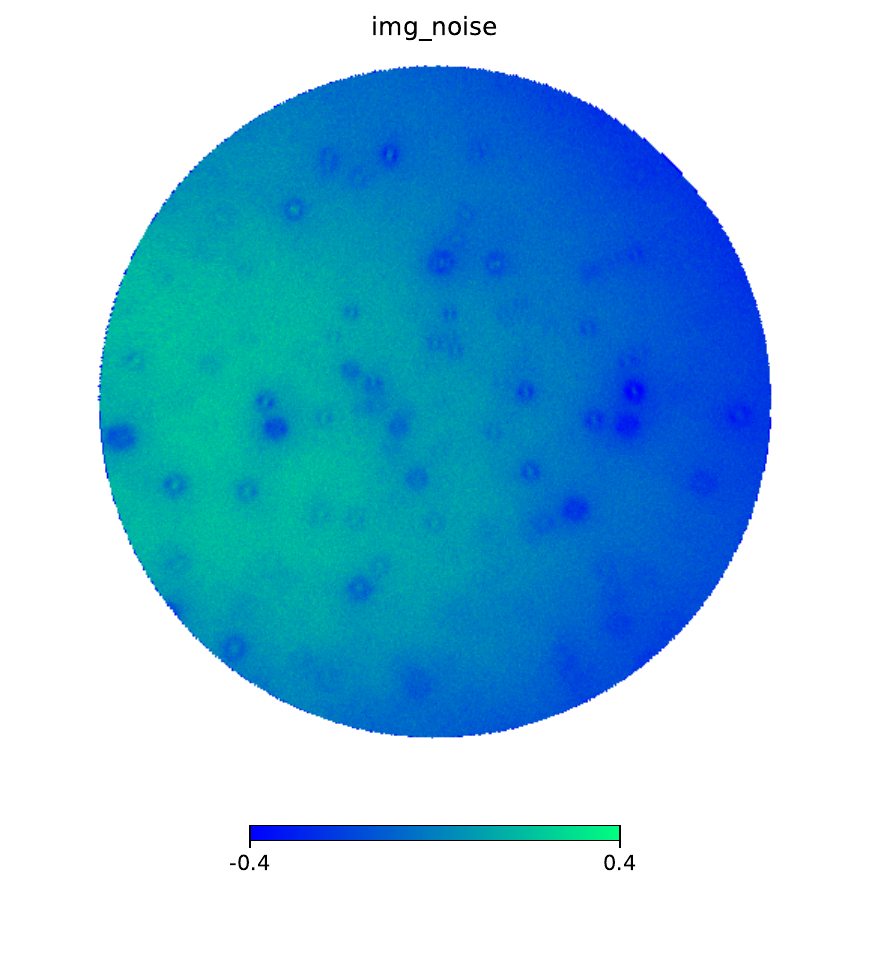}}
  \centering{
  \includegraphics[trim=25 5 25 5, clip, width=0.33\textwidth]{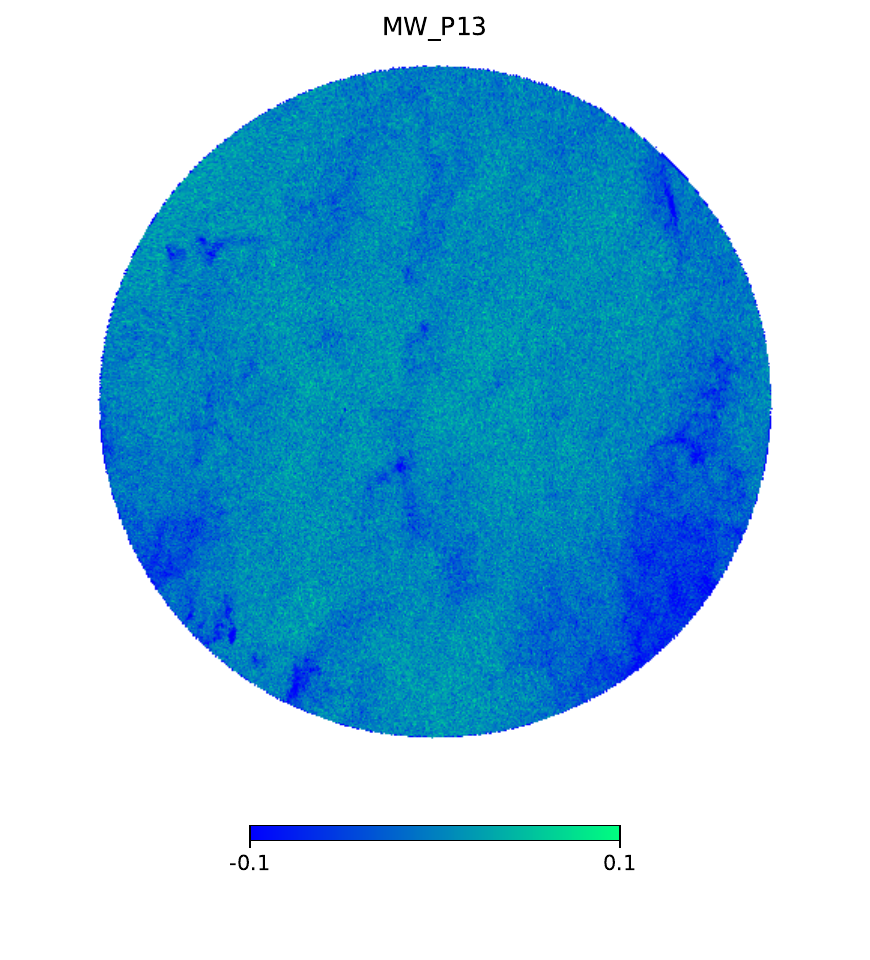}
  \includegraphics[trim=25 5 25 5, clip, width=0.33\textwidth]{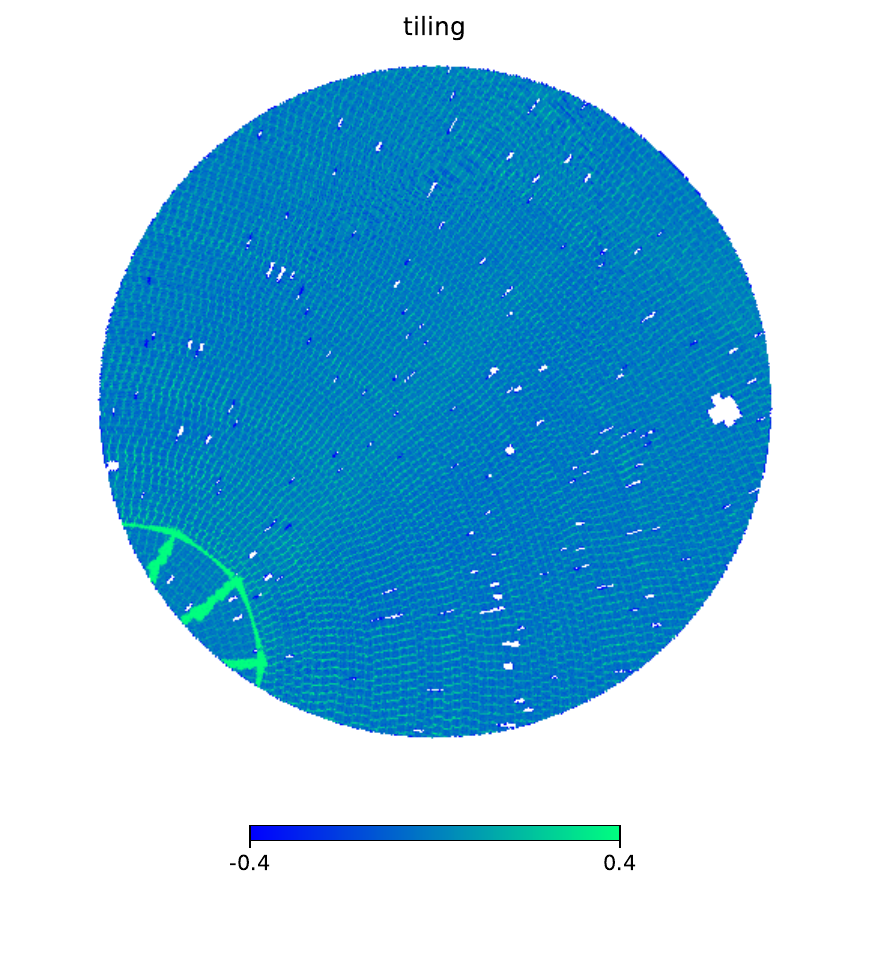}
  \includegraphics[trim=25 5 25 5, clip, width=0.33\textwidth]{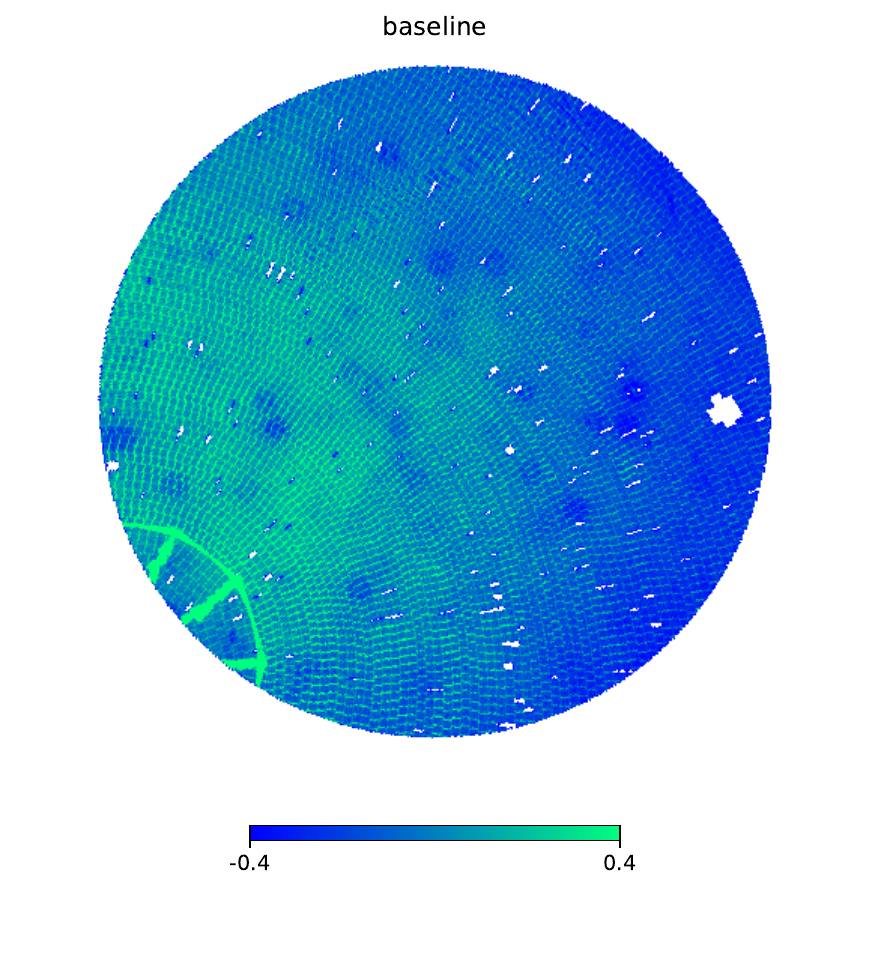}}
  \caption{Projected density contrasts of the random catalogues that are subject to
    the MVMs of the systematic cases listed in
    Table~\ref{table:cases} (labelled above each panel). The white spots in the last two maps denote unobserved regions.
    We note that the limits of the colour bar for the {\it MW\_P13} case are not the same
    as in the other the maps.}
  \label{fig:maps_shuffling}
\end{figure*}

To better visualise these MVMs, from the selected random catalogues we count the random
galaxies from $z=0.9$ to $1.8$ that fall onto a sky pixels for a \healpix tessellation
with $N_{\rm side}=512$, and from them we compute density contrasts simply as $(N_{\rm
  pix} - \overline{N})/\overline{N}$. Figure~\ref{fig:maps_shuffling} shows these maps;
all of them show large-scale features, which are especially strong in the very smooth
zodiacal light, while exposure time is the nuisance map that gives most small-scale power
to the {\it baseline} selection.

\begin{figure}
  \centering{
  \includegraphics[width=0.45\textwidth]{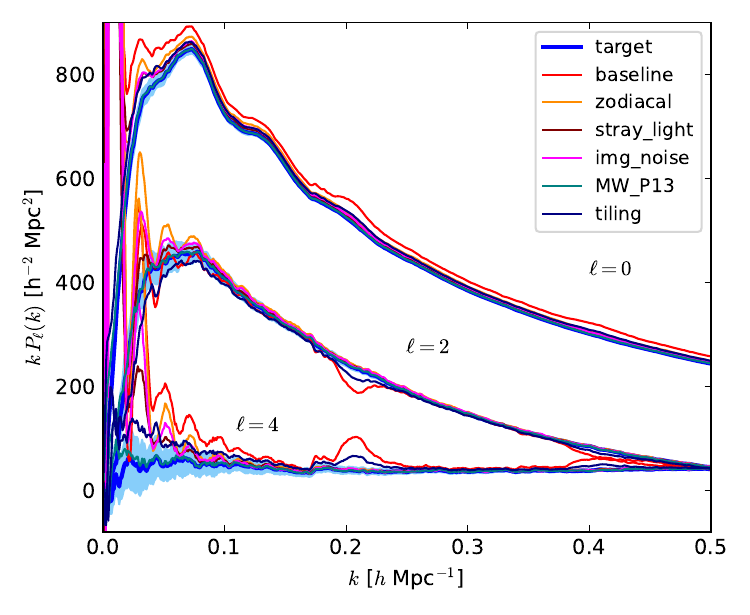}}
  \caption{Power spectrum multipoles for the sample of target galaxies and for the galaxy
    catalogues subject to systematics for the six cases reported in
    Table~\ref{table:cases}, when no mitigation of angular systematics is performed; we
    used the random catalogue with the ad hoc uniform selection. The line colours are given
    in the legend. The lighter blue shaded area gives the variance of the target
    measurement, rescaled to the average of five catalogues.}
  \label{fig:adhoc}
\end{figure}

We first show, in Fig.~\ref{fig:adhoc}, the power spectrum of target galaxies, from
Fig.~\ref{fig:pkmodel}, compared with the power spectra obtained for the six cases of
systematics when using the ad hoc random catalogue that has a uniform density on the sky,
so that no mitigation of angular systematics is performed. As expected, all the power
spectra show very significant deviations from the one of target galaxies, with the largest
effect present on large scales for the {\it zodiacal} and {\it MW\_P13} cases, while {\it
  tiling} and the {\it baseline} selections show peaks that beat with the projected
dimension of the NISP field.

\begin{figure}
  \centering{\includegraphics[width=0.45\textwidth]{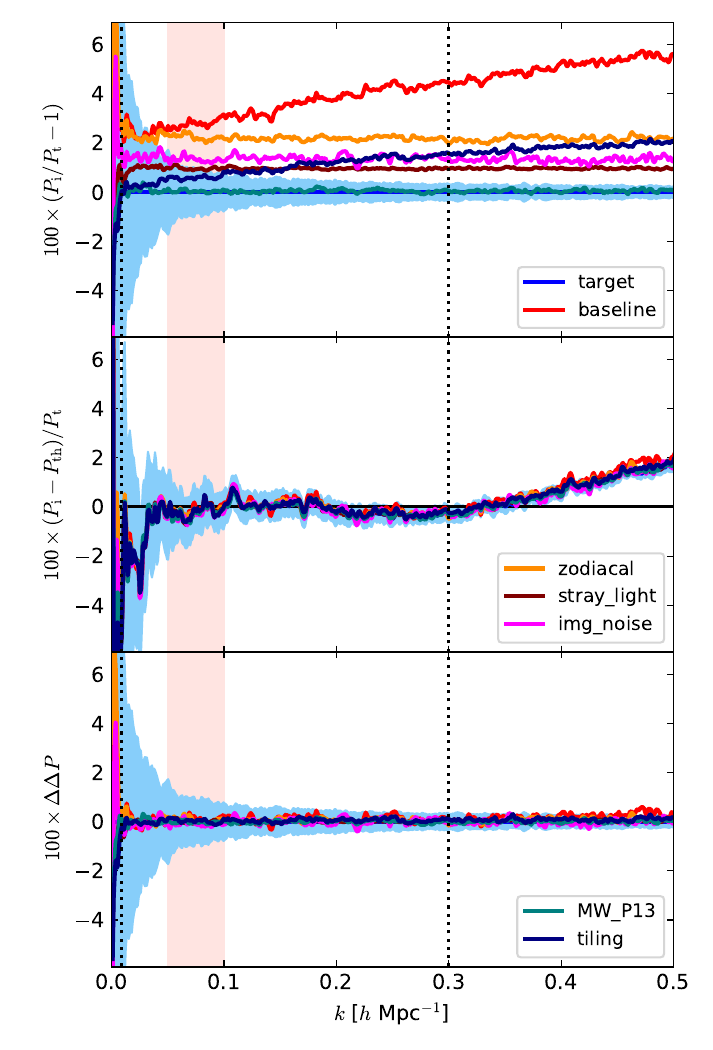}}
  \caption{Comparison of the measurements of the power spectrum monopole, for the sample
    of target galaxies ($P_{\rm t}$) and for the galaxy catalogues subject to systematics
    for the six cases reported in Table~\ref{table:cases} (denoted as $P_{\rm i}$), when
    ideal mitigation of angular systematics is performed (the data and random catalogues
    are selected with the same MVM). The legend is distributed over the various panels for
    better readability, the lighter blue shaded areas give the variance of the target
    measurement, rescaled to the average of five catalogues. In the top panel we show the
    relative difference, in per cent, of $P_{\rm i}$ with respect to $P_{\rm t}$; the
    measurements differ, despite their ideal mitigation, because the different random
    catalogues define different window functions. The middle panel gives, in per cent, the
    quantity $(P_{\rm i}-P_{\rm th})/P_{\rm t}$, where $P_{\rm th}$ is the best-fit model
    of the sample of target galaxies, convolved with the corresponding window. For the
    target sample, this is just the residuals of the fit, but the other lines give the
    difference between the measurement and the same theory model convolved with the
    relative window with systematics. Their agreement confirms that the differences
    observed in the upper panel are due to the different window functions. The bottom
    panel shows, again in per cent, the quantity $\Delta\Delta P$ defined in
    Eq.~(\ref{eq:deltadeltaP}), which gives the difference between the residuals for a given
    systematic and the residuals of the fit of the target measurement.}
  \label{fig:pkresiduals}
\end{figure}

Ideal mitigation can remove these features. This is shown in detail, separately for each
case of systematics, in Appendix~\ref{app:pk_shuffled}. Here we investigate, in
Fig.~\ref{fig:pkresiduals} and only for the power spectrum monopole, the agreement between
the measurements with ideal mitigation and the model convolved with the corresponding
window function. In the top panel we report the ratio between the ideal and target
measurements, $P_{\rm i}/P_{\rm t}-1$, quantified in per cent. The measurements differ by
up to 6\%, and this difference should be accounted for by the convolution with the window.
In the middle panel we show the residuals of measurement minus model, $P_{\rm i}-P_{\rm
  th}$, divided again by the target measurement $P_{\rm t}$. Here the theory prediction is
convolved each time with a different window; for the target measurement this line is
identical to what was shown in Fig.~\ref{fig:pkmodel} so its consistency with zero is a
result of the fit; however, for the other lines the theory prediction is obtained by
convolving the same model with the relative window function. All the lines fall on top of
each other, showing that the level of agreement of model and (ideal) measurement is the
same for a large range of MVMs, to sub-per cent accuracy. The bottom panel shows another
way of quantifying this excellent agreement. We use the difference of the
measurement-minus-theory residuals between the given systematics and the target result,
which we denote as

\be
\Delta\Delta P := \frac{(P_{\rm i} - P_{\rm th,i}) - (P_{\rm t} - P_{\rm
      th,t})}{ P_{\rm t}}\, ,
\label{eq:deltadeltaP}\ee

\noindent
where the theory model is convolved with the window of the specific systematic, $P_{\rm
  th,i}$, or with the {\tdeg} cone window of the target measurement, $P_{\rm th,t}$. This
statistics quantifies the bias induced by the convolution process, and it is consistent
with zero to sub-per cent level. This result provides a very strong validation of the whole
analysis pipeline.

We draw the following conclusions. (i) In all cases, the convolved model fits the power
spectrum of the ideal mitigation cases to within sub-per cent accuracy in the monopole and
correspondingly good accuracy for the quadrupole and hexadecapole (see
Fig.~\ref{fig:pk_shuffled}). (ii) In the no-mitigation cases, the window convolution fails
to achieve adequate accuracy. (iii) Fluctuations induced by zodiacal light and stray light
have most power on large scales, and their effect on the measurements is a more or less
constant relative shift on the monopole and quadrupole. (iv) MW extinction has effects
only on the largest scales, and produces no shift at small scales. (v) The exposure time
map induces power on small scales, and produces the most complicated features. (vi) The
baseline MVM shows a combination of all the above features, again reproduced by the
convolved model with excellent accuracy.

\subsubsection{Realistic mitigation}
\label{sec:realistic}

We now investigate to what level a realistic mitigation can recover the target power
spectrum. We can perturb the three nuisance maps (exposure time, noise, and MW
extinction), or we can perturb the detection model. We can assume that the exposure time
is known exactly.

\begin{figure*}
  \centering{
    \includegraphics[trim=25 5 25 5, clip, width=0.33\textwidth]{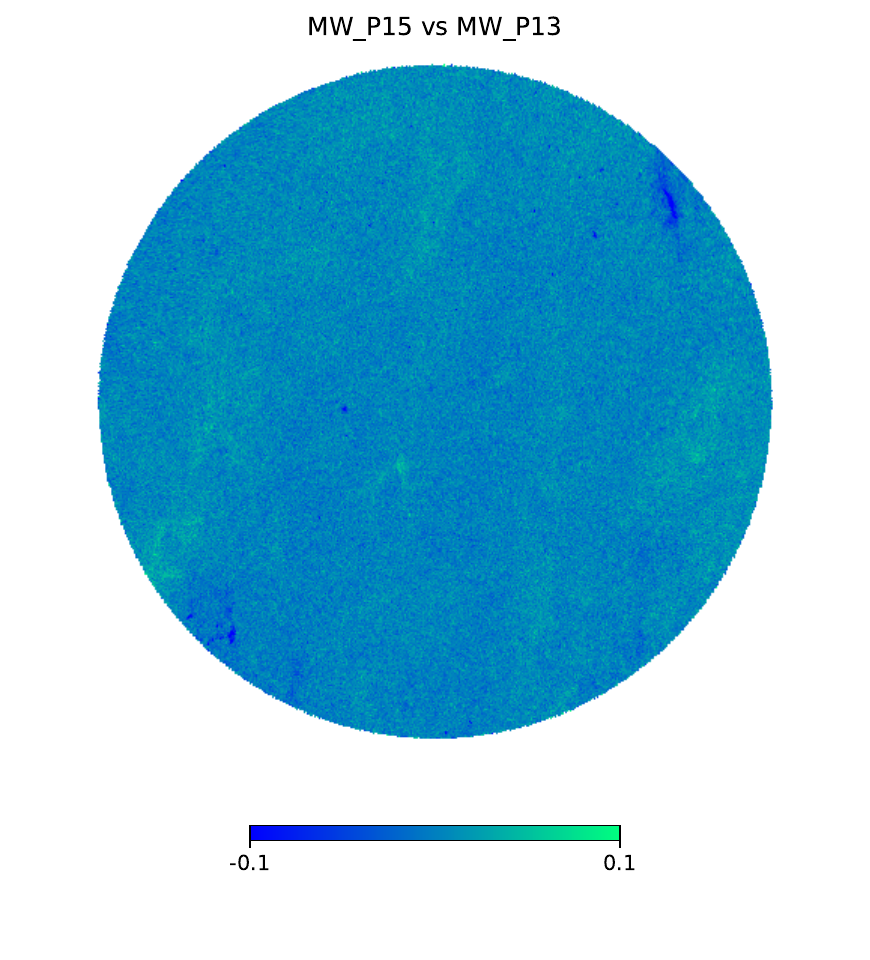}
    \includegraphics[trim=25 5 25 5, clip, width=0.33\textwidth]{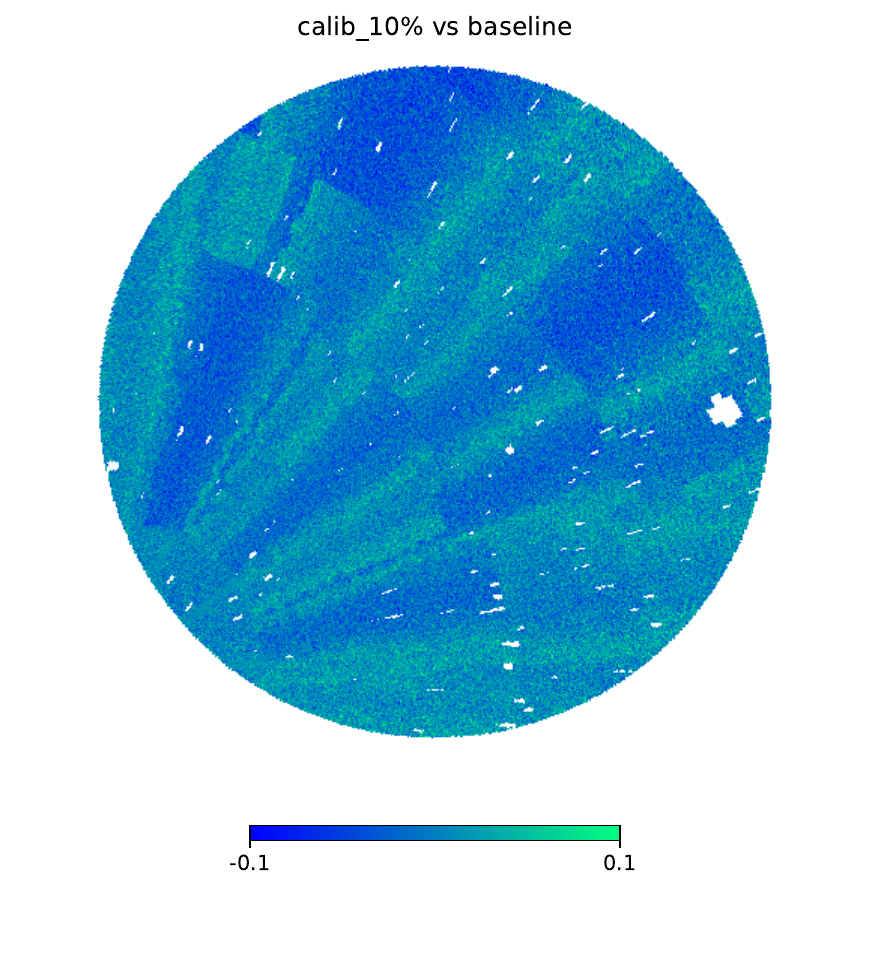}
    \includegraphics[trim=25 5 25 5, clip, width=0.33\textwidth]{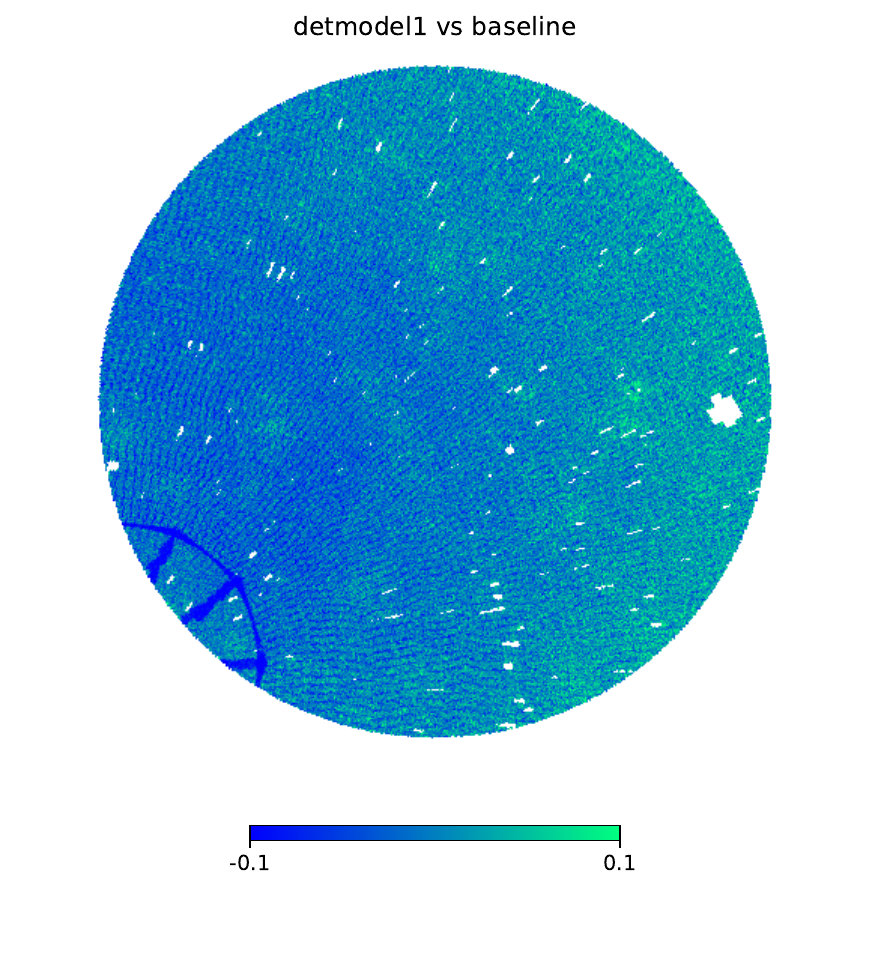}}
  \centering{
    \includegraphics[trim=25 5 25 5, clip, width=0.33\textwidth]{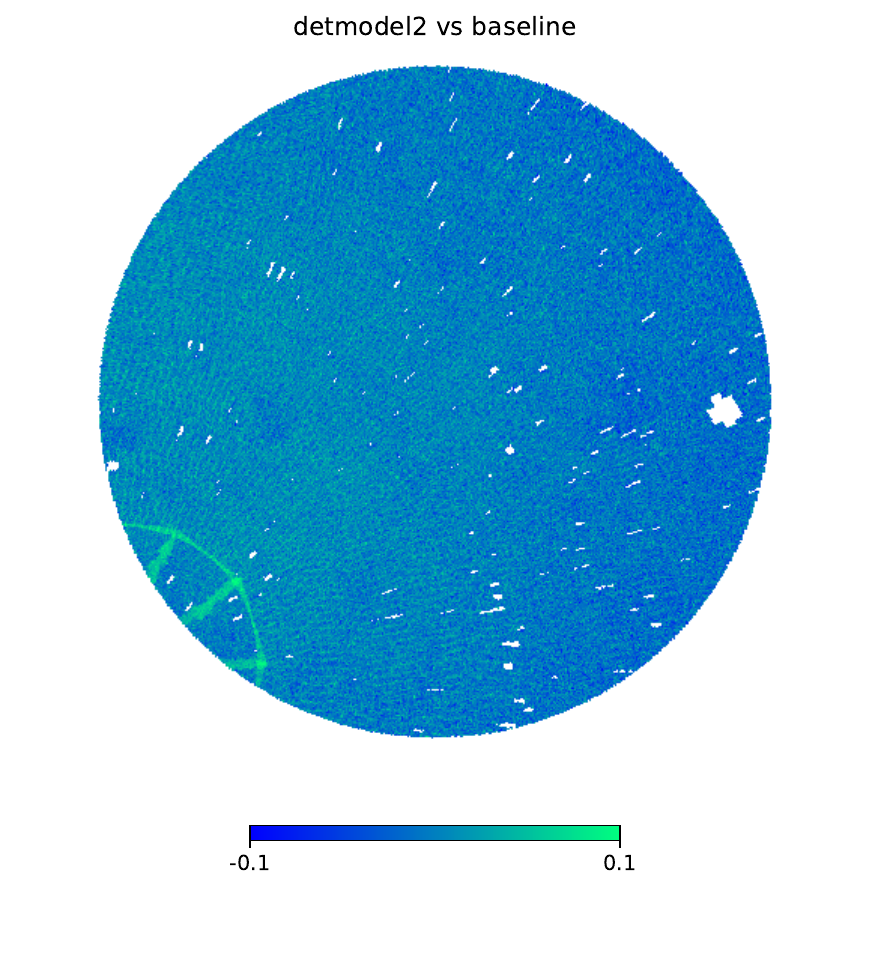}
    \includegraphics[trim=25 5 25 5, clip, width=0.33\textwidth]{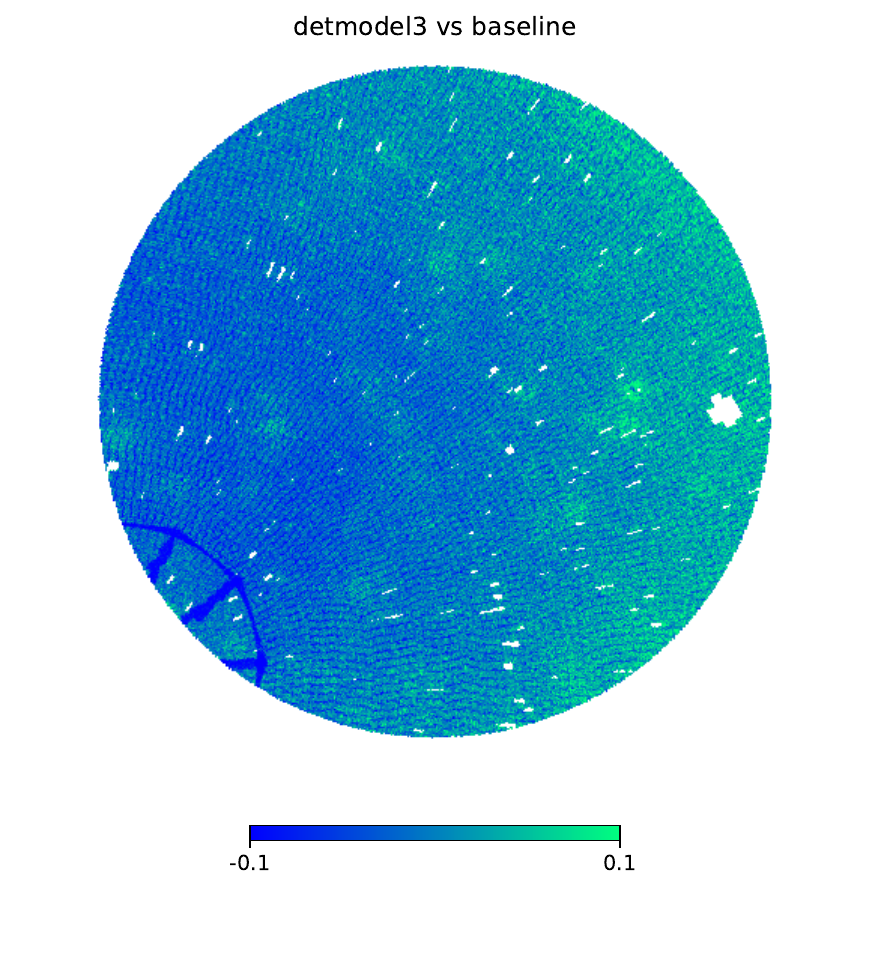}
    \includegraphics[trim=25 5 25 5, clip, width=0.33\textwidth]{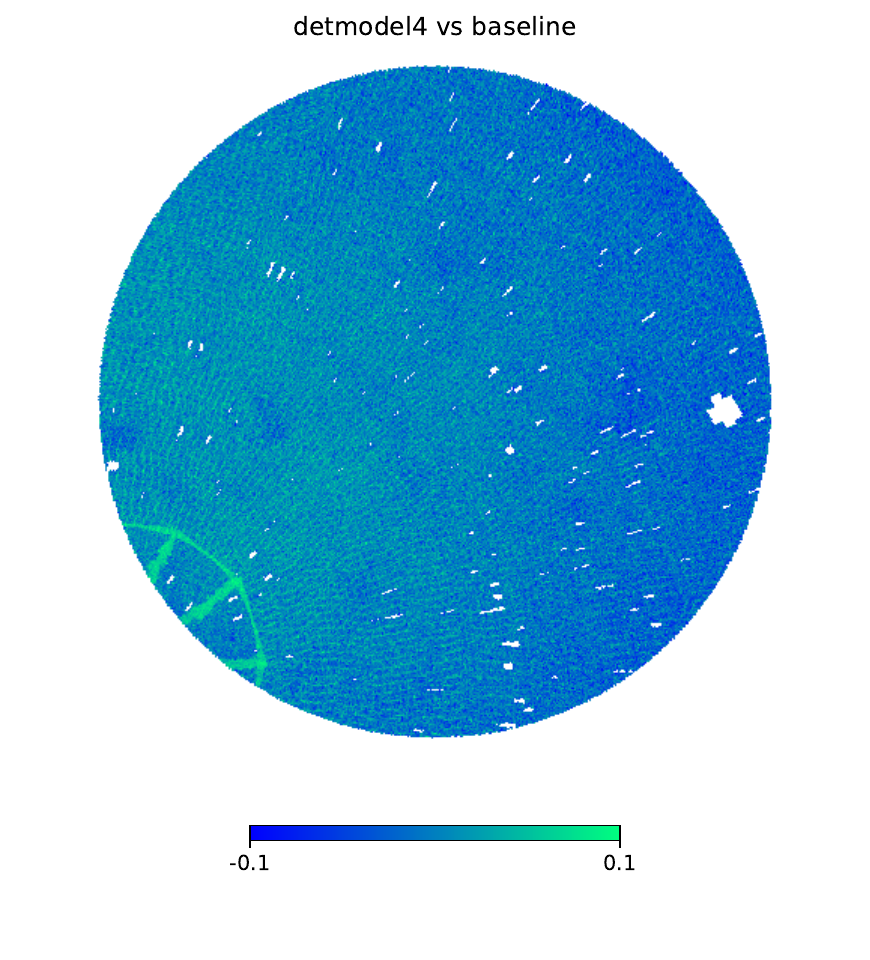}}
  \caption{Illustration of how an error in the calibration of the MVM emerges as a pattern
    on the sky. The maps report projected density contrasts obtained using two random
    catalogues, one representing the true MVM, $\Comp$, and the other the perturbed MVM,
    $\Compa$. For each random catalogue we compute its projected number density $N$ on a
    \texttt{healpix} map, and show $\Delta_\Comp = (N_{\Compa} - N_{\Comp})/N_{\Comp}$.
    The panels show (from upper left to the lower right) the cases selected from
    Table~\ref{table:cases}: {\it MW\_P15} vs. {\it MW\_P13}; {\it calib\_10\%} against
    {\it baseline}; {\it detmodel1} to {\it detmodel4} against {\it baseline}.}
  \label{fig:maps_realistic}
\end{figure*}

\begin{figure*}
  \centering{\includegraphics[width=0.9\textwidth]{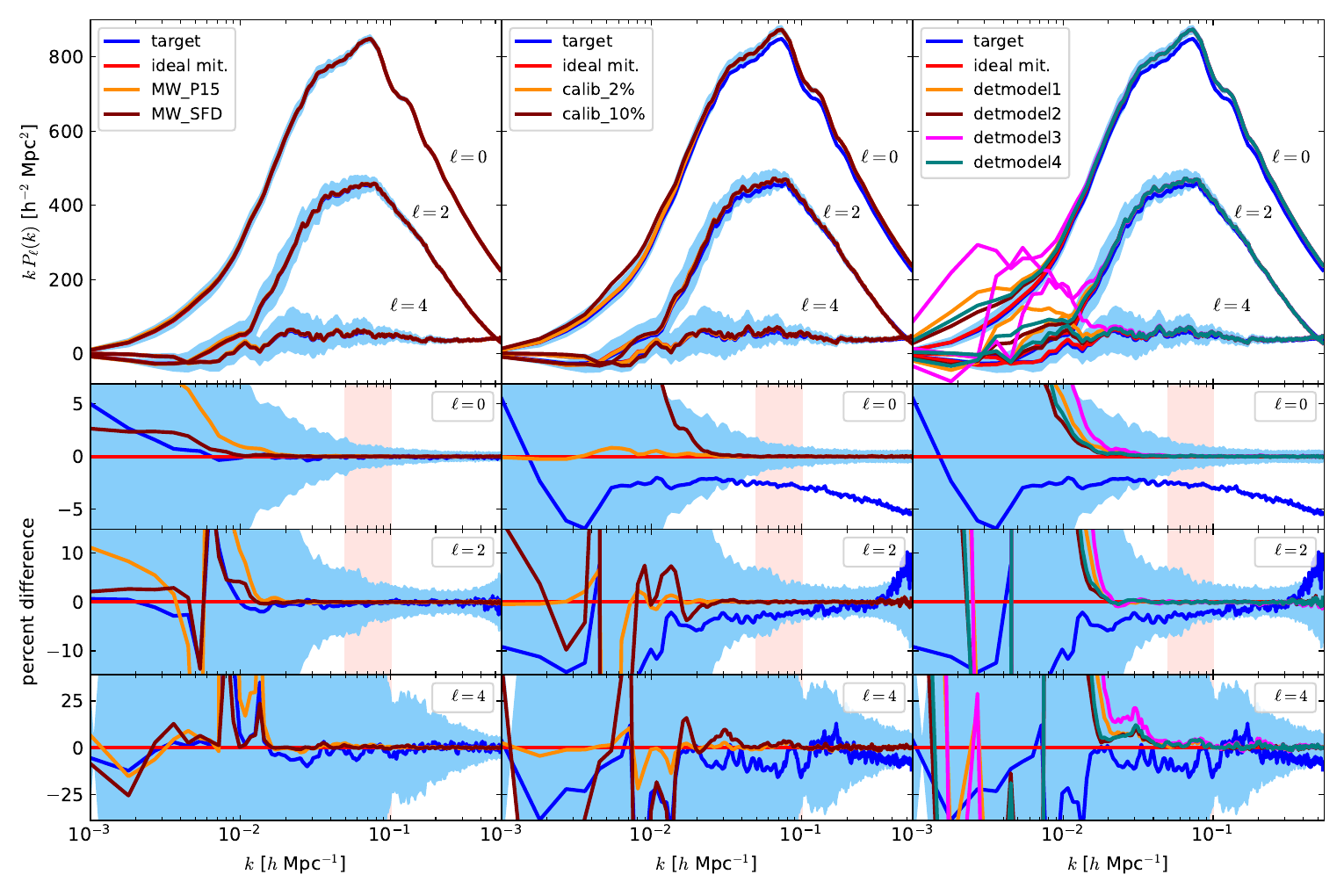}}
  \caption{Monopole, quadrupole, and hexadecapole of the galaxy power spectrum in the case
    of realistic mitigation, for MW extinction on the left, calibration error in the
    middle, and error on the detection model on the right. The lighter blue shaded areas
    give the variance of the ideal measurement, rescaled to the average of five
    catalogues. Here the $x$-axis scale is logarithmic. The residual panels report the
    ratio of the measurements with respect to the ideal mitigation, all in per cent.}
  \label{fig:pk_realistic}
\end{figure*}

As discussed in Appendix~\ref{app:MW}, assessing the uncertainty on MW extinction is not
easy. We represent it here using the difference between the three maps: in addition to P13, we
use another map presented by the Planck Collaboration \citep[hereafter
  P15;][]{PlanckDust2016} and the older \cite{Schlegel1998} reddening map (hereafter SFD).
These are reported in Table~\ref{table:cases}. To show the effect of the uncertainty in
this systematic, we compute (similarly to what we did in Fig.~\ref{fig:maps_shuffling}) a
projected density contrast based on the true MVM, that is the true completeness function
$\Comp$, and its approximation $\Compa$ (see Sect.~\ref{sec:contrast}), as $\Delta_\Comp =
(N_{\Compa} - N_{\Comp})/N_{\Comp}$, where $N$ is again the number of random galaxies in a
sky pixel. The upper left panel of Fig.~\ref{fig:maps_realistic} shows this quantity
computed using {\it MW\_P13} and {\it MW\_P15}. The strongest difference is due to the
modelling of dust spurs, but this difference induces modest projected density contrasts.
In Fig.~\ref{fig:pk_realistic}, left panel, we show the power spectrum multipoles, with a
logarithmic scale on the $x$-axis to better highlight the behaviour on large scales. Here
the data catalogues have been selected using P13, while the random catalogues have
  been processed using the three reddening maps, thus obtaining one ideal and two
  realistic mitigations; the extinction curve is the same in all cases. In this figure we
  report the target measurement, but we do not report again the ad hoc mitigation; we give
  in the residual panels the ratio of the measurements with respect to the ideal
  mitigation. The scale of the $x$-axis is logarithmic to better show the large-scale
  behaviour. 
While the BAO scale is again accurate to the sub-per cent level, large scales are
somehow sensitive to the specific map, and while SFD and P13 give fairly consistent
results, P15 deviates significantly. 

Another source of uncertainty that should be addressed comes from an extinction curve that
varies on the sky \citep[e.g.][]{Schlafly2016}, due for instance to a variation of dust
composition in the Solar neighbourhood. This case is discussed in
Appendix~\ref{app:entrypoints}, we expect this effect to be important on the largest
scales and do not test it in this paper.

The noise map will be read from the dispersed NISP images, so its measurement will be
subject to a statistical uncertainty that will be localised in the pixel or, at most, in
the surrounding pixels. Because the quantity that VMSP uses is the error on pixel electron
counts, to perturb this we should know the error on the error. While it is possible to
compute this quantity and perturb the pixel-level noise, we can anticipate that the impact
on the results would be negligible: this perturbation is degenerate with the effect of the
scatter in the detection model, which will be shown below to be negligible. We  thus
neglect this perturbation term.

A more important and potentially relevant effect is the accuracy in the calibration of the
photometric zero-point in the NISP images, discussed in Appendix~\ref{app:calibration}. A
drift of this zero-point would bias the noise map input to the tabulated detection
probability, and since it couples with the survey timeline, it would create large-scale
features in the MVM. The calibration of NISP images, described in
Appendix~\ref{app:calibration}, relies on repeated, monthly visits of a field named
SelfCal, aided by repeated visits of the EDS fields. We present here an experiment where
the calibration has a systematic variation that is reset at any visit of the SelfCal field
or of the EDS. Because the time span between two successive visits of a calibration field
can be a month, a drift of 0.7\% (as required for the stability of calibration of nearby
fields) between successive ROSs would give catastrophically large offsets, so we do not
use this figure but assume that the offset in zero point of flux measurement can have a
maximum value. We construct a toy model for the calibration error by assuming that: (i)
the zero point drifts as a parabola (as a function of time) between two successive visits
of either the SelfCal field or one of the EDS fields; (ii) the slope of this parabola at
the recalibration time is always the same; and (iii) the largest possible value of the
zero-point (computed for the largest time span between visits of the calibration fields)
is either 2\% or 10\% (Table~\ref{table:cases}), the first value being pessimistic and the
second unrealistic, used just to stress-test the scheme. The value of noise used in the
tabulated detection probability is then increased by this relative amount.

The upper middle panel of Fig.~\ref{fig:maps_realistic} shows the map obtained using the
{\it baseline} case for $\Comp$, and {\it calib\_10\%} for $\Compa$ to make the effect
more visible. This calibration error, coupled with the survey timeline, creates features
that are easy to recognise, yet relatively weak. In Fig.~\ref{fig:pk_realistic}, middle
panel, we show the power spectra of the target and ideal measurements, compared with
realistic mitigation for 2\% and 10\% calibration errors. 
It is evident that a pessimistic 2\% calibration error hardly changes the
measurement, while a 10\% error creates some significant features on large scales, but
gives no effect at or below the BAO scale. We conclude that the calibration error is not
expected to lead to significant effects on the clustering measurements.
This time we see significant deviations of the ideal mitigation with respect to the target
  one, due to the different sample selection.

To assess the impact of an error on the detection model, we follow the quantification of
its uncertainty presented in Sect.~\ref{sec:pipeline} and perturb the sigmoid function
adding $\pm1\sigma$ to the value of each of the fitting parameters. We thus obtain four
perturbed MVMs, which are also listed in Table~\ref{table:cases}. To construct these MVMs
we take the tabulated detection probability described in Sect.~\ref{sec:lookup_table}, for
each bin in $f$, $z$ and reddening we work out an effective S/N using the true detection
model, then feed it to the perturbed detection model to recompute the detection
probability. This algorithm should in principle be applied to the galaxies processed by
{\pypelid} and not to the probability averaged over many galaxies, but the fine binning in
redshift and flux used to compute the tabulated detection probability should guarantee
that all galaxies that end up in a bin have similar S/N, so this approximation is
considered adequate for this test.

Figure~\ref{fig:maps_realistic} shows the projected density contrast maps of the four
perturbed random catalogues with respect to the {\it baseline} one, and
Fig.~\ref{fig:pk_realistic} (right panel) shows the impact of the perturbations of the
detection model on the power spectrum multipoles. Errors in the detection model mostly
impact large scales, while all scales smaller than the BAO scale are largely unaffected.
However, from Fig.~\ref{fig:maps_realistic} it is clear that these systematics leave in
the realistic random catalogue a very predictable residual map. This will be recognisable
in the density contrast field traced by $n_{\rm o}-\alpha\, n_{\rm r}$, opening the
possibility of calibrating the detection model by requiring no correlation of the
mitigated density contrast field of the spectroscopic catalogue with this potential
residual map. In addition, these perturbations of the detection model induce a relative
change in the number density of the random catalogue of the order of 20--30\% with respect
to the {\it baseline} case, while the accuracy of the LF $\Phi(f|z)$ from the EDS will be
1 to 2\% \citepbruton. This choice would lead to a mismatch between random catalogue and
catalogue number densities, already discussed in Sect.~\ref{sec:random}, which could be
used to fine-tune the calibration of the detection model. This confirms that the
variations of the detection model we are implementing here are pessimistic.

A final source of systematics lies in the intrinsic scatter of the detection model. We
assume here that this scatter has no correlation with galaxy properties, an assumption
that will be directly verified in the calibrator set. We implement this scatter at the
catalogue level by modulating the probability to detect a galaxy (in the {\it baseline}
case) with a Gaussian-distributed multiplicative term with variance provided by the
fitting function shown in Fig.~\ref{fig:detmodel}. The random catalogue is consistently
selected with the {\it baseline} case, since the scatter is intrinsic to the data but
cannot be modelled in the MVM. The resulting catalogues show a power spectrum that is
nearly indistinguishable from that obtained using the {\it baseline} selection. Clearly,
such a pixel-level noise does not propagate to clustering on cosmological scales. We
finally notice that, as anticipated above, adding a scatter to the detection probability
is degenerate with adding an error to the noise map, and more generally to any pixel-level
source of uncertainty that adds scatter but whose average is well represented in the
random catalogue. These uncertainties do not propagate to clustering measurements, at
least on the scales that are used to infer cosmology.

\subsection{The effect of systematics on cosmological inference}
\label{sec:fullshape}

When the MVM is applied to non-shuffled fluxes, the intrinsic clustering signal changes as
a function of selection, due to the luminosity dependence of galaxy bias. We then test the
impact of realistic mitigation directly on the inferred cosmological parameters; the
metric for the success of a mitigation strategy will be how well the inference procedure
recovers the cosmological parameters as compared to the analysis of target galaxies.

We fit the four redshift bins simultaneously, using the model described in
Sect.~\ref{sec:inference} and a numerical covariance obtained by applying the {\it
  baseline} combination of systematics to all the 1000 EuclidLargeMocks and measuring its
power spectra in the four redshift bins. The fit is performed using the monopole and quadrupole moments in the range
$k_{\rm min}=0.009\,\hmpc < k < 0.20\,\hmpc$. We process five different cases: the target
catalogues, the ideal mitigation in the {\it baseline} case, and three of the pessimistic
scenarios listed in Table~\ref{table:cases}, namely {\it detmodel1}, {\it detmodel3}, and
{\it calib\_10\%}. For each data vector in the fit, we average over five mocks and
similarly rescale the covariance (see Sect.~\ref{sec:shuffled}). We use the EFT model,
with the priors listed in the second column of Table~\ref{table:Priors_fit}. To speed up
parameter inference, we apply analytical marginalisation as in \cite {Carrilho_2023}, a
technique that has been shown to be effective without biasing the un-marginalised
parameters \citep[see figure 8 in][]{pezzotta+2025} and can be applied to any parameter
that enters the model linearly and has analytically integrable priors. With this technique
and adopting wide Gaussian priors we analytically marginalise over $\theta_{\rm AM} =
\{c_0,c_2,c_4,c_{\rm nlo},N^P_0,N^P_{20}\}$.

\begin{figure}  
\centering{\includegraphics[width=1\columnwidth]{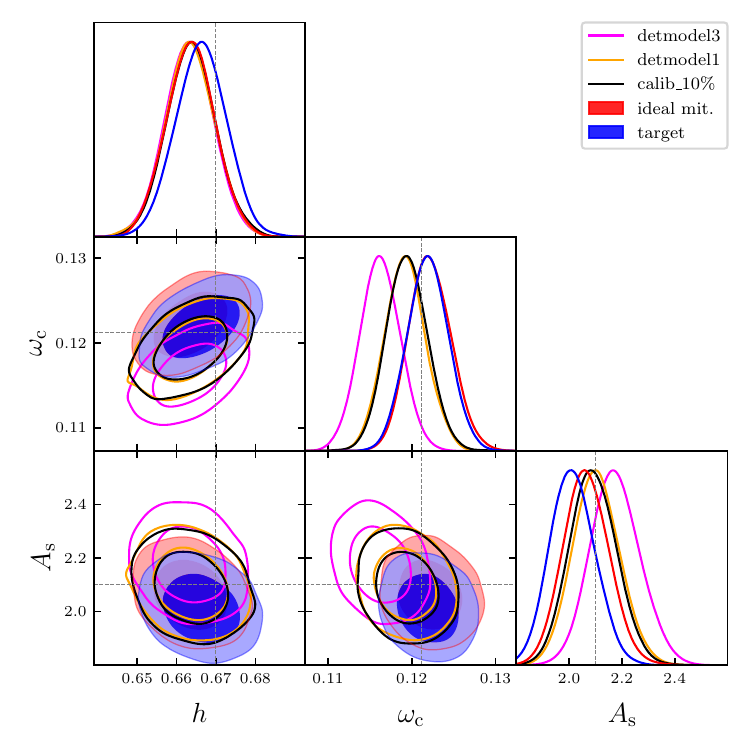}}
\centering{\includegraphics[width=1\columnwidth]{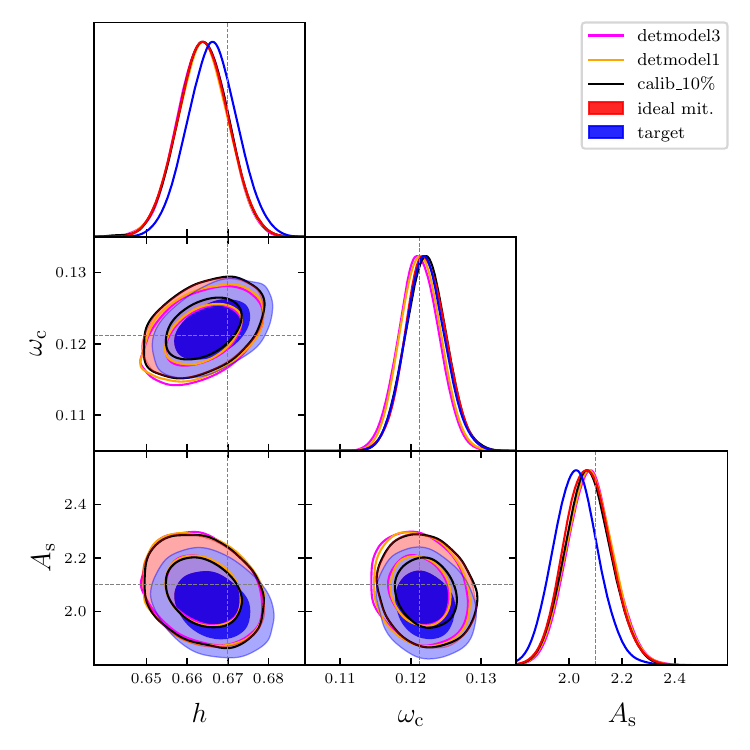}}
  \caption{Posteriors of the derived cosmological constraints due to angular systematics under
    various mitigation strategies. The colour scheme matches that of
    Fig.~\ref{fig:pk_realistic}. The upper panel uses $k_{\rm min} = 0.009\, \hmpc$, the lower panel $k_{\rm min} = 0.025\, \hmpc$. }
  \label{fig:corner_plot}
\end{figure}

In the upper panel of Fig.~\ref{fig:corner_plot}, we present the posteriors of cosmological parameters
obtained using a single set of five simulations. The colour scheme matches that of
Fig.~\ref{fig:pk_realistic}. The ideal mitigation achieved using the {\it baseline} random
catalogue almost perfectly matches the constraints of the target catalogue. Additionally,
we observe that realistic mitigations primarily affect the $\omega_{\rm c}$ parameter,
while their impact on $h$ is minimal and on $A_{\rm s}$ is limited. 
To test whether the induced bias in $\omega_{\rm c}$ can be reduced by excluding the
largest scales, where the impact of realistic mitigation on the power spectrum is
strongest (see Fig.~\ref{fig:pk_realistic}), we repeat the fit by setting the minimum
scale to $k_{\rm min} = 0.025\,h\,\mathrm{Mpc}^{-1}$. The results are shown in the lower
panel of Fig.~\ref{fig:corner_plot}, where the biases induced in $\omega_{\rm c}$ and
$A_{\rm s}$ are largely removed, and all models are consistent with the ideal mitigation
case. 

\begin{figure*}
\centering{
\includegraphics[width=0.9\columnwidth,trim={0 1cm 0 .9cm},clip]{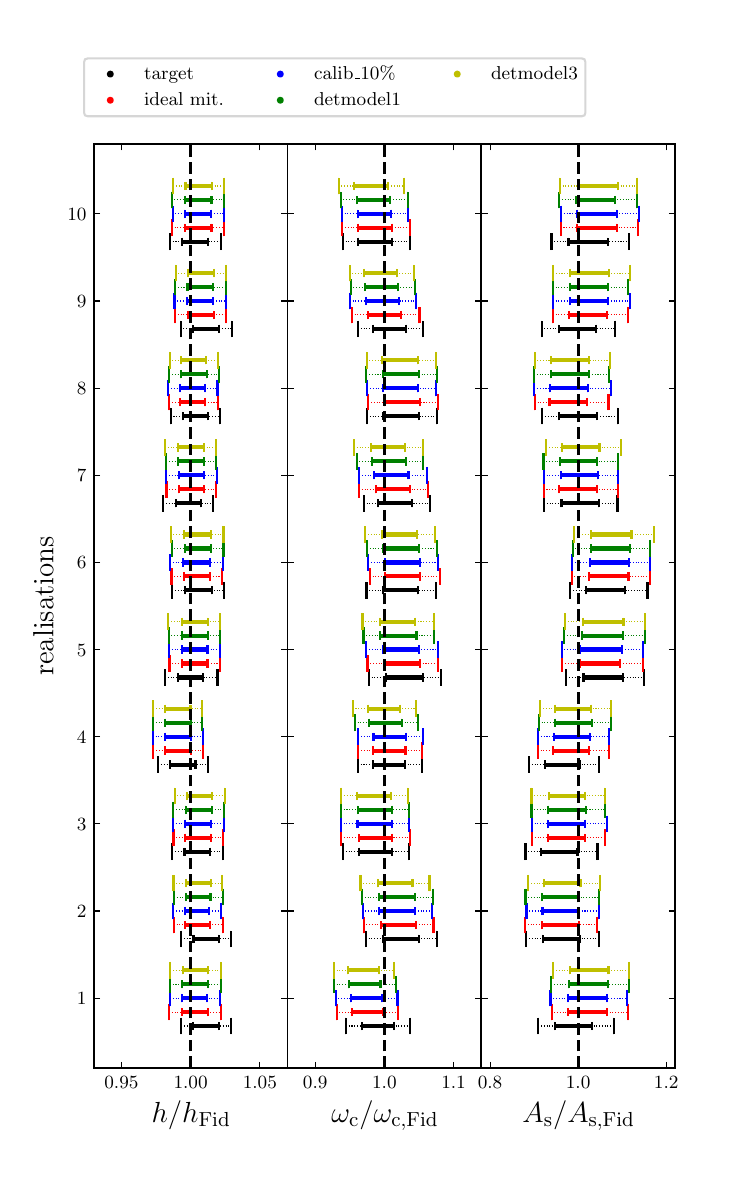}
\includegraphics[width=0.9\columnwidth,trim={0 1cm 0 .9cm},clip]{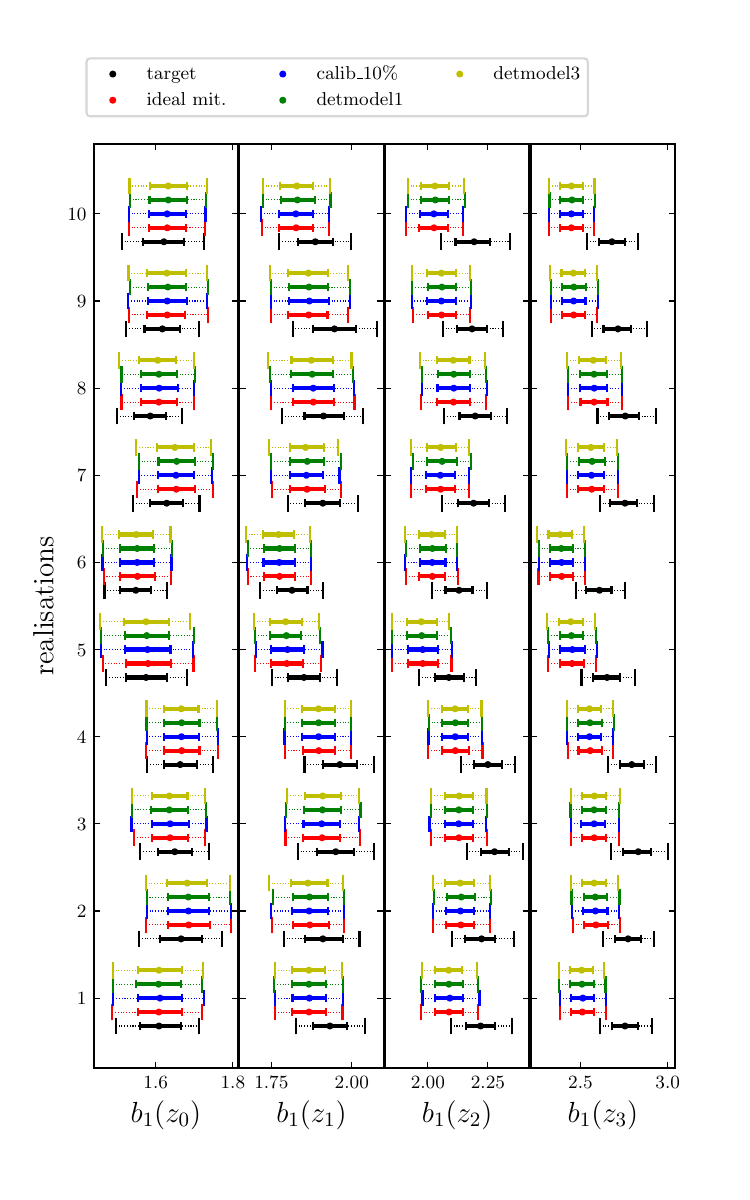}
}
  \caption{Averages of the posteriors obtained (using $k_{\rm min} = 0.025 \hmpc$) for the three cosmological parameters
    (left) and of linear bias parameter (right) included in our analysis, for all
    realisations across the five cases considered. The 68\% and 95\% confidence intervals
    are overplotted as thick and thin bars, respectively. }
  \label{fig:bar_plots_cosmo}
\end{figure*}

To show results across different realisations, in Fig.~\ref{fig:bar_plots_cosmo}, left
panel, we report the 1 and 2\,$\sigma$ confidence intervals on the cosmological
parameters, represented by thick and thin bars, respectively, for all ten sets of
realisations using $k_{\rm min} = 0.025 \hmpc$. 
The figure shows that increasing the $k_{\rm min}$ is sufficient to recover the cosmological parameters for all models with accuracy similar to those of the target sample. The right panel of the same figure reports
the inferred linear bias parameter $b_1$ for the four redshift bins. Due to the different
selections and to luminosity-dependent galaxy bias, the $b_1$ value of target galaxies
differs from the other selections, with this discrepancy becoming more pronounced at
higher redshifts. It is evident though that the model is effective in marginalising over
the differences in galaxy bias.

These results demonstrate that the strategy described in this paper to mitigate angular
systematics through a random catalogue allows unbiased inference of cosmological
parameters when the systematics are perfectly modelled. This work lays down a roadmap for
assessing the error budget for the \Euclid spectroscopic sample, which is dominated by the
uncertainty in the calibration of the detection model. Appendix~\ref{app:metrics} gives a
more detailed assessment of the metrics of the goodness of fit in the five cases
considered here.

\section{Conclusions}
\label{sec:conclusions}

We presented the strategy devised by the Euclid Collaboration for dealing with data
systematics in the cosmological exploitation of the \Euclid spectroscopic sample, starting
from the definition of the observable, the galaxy density $n_{\rm g}$ and its density
contrast $\delta_{\rm g}$. We surveyed the processing pipeline to have a wide and
organised view of all the possible contributions to the cosmological error budget.
Differently from previous surveys, the spectroscopic Euclid Wide Survey will have minimal
biases coming from the photometric pre-selection of targets, at the cost of having a
higher number of catastrophic redshift errors, leading to a sizeable fraction of line and
noise interlopers in the observed sample. We classified the systematics into two main classes,
angular systematics that imprint spurious clustering at all redshifts, and redshift errors
that mix clustering coming from different ranges along the line of sight. To a first
approximation, angular systematics are multiplicative in the galaxy density $n_{\rm g}$,
while redshift errors give additive terms; however,  both are more complicated than additive or
multiplicative for the density contrast $\delta_{\rm g}$.

We then focused on angular systematics, where mitigation will rely on the construction of
a visibility mask in the form of a random catalogue. This can be seen as a forward model
of the selection function, based on a detection model that gives the probability of
detecting yet another galaxy in the observed images as a function of the S/N of its {\ha}
line. The random catalogue is constructed with a (bypassed) source injection algorithm,
while weights (alias detection probabilities) are assigned to random galaxies and not to
observed ones.

We used 50 EuclidLargeMocks \citep{EP-Monaco1}, split into in ten groups of five mocks
that cover the expected final EWS area, to test our ability to recover unbiased power
spectrum measurements and cosmological parameters in the presence of angular systematics.
This was done in the case of ideal mitigation by applying to the mocks the same
visibility mask represented by the random catalogue, and in the case of more realistic
mitigation, where we took into account the uncertainty of the detection model. Weighting
random galaxies implies an average density that varies on the sky, which can be taken into
account in Fourier space, at the sub-per cent level in the power spectrum monopole, by
convolving the theory model with a window function obtained via a Fourier transform of the
random catalogue itself.  The resulting cosmological parameters are very consistent for
ideal mitigation, as revealed by our FoB, FoM, credibility interval width, and $p$-value
analyses (see Appendix~\ref{app:metrics}). The model can effectively marginalise over
galaxy bias parameters, even though their values depend on galaxy selection. While
photometric calibration does not seem to have a strong impact, when perturbing the
detection model we stress-tested the parameter estimation, obtaining in one case a worse FoB and
a bad $p$-value, together with a smaller FoM. However, such miscalibration of the
detection model would be easily recognisable as a residual correlation of galaxy density
contrast with nuisance maps.

Most of the work presented in this paper was done before the launch of \Euclid, although
it was finalised after Quick Data Release 1, and while DR1 data analysis was in
preparation. Early data have fully confirmed the extent of the challenge of extracting
spectra from thousands of square degrees of dispersed images, and the pipelines have
quickly evolved to meet the challenge and are now ready for the first wave of cosmological
results. The EDS is progressing, but, as planned in \cite{Scaramella-EP1}, none of the
surveyed sky patches is presently available at full depth. This means that the present
strategy will be fully usable starting from the next data release (DR2), when at least
one-half of the EDS will be surveyed at full depth.

\begin{acknowledgements}
 
{\AckEC} 

This work has made use of CosmoHub \citep{Carretero:17,Tallada:20}
  developed by PIC (maintained by IFAE and CIEMAT) in collaboration
  with ICE-CSIC. CosmoHub received funding from the Spanish government
  (MCIN/AEI/10.13039/501100011033), the EU NextGeneration/PRTR
  (PRTR-C17.I1), and the Generalitat de Catalunya.
This paper is supported by the Italian Research Centre on High Performance Computing Big
Data and Quantum Computing (ICSC) and by the Fondazione ICSC National Recovery and
Resilience Plan (PNRR) Project ID CN-00000013 ``Italian Research Centre on
High-Performance Computing, Big Data and Quantum Computing'' funded by MUR Missione 4
Componente 2 Investimento 1.4: ``Potenziamento strutture di ricerca e creazione di
campioni nazionali di R$\&$S (M4C2-19)'' - Next Generation EU (NGEU); by the PRIN 2022
PNRR project ``Space-based cosmology with \Euclid: the role of High-Performance
Computing'' (code no. P202259YAF), funded by ``European Union – Next Generation EU''; and
by MUR PRIN 2022 (grant 2022NY2ZRS 001). We acknowledge usage of Pleiadi computing system
of INAF \citep{Taffoni2020,Bertocco2020}. PM thanks Yi-Kuan Chiang for discussions on MW
extinction.

\end{acknowledgements}

\bibliography{mybiblio}

\begin{appendix}

\section{Response of galaxy density to nuisance maps}
\label{app:flux_modulations}

If the galaxy bias is independent of luminosity, and the galaxy ({\ha} line) luminosity
function $\Phi(f|z)$ is universal (meaning that its shape is the same in all space
location, independent of galaxy density), it is useful to define the effective flux limit
$f_{\rm lim}$ of the survey such that the number density of galaxies brighter than this
flux is equal to the number density of observed galaxies,

\be
f_{\rm lim}(\nh,z,\{N_i\}): \ \ \ \int_{f_{\rm lim}}^\infty \Phi(f|z)\, {\rm d}f = \int_0^\infty \Comp(f,z,\{N_i\}) \, \Phi(f|z) \, {\rm d}f\, ,
\label{eq:flux_limit}\ee

\noindent
where, as in Sect.~\ref{sec:formalism}, $\nh$ gives the angular position on the sky,
$\{N_i\}$ is the set of nuisance maps and $\Comp$ is the completeness function, alias MVM.
This flux limit acquires a dependence on sky coordinates $\nh$ through the nuisance maps
$\{N_i\}$. The observed number density of correct galaxies is then

\be
n_{\rm oc} = (1+\delta_{\rm g}) \int_{f_{\rm lim}}^\infty \Phi(f|z)\, {\rm d}f\, ,
\ee

\noindent
while the number density of target galaxies $n_{\rm t}$, given in
Eq.~(\ref{eq:target_ng}), is expressed by the same integral starting from $f_0$. The
relative residual ${\cal R}$ of $n_{\rm oc}$ with respect to $n_{\rm t}$, can be written
as

\be
{\cal R} := \frac{n_{\rm oc}-n_{\rm t}}{n_{\rm t}} = -\frac{1}{\overline{n}_{\rm t}} \int^{f_{\rm lim}}_{f_0} \Phi(f|z)\, {\rm d}f \, .
\label{eq:residual} \ee

\noindent
This shows that a variation of flux limit gives rise to a multiplicative systematic effect
on the density.

As discussed in \cite{Monaco2019}, it is convenient to Taylor-expand the integral in flux
limit from $f_0$ to $f_{\rm lim}$; because the modulations of $f_{\rm lim}$ are not
necessarily expected to be small, the expansion cannot be limited to the first order. We
call $\epsilon := (f_{\rm lim} - f_0)/f_0$ the relative variation of the flux limit with
respect to the fiducial one, and define two redshift-dependent coefficients that are
determined by the shape of number counts: $S_\Phi := f_0\,\Phi(f_0)/\overline{n}_{\rm t}$
and the logarithmic slope $\alpha_\Phi:=\partial\ln\Phi/\partial\ln f\, (f_0)$. By
expanding the integral to second order, the relative residual becomes

\be
{\cal R} = -S_\Phi \left( \epsilon + \frac{1}{2} \alpha_\Phi \epsilon^2 + {\cal O}(\epsilon^3) \right)\, .
\label{eq:expansion}\ee

To compute the response of density residuals to systematics we need to relate $f_{\rm
  lim}$ to the nuisance maps. We call the exposure time ${\cal T}$ and assume that the
noise map represents the variance of the pixel-level counts, and is the sum of a series of
noise terms ${\cal N}_i$ (with calligraphic font to distinguish this from the list of
nuisance maps), so that the total noise is ${\cal N}:=\Sigma {\cal N}_i$. For the
reddening $E(B-V)$, we define the relative decrease of the observed flux as

\be {\cal E}:=10^{-0.4\, R(z)\, E(B-V)}\, , \ee

\noindent
where $R(z)=R_\lambda\,[\lambda_{{\rm H}\alpha}(1+z)]$. Then the S/N $S$ of the {\ha} line
can be related to is flux as

\be
S=\kappa_1\, f\, {\cal E} \sqrt{\frac{\cal T}{\cal N}}\, ,
\label{eq:snr}
\ee

\noindent
where $\kappa_1$ is a suitable constant that translates the right-hand side of that
equation into a dimensionless number. The completeness function $\Comp$ in
Eq.~(\ref{eq:flux_limit}) can be expressed as the marginalised detection probability
$P_{\rm det}$ (Eq.~\ref{eq:sigmoid}), which is a function of $S$ alone. This function goes
from 0 to 1 around the value $S_0$, so it is clear that

\be
f_{\rm lim} = \kappa_2(z) \frac{S_0}{\kappa_1{\cal E}} \sqrt{\frac{\cal N}{\cal T}}\, ,
\label{eq:flim} \ee

\noindent
where $\kappa_2(z)$ will be determined by the solution of the integral in
Eq.~(\ref{eq:flux_limit}) and will depend on the shape of the number counts $\Phi(f|z)$,
which is again redshift dependent.

The response of galaxy density to a specific nuisance can be computed as the partial
derivative of $\cal R$ with respect to that nuisance. For a noise term ${\cal N}_i$,

\be
\frac{\partial {\cal R}}{\partial {\cal N}_i} = \frac{\partial {\cal R}}{\partial \epsilon} \frac{\partial \epsilon}{\partial {\cal N}_i} = - \frac{1}{2} S_\Phi \left(1+\alpha_\Phi\epsilon +{\cal O}(\epsilon^2)\right)  \frac{1-\epsilon}{{\cal N}}\, ,
\label{eq:response_n}\ee

\noindent
where the (total) noise ${\cal N}$ is computed at the unperturbed value. The response to a
variation in exposure time can be computed similarly,

\be
\frac{\partial {\cal R}}{\partial {\cal T}} = \frac{\partial {\cal R}}{\partial \epsilon} \frac{\partial \epsilon}{\partial {\cal T}} = \frac{1}{2} S_\Phi \left(1+\alpha_\Phi\epsilon +{\cal O}(\epsilon^2)\right)  \frac{1-\epsilon}{\cal T}\, ,
\label{eq:response_t}\ee

\noindent
while the response to a variation in reddening (in ${\cal E}$) would be

\be
\frac{\partial {\cal R}}{\partial {\cal E}} = \frac{\partial {\cal R}}{\partial \epsilon} \frac{\partial \epsilon}{\partial{\cal E}} = S_\Phi \left(1+\alpha_\Phi\epsilon +{\cal O}(\epsilon^2)\right)  \frac{1-\epsilon}{\cal E}\, .
\label{eq:response_e}\ee

Clearly the response to a nuisance is generally non-linear and depends on the shape of the
number counts and through them on redshift. This is why an approach based on mock
catalogues, paired with a random catalogue created from a (bypassed) source injection
procedure, is to be preferred, since it is able to provide unbiased and robust
measurements. A mitigation based on response to a set of potential nuisance fields would
be equivalent only if the response is properly modelled.

An important assumption that we make here is that the physical process including a
systematic effect on the galaxy density is statistically independent of the galaxy density
field itself. This is not always true: (i) the map (based on FIR observations from {\it
  Planck}) used to correct extinction from MW dust may have contamination from the cosmic
infrared background \citep[CIB; see][]{Chiang2019}, so the correction for extinction may
contain a (spurious) term that couples with the density field (this is discussed below in
Appendix~\ref{app:MW}); moreover, (ii) the effect of confusion in the spectral images is
independent of the density field only as long as the main contaminating sources, which are
bright enough to have a significant continuum in the dispersed image, are at $z<0.9$. Such
cases, if not deemed to be negligible, deserve a dedicated study.

Luminosity-dependent bias makes the scheme even more complicated. We assume again that we
can treat the completeness function as a Heaviside $\theta$-function stepping from 0 to 1
at $f_{\rm lim}$. We also call $\beta_1:=\partial b_1/\partial f$ the linear bias
parameter of galaxies in a small range of flux ($b_1$ is reserved for the bias of the
total galaxy sample), and we assume that it depends on flux, $\beta_1=\beta_1(f|z)$. Then

\be
n_{\rm oc} = \int_{f_{\rm lim}}^\infty [1+\beta_1(f)\,\delta] \,\Phi(f|z)\, {\rm d}f\, .
\ee

\noindent
The Taylor-expanded expression of $n_{\rm oc}-n_{\rm t}$ will then be

\begin{align}
    n_{\rm oc}-n_{\rm t} =& f_0\Phi(f_0)\left\{ [1+\beta_1(f_0)\,\delta]\, \epsilon + \frac{1}{2} \alpha_\Phi [1+\beta_1(f_0)\,\delta]\, \epsilon^2 \nonumber \right.\\& \left. +\ \frac{1}{2} \beta_1(f_0)\, \alpha_\beta\, \delta\, \epsilon^2 + {\cal O}(\epsilon^3) \right\}\, ,
\end{align}

\noindent
where we have defined $\alpha_\beta:=\partial \ln \beta_1 /\partial \ln f$ as the
logarithmic slope of the $\beta_1(f)$ function.

Luminosity-dependent bias has two main effects: to add a further term of order
$\epsilon^2$ that is proportional to $\alpha_\beta$, and to break the simple dependence on
matter density that allows us to divide this quantity by $n_{\rm t}$ keeping the
expression treatable. This confirms that the response of relative residuals to nuisances
given in Eqs.~(\ref{eq:response_n}) to (\ref{eq:response_e}) is only an approximation.

\section{Entry points of systematics}
\label{app:entrypoints}

We present here a more detailed discussion of the SGS pipeline, described in
Sect.~\ref{sec:pipeline} (see also Fig.~\ref{fig:OUpipeline}). We highlight the entry
points, which are the steps where systematics may contaminate the galaxy catalogue. This
discussion is at the basis of several of the entries of Table~\ref{table:systematics}.

\subsection{The photometric sample}
\label{app:photometric}

Slitless spectroscopy allows us, in principle, to define a sample purely selected in line
flux. This is not exactly true in the \Euclid pipeline analysis, since the spectra will be
extracted starting from a list of photometric sources. The risk is then to contaminate the
spectroscopic sample with fluctuations in survey depth of the photometric sample; although
space has a definite advantage over Earth in the cleanness of photometric selection, the
characterisation of the depth of a photometric sample obtained from the merging of VIS and
NISP photometric catalogues is tricky, since images from the two instruments are affected
at a varying degree by many of the systematics that will be described below. Moreover, the
deblending algorithm of NISP sources adds some complexity to the process, due to the
poorer PSF in the NIR and to persistence effects. However, photometry goes far deeper than
spectroscopy, and this is already evident from Table~\ref{table:number_of_sources}, where
the ratio between $\HE<24$ sources and ELGs is of order 50:1. If the number of ELGs with
photometric fluxes near or below the threshold is negligible, fluctuations of this
threshold will not influence the resulting catalogue. The number of ELGs with {\ha} line
flux above $f_0$ and with $\HE>24$ was estimated by \cite{Bagley2020} to be around 2\%, so
a modulation of this flux limit would have an influence on a few per cent of the sample,
inducing negligible effects on the galaxy density. This claim is a bit optimistic,
however, since the threshold for line detection is not a Heaviside theta function of the
true line flux, so a significant part of the survey will be at lower fluxes, where the
probability of being hosted in an $\HE\ge24$ galaxy is higher.

Photometric information, both from \Euclid and ground-based images, can be used by SEL to
significantly improve our ability to reduce both line and (especially) noise interlopers,
thanks to the broader spectral coverage \citep{Cagliari24}. However, this is done at the
cost of introducing subtle systematics, especially when ground-based data are used; in
this case a pure forward model of systematics would be impossible. Systematics in the
ground-based photometry may modulate on the sky the effectiveness of the removal of
interlopers; this modulation would impact on a ten-per cent-level contribution, so its
effect would possibly be small. However, it would also remove a significant fraction of
faint correct galaxies, giving a potentially larger effect. Careful tests will be needed
when assessing the effects of any photometric-based selection. A similar concern applies
to usage of photometric redshifts to inform the choice among different redshift solutions,
aimed at limiting interlopers.

Photometry is used by SIR to decontaminate the spectra of ELGs from those of brighter,
continuum-dominated sources such as stars and nearby galaxies. Here a degradation of
photometry could bias spectral decontamination and affect galaxy spectra in a subtle way,
but such bright sources will hardly be affected by modulations of a flux limit that is at
least four magnitudes fainter.

Another issue of photometry that can affect the spectroscopic sample is deblending of
sources, which can fail in crowded fields. Deblending will mostly rely on VIS images,
which have a better PSF, but very red sources may be missing in VIS, and occasional
problems of some VIS pointings may worsen the ability of deblending. Under-deblending will
result in a missing source in the photometric catalogue, and in a mis-centring of the main
blended galaxy (which will likely be at low redshift), impacting on sample completeness in
a way that is not represented by the random catalogue. Conversely, over-deblending will
result in fake sources, contributing to noise interlopers; in some cases it may induce a
double-counting of ELGs, but this would only happen to very near and extended galaxies.
These effects will strongly depend on source surface density, but will likely be
negligible with respect to confusion in spectroscopic images.

\subsection{Astrophysical foregrounds}
\label{app:foregrounds}

Background noise in the dispersed images will have an important contribution from
astrophysical foregrounds, mostly zodiacal light and, to a lesser extent, nearby galaxies.
We do not consider here stars in this class because they mostly impact through stray light
and spectral confusion, which will be addressed below. Zodiacal light is, along with MW
extinction, the main astrophysical killer of detection probability, and one of the main
drivers of the EWS design. However, its behaviour is very smooth, both in angular
coordinates and in SED, and it is moreover very predictable, so we expect to be able to
remove its impact through the random catalogue. As a consequence, in
Table~\ref{table:systematics} zodiacal light has a very high impact, but a very low risk.

Nearby galaxies have a much smaller angular footprint, but their collective area can hide
galaxies in the background. In principle VMSP will take account of their obscuration,
unless they are due to a failure of the MER deblending algorithm: as an example, VMSP may
place a random galaxy in the outskirts of a nearby galaxy, and the {\ha} line may fall in
a region where it is detectable, but the deblending algorithm may fail to recognise it as
a target galaxy and not as a star-forming region of the foreground object; as a
consequence VMSP would inject a random galaxy where a true one would go undetected, thus
creating a fake underdensity. This may be recognised by correlating the galaxy density
field with a map of nearby galaxies. However, this effect is expected to be small.

\subsection{MW extinction}
\label{app:MW}

Light coming from extragalactic sources is absorbed or scattered by dust in the MW in a
way that is very hard to control. Galactic dust is heated to a temperature of around 20 K
by the local radiation field of stars, and it re-emits this energy in the FIR as a
modified blackbody, peaking at 150 {\micron}. This emission is an important foreground for
CMB observations, and it can be effectively separated from the CMB signal due to its
different spectrum. The most accurate measurement of dust emissivity has been presented by
Planck Collaboration in 2013 (P13), which thanks to its high-frequency instrument HFI,
augmented with IRAS data at 100 {\micron}, has been able to measure all-sky dust radiance
(the integral in frequency $\nu$ of the monochromatic surface brightness $I_\nu$). In that
paper an SED was constructed for each 5{\arcminute} sky pixel of a \healpix tessellation,
and dust emissivity in four bands was fit by a single modified blackbody, characterised by
a temperature $T$, an index $\beta$, and an optical depth $\tau_{353}$ at 353 GHz. Dust
radiance was found to be robust against the known degeneracy of $T$ and $\beta$. The
Planck Collaboration provided best-fit parameters and dust radiance as \healpix maps,
along with their uncertainty and with an estimation of induced reddening, $E(B-V)$.

While the measurement of dust radiance can be considered robust, its relation with
reddening and extinction is much more uncertain. In P13 it was assumed that radiance is
correlated with reddening, and the correlation coefficient was found using a set of
54\,492 quasars from the Sloan Digital Sky Survey, for which reddening was estimated by
comparing their spectra with a template. This procedure is valid as long as dust physical
properties and radiation field are constant in the volume of the dust disc that intersects
the survey line of sight. We can estimate the size of this volume by assuming an height of
100 pc for the dust disc and a zone of avoidance at Galactic declination $b<20$\degree;
the intersection is thus a cylinder of diameter $\la500$ pc. A gradient in dust
composition, temperature or in stellar radiation field in this volume will influence the
coefficient that translates dust radiance to reddening.

In P15, a more extended analysis of dust emission was presented, adding more bands at
shorter wavelengths from WISE data and fitting the pixel-level SEDs with the model of
\cite{Draine2007}, to produce more physically motivated estimates of dust masses and
$V$-band extinction, $A_V$. However, extinction values turned out to be off by a factor of
two, and were recalibrated by using a more extended sample of SDSS quasars. Considering
the high quality of the dust model used, such a recalibration casts legitimate doubt on
the level of control that we can achieve on the physics behind dust extinction.

Calling $A_V=R_V\, E(B-V)$ the extinction in the Johnson $V$ band, where $R_V$ is a
scaling parameter to which we assign the standard value of $R_V=3.086$, dust extinction at
wavelength $\lambda$ is computed as $A_\lambda = k(\lambda)\, E(B-V)$, where $k(\lambda)$
is an extinction curve whose integral over the $V$-band wavelength range gives $R_V$.
There is ample evidence that both $R_V$ and the extinction curve are not universal but
change with the line of sight \citep[e.g.][]{Schlafly2016,Ferreras2021,Green2024}. Then,
even a perfect knowledge of reddening $E(B-V)$ does not guarantee a precise knowledge of
extinction at the wavelengths of the red grism. Moreover, a straightforward propagation of
the nominal error in the extinction would not represent the true uncertainty, since for a
two-point clustering measurement, and even more for its covariance, the important quantity
is not the 1-point PDF of the error but its correlations on the sky. This point was
addressed in \cite{Monaco2019}, where a more extended discussion can be found.

A known obstacle to the measurement of dust properties is given by the already mentioned
cosmic infrared background, or CIB, due to FIR emission of background galaxies. If not
properly subtracted, this background can modulate the measurement of dust temperature $T$,
especially through its degeneracy with $\beta$. While the CIB monopole is easy to be
characterised and subtracted, its anisotropies due to galaxy clustering can affect dust
extinction estimates. CIB contamination of dust maps was carefully assessed by
\cite{Chiang2019} and \cite{Chiang2023}, who proposed a data-driven, map-level
reconstruction method to subtract CIB contribution from dust radiance maps. If not
properly subtracted, CIB provides a very worrying systematic effect, as it correlates with
the cosmological signal $\delta_{\rm g}$ we are measuring, breaking the basic assumption
that $\Comp$ and $\delta_{\rm g}$ are statistically independent.

A second problem is related to the fact that dust emission has structure on small scales
that are not sampled at a resolution of about \ang{;5;}; such intra-field variations are
not mitigated by applying extinction from the {\it Planck} maps, or from maps at similar
resolution. To assess the impact of this effect, it is useful to analyse the angular power
spectrum of dust reddening, or equivalently of a random catalogue subject to extinction.
This spectrum is shown in figures 3 and 6 of \cite{Monaco2019}, and is well represented by
a decreasing power law with slope $-1$. At the same time, the galaxy angular power
spectrum is raising and dominates over the dust power spectrum at $\ell\ga100$. An
extrapolation of this power law to smaller scales would give a sub-dominant contribution
to galaxy clustering from any intra-field dust. This is confirmed by the analysis
presented in Sect.~\ref{sec:results}: MW extinction adds spurious power on very large
scales, so a small-scale contribution due to an extrapolation of the dust power spectrum
would give a negligible effect.

This argument can be extended to estimate the impact of CIB anisotropies: a strong
contamination from galaxies would give rise to a growing power spectrum at high $\ell$
that is not observed. At the same time, a contamination, as quantified in
\cite{Chiang2019}, which does not perturb much the angular power spectrum of dust adds
power on scales that have little effect on the measurement of galaxy clustering. However,
on the long term mitigation of MW extinction should be done using maps where CIB is
properly subtracted.

MW extinction is expected to give rise to fake power on the largest scales, so it is a
killer systematic effect for any science case based on scales at or below (in $k$-space)
the peak of the power spectrum, such as scale-dependent bias from primordial non-Gaussianity
or relativistic effects near the horizon. As an example, \cite{Karim2025} found that
parameter estimation in the cross correlation of DESI and CMB lensing data depends on the
assumed extinction map. Nonetheless, there is no reason  not to use a
reddening map to correct for extinction, although a straightforward propagation of the
quoted errors cannot represent its true uncertainty. Following \cite{Monaco2019}, the most
effective mitigation of this uncertainty is to consider the residuals after correction
with a map as an unknown residual systematic that must be controlled a posteriori, with
methods based on the cross correlation of different redshift bins

As a final remark, dust has emission down to NIR wavelengths \citep{Lim2023}, so it can
contribute to the general background at low Galactic latitudes. This contribution is
expected to be subdominant with respect to zodiacal light and stray light, and will sum to
the measured background that VMSP uses to construct the random catalogue. Its intra-field
variations will be taken into account by the usage of a local measurement of noise in the
detection model.

\subsection{Systematics in NISP dispersed images}
\label{app:images}

The first crucial step in the detection of spectroscopic galaxies is performed by SIR,
which extracts, decontaminates, and coadds the spectra from the four NISP dispersed
dithers. Pixel-level noise of the signal is later read by VMSP to compute $P_{\rm det}$
using the detection model (Sect.~\ref{sec:random}). NISP dispersed images are affected by
systematics that are in common with all exposures: (i) detector noise; (ii) bad pixels;
(iii) masked regions around saturated stars; (iv) cosmic rays; (v) ghosts from saturated
objects; (vi) zeroth- and second-order spectra from bright objects (in dispersed images);
(vii) stray light; (viii) persistence from previous images; and (ix) zero-point
calibration errors. Spectra are then separated using a decontamination algorithm, taking
advantage of the four dithers dispersing in four different directions. The final S/N of
the spectrum will further depend on modulations of noise due to astrophysical foregrounds
(e.g. zodiacal light and nearby galaxies) and MW extinction.

For detector noise, bad pixels, saturated stars, cosmic rays, ghosts, and zeroth-order
spectra, all these effects influence the image quality, and their impact is encoded into
pixel-level masks, which allow us to exclude contaminated pixels; moreover, they modulate
the pixel-level variance of the signal, in case the feature is modelled and subtracted. We
can envisage that the handling of these effects will have two phases. In the first phase
of data analysis, the algorithms to remove these signals will be sub-optimal, leaving
angular imprints in the galaxy density. These can be identified by creating maps of
potential contaminants and seek for a positive cross-correlation of the galaxy density
with these maps. As an example, an incomplete treatment of saturated stars will create an
anti-correlation of the galaxy density with the density field of bright stars; problems in
the removal of cosmic rays will create a correlation with a measure of (time-dependent)
cosmic ray hits, and so on. The absence of such cross correlations will be a very useful
validation test for the algorithms. In a second phase, when images are reprocessed with
optimised algorithms, all these effects will be encoded in a pixel-level map, which will
be used to compute the variance of the signal. The VMSP algorithm will then be in a
position to propagate this uncertainty to the MVM.

One of the most important contribution to noise in SIR images will be the stray light from
bright stars scattered inside the instrument. This contribution will be contained in the
measurement of the pixel-level signal, used by VMSP to construct the random catalogue, so
the impact of stray light is high in the determination of the S/N, but it can be fully
mitigated by the random catalogue so the associated risk is considered low.

\subsection{Persistence}
\label{app:persistence}

When the photons of a bright source hit a pixel of the NISP detector, some of the
electrons produced are not processed immediately by the readout hardware but are trapped
between the semiconductor layers and later released to create a spurious signal that
slowly dies off. As a consequence, a bright source imaged in a pointing remains visible in
the successive exposures, in a way that is hard to characterise \citep{EU-Kubik}. This
gives rise to four main effects. (i) Spectro-to-photo: dispersed trails of bright sources
are visible in the next photometric pointings as long stripes. This has a strong impact
because the MER deblending code (discussed above, Appendix~\ref{app:photometric}), if not
perfectly calibrated, can fragment these stripes into a stream of fake sources. (ii)
Photo-to-spectro: a bright source in a photometric pointing can contaminate a dispersed
image and mimic an emission line in the spectrum of a random source in one of the dithers.
(iii) Photo-to-photo: the three consecutive photometric pointings {\JE}, {\HE}, and {\YE},
obtained without moving the telescope (with the exceptions of small movements due to the
rotation of filter wheels, where an imperfect compensation can lead to pixel-level
shifts), will be subject to self-persistence, where each flux in the {\HE} and {\YE}
exposures will contain a 1--2\% systematic contribution from the previous pointing. (iv)
Spectro-to-spectro: signal in very saturated pixels may persist from a dispersed
spectroscopic image to the next, creating a feature that is not corrected for by the
decontamination algorithm. In addition to these four effects, since the detector is not
closed during the rotation of the grism wheel, persistence can produce streaks
corresponding to the apparent movement of bright sources when the pointing is adjusted;
this is a minor contribution which, however, cannot be removed using the previous
pointings.

A detailed characterisation of persistence was performed before launch \citep{EU-Kubik},
but the details of the detectors behaviour in space showed that laboratory tests do not
represent the full complexity of actual images. Modelling  this effect is an
important ongoing activity, and a further improved characterisation and subtraction of the
persistent signal is expected for the data releases beyond DR1.

\subsection{Calibration error}
\label{app:calibration}

Another potential killer systematic effect on large scales is the stability of the zero
point of the photometric system. This is especially true for ground-based surveys, where
target selection is based on available photometry. The spectroscopic sample is, to
first-order, free from this effect (see Appendix~\ref{app:photometric}); moreover, once an
emission line is detected, its selection is mostly based on the reliability flag of the
redshift measurement (Appendix~\ref{app:redshift} below) and not on the absolute
measurement of its flux. However, there is a point where the absolute measurement of the
signal is used in the pipeline: VMSP will base its estimation of the detection probability
of a random galaxy on a measurement of the pixel-level noise at the location of the galaxy
spectrum, and this is where the absolute flux calibration in dispersed images is relevant.

Even in this case, the calibration error is expected to be low. Observations will be done
by the same instrument throughout the survey, in absence of varying atmospheric
conditions, and the survey will continually visit the SelfCal calibration field
\citep{Scaramella-EP1}, making it possible to have very stable calibration throughout the
six years of the survey. The baseline requirement for fluctuations in calibration
zero-point is 0.7\% between different NISP fields, and a set of algorithms are being
devised to guarantee this calibration at the level of a single detector, of the whole
field, and at the level of `UberCalibration' \citep{Markovic2017}, where the overlap of
adjacent pointings is used to further check the stability of calibration on larger scales.
At the present stage, we can say that it is unlikely that the UberCalibration will be
worse than 1--2\%.

\subsection{Extraction of 2D spectra and wavelength calibration}
\label{app:2Dspectra}

The extraction of 2D spectra performed by SIR \citep{Q1-TP006} requires the mapping of
pixels to wavelengths; this requires wavelength calibration and correction for
non-linearities in the grism response. The formal requirement on the precision of
wavelength calibration is $0.8$ pixels, corresponding to $10.8$ {\AA}, below the spectral
point-spread function. Calibration is performed by observing planetary nebulae, which have
a very well-known spectrum dominated by emission lines, achieving sub-pixel precision.
Sometimes, a dispersed image of $530\times5$ pixels can fall across a detector gap and
thus have signal in two different the detectors; in this case the reconstruction of the
spectrum relies on knowledge of the exact geometry of detectors. The accuracy of
wavelength calibration can also be tested a posteriori by comparing the measured spectra
with ground-based ones for archive sources, where ground-based spectra have much higher
resolution and robust wavelength calibration.

Since the uncertainty in wavelength is below the spectral point-spread function, the
impact of this uncertainty is expected to be very low. Its effect would be a distortion in
the anisotropic power spectrum, which may be considered equivalent to a (negligible) bias
in the estimation of Alcock--Paczynski parameters.

\subsection{Confusion and spectral decontamination}
\label{app:confusion}

Spectral confusion is one of the most important systematics in the extraction of \Euclid's
spectroscopic catalogues. Galaxies are dispersed in $530\times5$ pixel stripes, and
photometric sources will oversample the focal plane, so each galaxy will overlap with 10
to 30 other sources. As discussed above (Table~\ref{table:number_of_sources} and
\citeapassa), only galaxies with $\HE\lesssim20$ can have a detectable continuum, so
confusion will be driven by these relatively bright objects. However each galaxy will
overlap with different objects in the four dithers, making decontamination from brighter
sources possible. The decontamination algorithm, described in \cite{Q1-TP006}, will
separate the sources, but the bright spectra will act as a background noise term for the
fainter ones. The main effect of confusion will be to increase the spectral noise
so that a galaxy that would have barely been detected in isolation may be missed when
contaminated. A less effective decontamination method would just make it more probable
that a galaxy is missed, or could create noise interlopers, with a probability that
acquires a dependence on source surface density. Because {\JE} and {\YE} magnitudes are
used to aid decontamination, its effectiveness will be affected by spectro-to-photo and
photo-to-photo persistence.

We can separate the contaminating sources in two categories: (1) stars and foreground
galaxies at redshift $z<0.9$ (or beyond $z>1.8$), and (2) galaxies at $z>0.9$. The first
class of sources is statistically independent of the spectroscopic galaxies, so their
effect can be absorbed into the class of astrophysical foregrounds that create angular
modulations of the density. As a matter of fact, VMSP accounts for of these sources, as
the noise used to estimate the variance is read directly from the NISP dispersed image.
If, however, there is a significant contamination from sources at similar redshift as the
spectroscopic sample, the assumption of statistical independence of foreground and
cosmological signal is not valid. This may be important, say, in the cores of galaxy
clusters, creating a dangerous density-dependent bias. This point is addressed in the
companion paper \citepassa, where it is shown that the relative impact of contamination
from galaxies at $z>0.9$ on the observed galaxy density amounts to around 4\%.

\subsection{Fit of the 1D spectrum, line detection, and redshift measurement}
\label{app:redshift}

Once SIR has coadded the four 2D spectra into a single calibrated 1D spectrum, SPE
\citep{Q1-TP007} has the task of fitting it to classify the source, measure its redshift
and line fluxes, and assess the reliability of the redshift measurement. The main
condition for a source to get into the spectroscopic sample will be to have a reliable
redshift measurement in the nominal redshift range of {\ha} ELGs. To maximise sample
density, emission lines down to a relatively low S/N of 3.5 are selected. However, since
the photometric sources are much more numerous than the detectable ELGs, the probability
that a noise fluctuation is misinterpreted as a spectral line is not negligible. Moreover,
since the 2D spectra of photometric sources oversample the NISP field (see the discussion
at the beginning of Sect.~\ref{sec:numberofsources}), any spurious feature in the
dispersed image that is not perfectly subtracted or masked will fall over the stripe of
some other source and may thus create a fake spectral feature. We then expect the
spectroscopic catalogue to be contaminated by a relatively high fraction of noise
interlopers.

Many sources will have a single (correctly) detected emission line. To maximise the
density of the spectroscopic sample, fits will use a prior that enhances the
interpretation that a single line is {\ha}, thus boosting completeness at the expenses of
purity.

Spectral templates play an important role in the redshift measurement. For instance, the
centroid of the {\ha}--{\nii} blended complex can be systematically offset from a standard
reference position, due to the asymmetry of the {\nii} doublet. Since the spectral
resolution is not sufficient to cleanly separate these lines, the blend can lead to a
systematic offset in the centroid, leading to a biased redshift (\citeapassa; Euclid
Collaboration: McCarthy et al., in prep.). The overall bias in the sample can be minimised
by using spectral templates that span a realistic range of {\nii}/{\ha} line ratios, and
thus of shapes of the unresolved triplet combination. The maximum centroid offset over
this range occurs when {\nii} contributes more than half of the total flux; at the extreme
we expect a redshift error of at most 0.5 pixels, of the same order as the wavelength
calibration error.

Since the fitting procedure provides a redshift for any extracted spectrum, a sample with
permissive cuts will have very low purity. Cleaning the catalogue of interlopers will be
firstly based on a spectral quality flag; this will be defined as the integral of the
normalised PDF of the redshift posterior around the main peak, a high value marking the
most reliable measurements. Furthermore, a machine-learning algorithm will be trained on
EDS galaxies to identify the region in the space of \Euclid magnitudes that corresponds to
the locus of potential {\ha} ELGs, thus cleaning out a significant fraction of
interlopers. As mentioned in Appendix~\ref{app:photometric}, usage of photometric
information can be the entry point for systematics in photometry, so this option will be
carefully tested before being used; already, \cite{Cagliari24} have shown that a cut at
$\HE<22.5$ is effective in increasing purity.

Correctly identified emission lines lead to a valid redshift measurement, which is
expected to have a statistical error with standard deviation of $\sigma_z \simeq 0.001
(1+z)$ \citep[see][]{Q1-TP007}; due to the grism response, this error does not
depend on redshift (see Sect.~\ref{sec:redshift_errors}). The effect of redshift errors is
to imprint an anisotropic smoothing to galaxy clustering, which must then be added to the
model; the exact width of the redshift error can be left as a nuisance parameter with a
tight prior, to be measured and marginalised over at the likelihood level. A
straightforward way to validate the estimation of the redshift error is to compare \Euclid
redshift measurements with ground-based archival data; this will also allow us to check
the shape of the PDF, which is not expected to be Gaussian \citep[see][]{Q1-TP007}.

\subsection{Dependence of detection on galaxy properties}
\label{app:galaxyproperties}

A whole sector of systematics comes in principle from the dependence of detection
probability on the intrinsic properties of the galaxy other than its line luminosity.
These properties determine the precise mapping from dark matter haloes to observed
galaxies, and thus determine the galaxy bias parameters. The state of a galaxy is mostly
determined by the host halo mass and its merger history, which means that the bias is
likely to have a component from assembly bias. However, our cosmological models are not
supposed to be predictive of galaxy bias parameters; these are treated as nuisance
parameters, and cosmological inference is obtained by marginalising over them. So the
question shifts to whether the parametrisation of galaxy bias used in the fitting model,
obtained from a perturbative expansion \citep{Desjacques2018} or from an HOD based on an
halo model \citep[e.g.][]{Avila2020}, is general enough to account for such an effect.
From this point of view, the most worrying potential systematics are those that are not
included in a standard bias model. One of these comes from the dependence of line
detection on the inclination of galaxy images with respect to the dispersion direction in
the grism. Galaxy inclination is known to be correlated with the large-scale tidal fields
\citep{Hirata2009,Obuljen2020,Lamman2023}, implying that there is a link between
selection, galaxy orientation, the larger scale density field, and galaxy velocity, which
is anisotropic with respect to the line of sight. This can then cause a non-local bias
term that is not present in standard models. Being this contingent on the selection,
accounting for this term requires careful modelling of the specific samples -- for \Euclid
this would involve coupling models of {\ha} emission as a function of galaxy orientation
with simulations that include the coupling between orientation and tidal fields.

\section{Effect of each case of systematics on the galaxy power spectrum}
\label{app:pk_shuffled}

\begin{figure*}
  \centering{\includegraphics[width=0.9\textwidth]{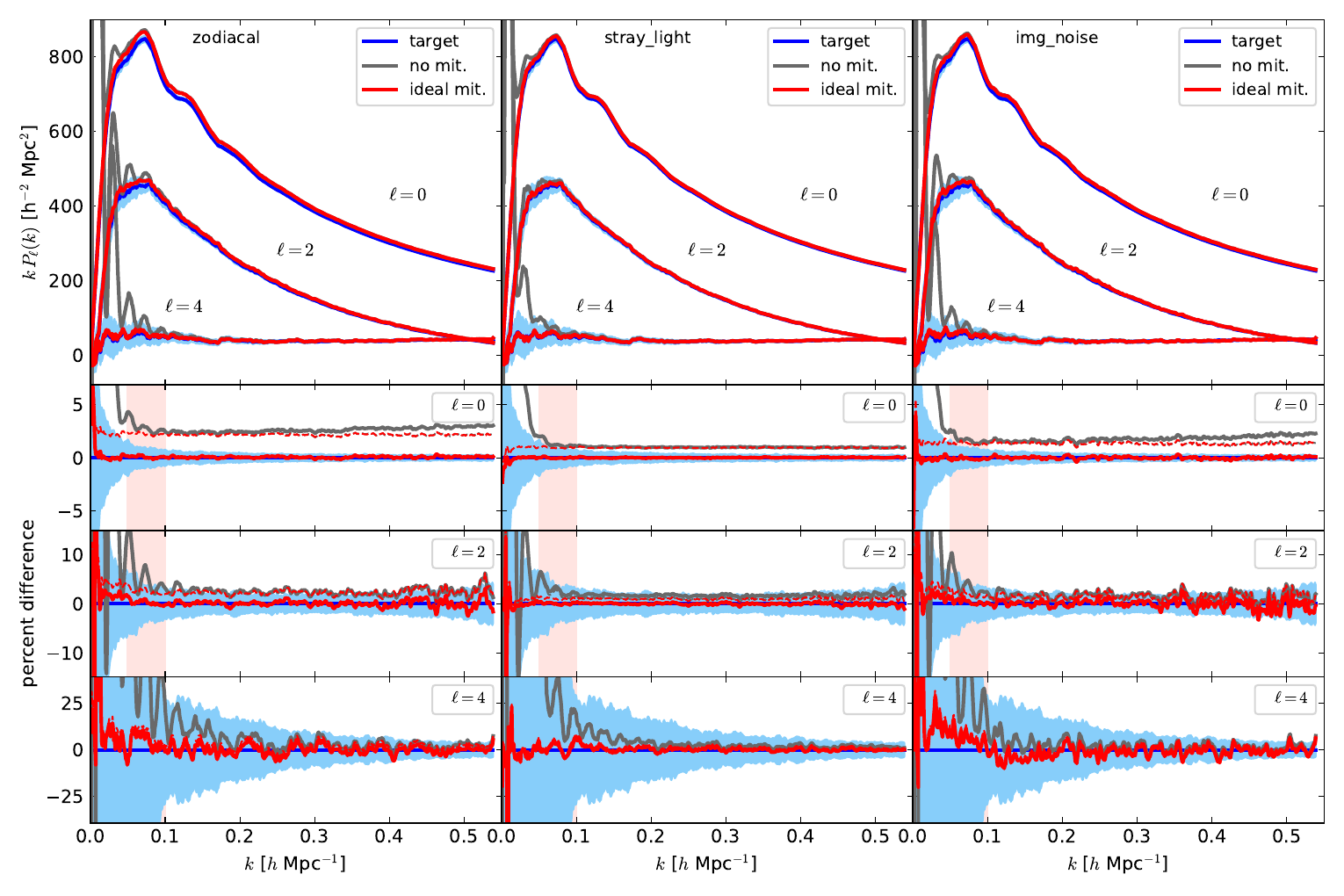}}
  \centering{\includegraphics[width=0.9\textwidth]{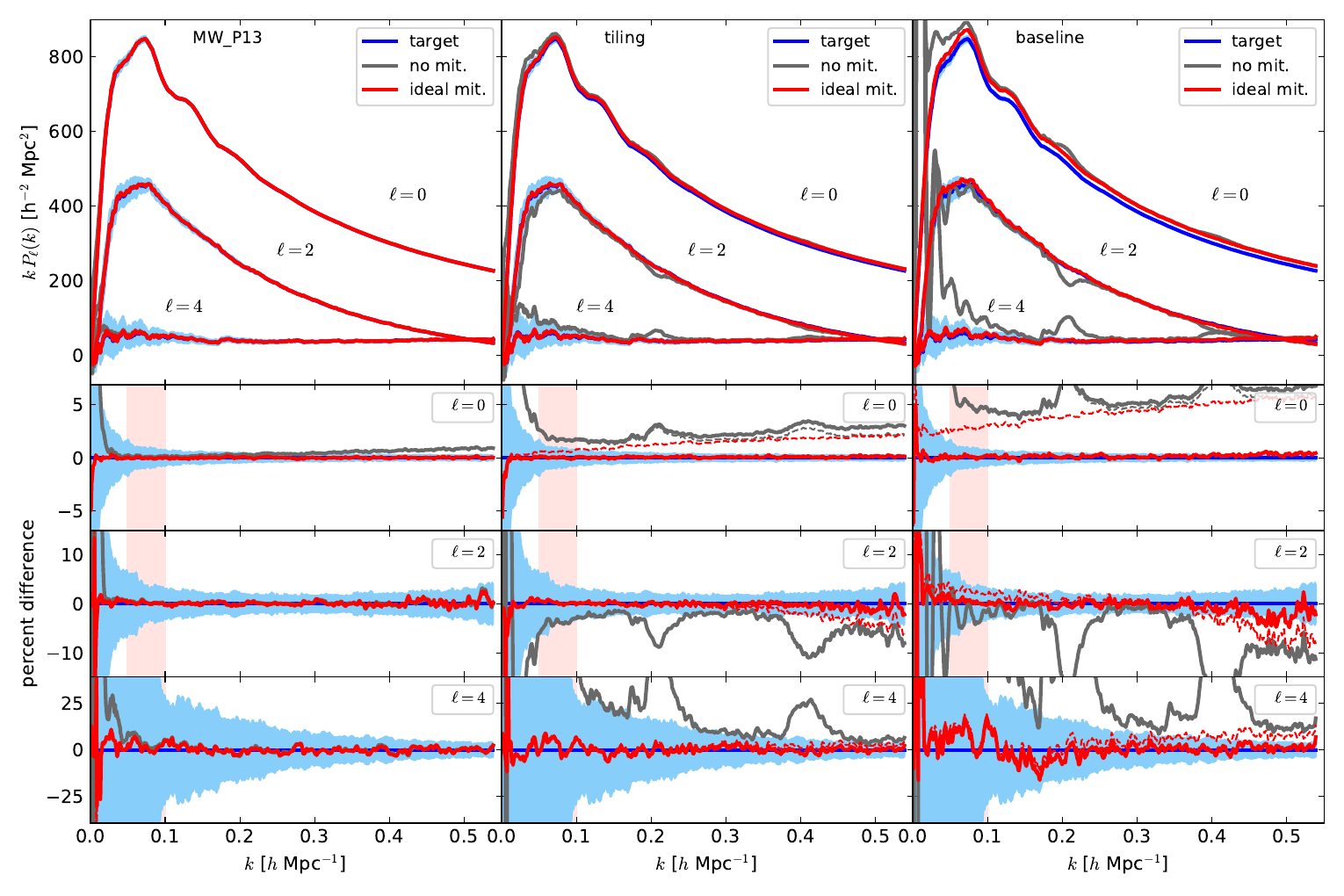}}
  \caption{Monopole, quadrupole, and hexadecapole of the galaxy power spectrum, for the six
    cases of systematics of Table~\ref{table:cases}. For each panel the layout is as in
    Fig.~\ref{fig:pkmodel}. The blue lines give the target measurement; the lighter blue
    shaded area gives the error on an average of five measurements. The red lines give the
    measurement with the systematic declared above each panel. The grey lines give the
    measurement relative to the ad hoc random catalogue with no mitigation of angular
    systematics. In the sub-panels with residuals, the thin dashed lines give the ratio of
    measurements, the thick lines the quantity $\Delta\Delta P$ (Eq.~\ref{eq:deltadeltaP}) as
    in the lower panel of Fig.~\ref{fig:pkresiduals}, all in per cent.}
  \label{fig:pk_shuffled}
\end{figure*}

In this Appendix, we present the galaxy power spectra for the six cases of systematics
identified in Sect.~\ref{sec:ideal} and listed in Table~\ref{table:cases}. They are shown
in Fig.~\ref{fig:pk_shuffled}. We use here catalogues selected using shuffled fluxes to
remove the luminosity-dependent bias. As a consequence, the convolution of the best-fit
model (Fig.~\ref{fig:pkmodel}) with the window function measured from the random catalogue
should fit the measurement obtained with ideal mitigation (as demonstrated in
Fig.~\ref{fig:pkresiduals}). Each panel shows measurements of the first three even
multipoles of $P(k)$, averaged over 50 mocks, with shot noise subtracted. All panels
report the target measurement in blue, with the lighter blue shaded area giving the
relative error of the average of groups of five mocks. The red lines give measurements
based on ideal mitigation, grey line those relative to no mitigation (also shown in
Fig.~\ref{fig:adhoc}). The three sub-panels below the main panel report, for the three
multipoles, the relative variance of the target measurements as the shaded area, the ratio
of the measurements $P_{\rm i}/P_{\rm t}-1$ (e.g. in the top panel of
Fig.~\ref{fig:pkresiduals}) as thin dashed lines, and the quantity $\Delta\Delta P$
discussed above as thicker lines, this time shown for all multipoles.

\section{Metrics for the goodness of fit in the presence of systematics}
\label{app:metrics}

To quantitatively estimate the impact of different mitigation strategies on cosmological
parameters, we employ the following metrics.

\paragraph{Figure of bias (FoB):}
The first metric we use to evaluate the performance of each mitigation procedure is the
consistency of the inferred parameters with their true values. To quantify this
consistency, we define the figure of bias as the covariance-weighted difference between
the posterior means and their fiducial values,

\begin{equation}
    {\rm FoB} = 
    \left[(\vec{\overline{\theta}}-\vec{\theta}_{\rm fid})\,S^{-1}\, (\vec{\overline{\theta}}-\vec{\theta}_{\rm fid})^T\right]^{1/2}\,,
\end{equation}

\noindent
where $\vec{\overline{\theta}}$ and $\vec{\theta}_{\rm fid}$ are vectors containing the
posterior means of the three cosmological parameters and their fiducial values, and $S$ is
the parameter covariance matrix. We compute the parameter means and covariance for the
cosmological parameters using the \texttt{getdist} package. The covariance matrix converts
the parameter differences into levels of significance, such that a value of FoB$ =
1.88\,(2.83)$ can be interpreted as corresponding to the $68\%\,(95\%)$ confidence level.

\paragraph{Figure of merit (FoM):} 
This quantifies the constraining power of the measurement over the cosmological parameters
and is computed via

\begin{equation}
    {\rm FoM} = \frac{1}{\det{S}}\,.
\end{equation}

\paragraph{Credibility interval width:} 
This is the fractional change in the average width of the parameter 68\% credibility
intervals with respect to the target catalogues,

\begin{equation}
    \delta \sigma = \frac{(\rm FoM)^{1/3}}{(\rm FoM_{\rm target})^{1/3}} -1 \,. 
\end{equation}

\paragraph{Goodness of fit:}
We quantify it by estimating the probability that the power spectrum realisations are
generated by the best-fit model. This probability, usually dubbed $p$-value, is computed
using the $\chi^2$ statistic obtained for each realisation. Specifically, for a given
mitigation strategy that we denote with an index $a$ and a realisation $i$ yielding a
value $\chi^2_{i,a}$, the $p$-value is defined as

\begin{equation}
    p_a = 1 - \frac{1}{N_{\rm real}} \sum_{i=1}^{N_{\rm real}} \int_0^{\chi^2_{i,a}} \mathcal{P}(\,\chi^2, n_\text{d})\,{\rm d}\chi^2\,,
\end{equation}

\noindent
where $\mathcal{P}(\,\chi^2, n_\text{d})$ denotes the $\chi^2$ distribution with
$n_\text{d}$ degrees of freedom. We set $0.05$ as the threshold probability, below which
we reject the hypothesis that the model is the underlying truth of the measurement in a
particular case.

\begin{figure}
\centering\includegraphics[width=.9\columnwidth]{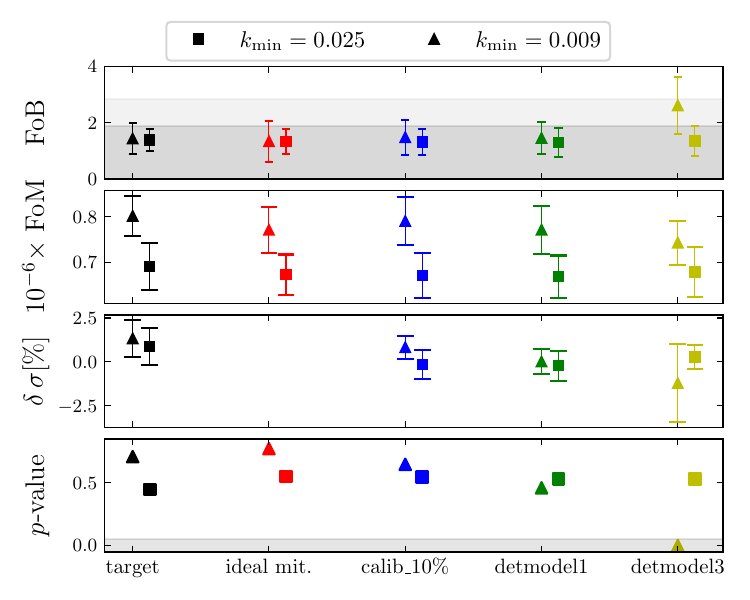}
  \caption{ \label{fig:FOB_FOM}From the top, first panel: Average and scatter of the
    FoB values for the cosmological parameters $\{h, A_{\rm s}, \omega_{\rm c}\}$,
    respectively, across the different models. We shaded the regions corresponding to 68\%
     and 95\% confidence levels. Second panel: Average and scatter of the FoM values.
    Third panel: Average and scatter of the $\delta \sigma$ values. Fourth panel: 
    $p$-value metric for the various cases; the shaded area shows the $p=0.05$ rejection
    threshold.}
\end{figure}

Figure~\ref{fig:FOB_FOM} presents the performance metrics for all five cases for both values of $k_{\rm min}$ used in this work (0.009 and 0.025 \kmpc, see Sect.~\ref{sec:fullshape}. In the top
panel, which displays the FoB values, we observe that the estimated parameters do not
exhibit significant bias for most of the mitigation strategies. However, in {\it
  detmodel3} with $k_{\rm min} = 0.009 \hmpc$, the bias exceeds the 68\% confidence threshold, indicating a systematic
deviation (evident in Fig.~\ref{fig:corner_plot}) that is absent after increasing $k_{\rm min}$. The second panel from the top
shows, through the FoM, that the mitigation strategy has little impact on the model’s
constraining power, however there is a slight decrease in the FOM using the higher $k_{\rm min}$. The third panel shows that parameter error bars are not strongly
influenced by the mitigation strategy, the largest deviations are below 5\%. Finally, in
the bottom panel the $p$-value metric indicates that the model successfully fits most
cases, except {\it detmodel3}, which is below the rejection
threshold in the low $k_{\rm min}$ case.

\end{appendix}

\label{LastPage}
\end{document}